\documentclass[nofootinbib,superscriptaddress,eqsecnum,tightenlines,11pt,longbibliography,notitlepage]{revtex4}

\usepackage{hyperref}
\usepackage{graphicx}
\usepackage{amsmath,amssymb,amsfonts,amsthm,stmaryrd,mathtools,mathtools}
\usepackage{yfonts}
\usepackage{multirow,bigdelim}
\usepackage{dsfont}
\usepackage{color}
\usepackage{graphicx,color,epstopdf,epsfig,psfrag}
\usepackage[dvipsnames]{xcolor}
\usepackage{ mathrsfs }
\usepackage{tikz}
\usetikzlibrary{calc}
\usetikzlibrary{decorations.pathmorphing}
\usetikzlibrary{shapes.geometric}
\usetikzlibrary{arrows,decorations.markings}

\hypersetup{
    colorlinks=true,       
    linkcolor=MidnightBlue,          
    citecolor=BrickRed,        
    filecolor=BrickRed,      
    urlcolor=BrickRed           
}
 
\makeatletter
\newcounter{subsubsubsection}[subsubsection]
\renewcommand\thesubsubsubsection{\thesubsubsection .\@alph\c@subsubsubsection}
\newcommand\subsubsubsection{\@startsection{subsubsubsection}{4}{\z@}%
                                     {-3.25ex\@plus -1ex \@minus -.2ex}%
                                     {1.5ex \@plus .2ex}%
                                     {\centering\normalfont\small\textit}}
\newcommand*\l@subsubsubsection{\@dashedtocline{3}{10.0em}{4.1em}}
\newcommand*{\subsubsubsectionmark}[1]{}
\makeatother

\newcommand{\acite}[1]{[{\color{MidnightBlue}\texttt{#1}}]}

\numberwithin{equation}{section}
\numberwithin{paragraph}{subsection}

\newcommand{\E}{\mathrm{e}}
\newcommand{\I}{\mathrm{i}}
\renewcommand{\Im}{{\mathrm{Im}}}

\newcommand{\C}{{\mathbb C}}
\newcommand{\N}{{\mathbb N}}
\newcommand{\Q}{{\mathbb Q}}
\newcommand{\R}{{\mathbb R}}

\newcommand{\cF}{{\mathcal F}}
\newcommand{\cG}{{\mathcal G}}
\newcommand{\cI}{{\mathcal I}}

\newcommand{\cH}{{\mathcal H}}

\newcommand{\cP}{{\mathcal P}}
\newcommand{\cT}{{\mathcal T}}

\newcommand{\cC}{{\mathcal C}}

\newcommand{\SU}{\mathrm{SU}}

\newcommand{\SL}{\mathrm{SL}}
\newcommand{\SO}{\mathrm{SO}}

\renewcommand{\d}{{\mathrm{d}}}

\newcommand{\be}{\begin{equation}}
\newcommand{\ee}{\end{equation}}
\newcommand{\beq}{\begin{eqnarray}}
\newcommand{\eeq}{\end{eqnarray}}
\newcommand{\bes}{\begin{eqnarray}}
\newcommand{\ees}{\end{eqnarray}}

\newcommand{\mat} [2] {\left ( \begin{array}{#1}#2\end{array} \right ) }

\newcommand{\su}{{\mathfrak{su}}}

\newcommand{\W}{W}

\newcommand{\la}{\langle}
\newcommand{\ra}{\rangle}

\newcommand{\Tr}{{\mathrm{Tr}}}
\newcommand{\f}{\frac}
\newcommand{\tl}{\widetilde}
\renewcommand{\bar}{\overline}
\def\nn{\nonumber}
\def\pp{\partial}

\def\eps{\epsilon}

\newcommand{\id}{\mathbb{I}}

\def\om{\omega}

\newcommand{\G}{{\Gamma}}

\newcommand{\up}{{\,\uparrow\,}}
\newcommand{\down}{{\,\downarrow\,}}

\renewcommand{\hat}{\widehat}

\newcommand{\ellpl}{\ell_\text{Pl}}
\newcommand{\ellcc}{\ell_\text{cc}}

\renewcommand{\sinh}{\mathrm{sh}}
\renewcommand{\cosh}{\mathrm{ch}}

\def\vJ{\vec{J}}

\theoremstyle{definition}

\theoremstyle{remark}

\begin{document}
\title{\LARGE Quasi-local holographic dualities\\ in non-perturbative 3d quantum gravity
 \\\vspace{.5em} \large  I - Convergence of multiple approaches and examples of Ponzano--Regge statistical duals}

\author{{\bf Bianca Dittrich}}\email{bdittrich@perimeterinstitute.ca}
\affiliation{Perimeter Institute, 31 Caroline St North, Waterloo ON, Canada N2L 2Y5}

\author{{\bf Christophe Goeller}}\email{christophe.goeller@ens-lyon.fr}
\affiliation{Laboratoire de Physique, ENS Lyon, CNRS-UMR 5672, 46 all\'ee d'Italie, Lyon 69007, France}
\affiliation{Perimeter Institute, 31 Caroline St North, Waterloo ON, Canada N2L 2Y5}

\author{{\bf Etera R. Livine}}\email{etera.livine@ens-lyon.fr}
\affiliation{Laboratoire de Physique, ENS Lyon, CNRS-UMR 5672, 46 all\'ee d'Italie, Lyon 69007, France}
\affiliation{Perimeter Institute, 31 Caroline St North, Waterloo ON, Canada N2L 2Y5}

\author{{\bf Aldo Riello}}\email{ariello@perimeterinstitute.ca}
\affiliation{Perimeter Institute, 31 Caroline St North, Waterloo ON, Canada N2L 2Y5}

\date{\today}

\begin{abstract}

\noindent
This is the first of a series of papers dedicated to the study of the partition function of  three-dimensional quantum gravity on the twisted solid torus with the aim to  deepen our understanding of  holographic dualities from a non-perturbative quantum gravity perspective. Our aim is to compare the Ponzano--Regge model for non-perturbative three-dimensional  quantum gravity with the previous perturbative calculations of this partition function. We begin by reviewing the results obtained in the past ten years via a wealth of different approaches, and  then introduce the Ponzano--Regge model  in a self-contained way. Thanks to the topological nature of three-dimensional quantum gravity we can solve exactly for the bulk degrees of freedom and identify dual boundary theories which depend on the choice of boundary states, that can also describe finite, non-asymptotic boundaries.
This series of papers aims precisely at the investigation of the role played by  the different quantum boundary conditions leading to different boundary theories. Here, we will describe the spin network boundary states for the Ponzano--Regge model on the twisted torus and derive the general expression for the corresponding partition functions. 
We identify a class of boundary states describing a tessellation with maximally fuzzy squares for which the partition function can be explicitly evaluated. In the limit case of a large, but finely discretized,  boundary  we find a dependence on the Dehn twist angle characteristic for the BMS$_3$ character. We furthermore show how certain choices of boundary states lead to known statistical models as dual field theories -- but with a twist. 

\end{abstract}

\maketitle

\newpage

\small
\tableofcontents
\normalsize
\newpage

\section{Introduction}

This is the first of a series of papers exploring the holographic nature of 3d quantum gravity.
We focus on developing a framework for understanding and studying holographic dualities in the  quasi-local and non-perturbative context provided by the Ponzano--Regge model.
\footnote{To streamline this introduction, we omit all citations, which will anyhow appear in the subsequent review section.}
First defined in 1968 [sic], the latter is an instantiation of $BF$ topological quantum field theory, and it has been rigorously related to other approaches to 3d quantum gravity, notably to the combinatorial quantization of Chern--Simons theory and to Loop Quantum Gravity.

Being formulated in terms of a local state-sum, the Ponzano--Regge model allows one to compute the amplitude of quantum gravitational processes within {\it finite}, i.e. quasi-local, regions. This is to be contrasted to the AdS/CFT framework,  which intrinsically refers to the asymptotic boundary of AdS. 
It also implies, again differently from the AdS/CFT philosophy, that each amplitude is associated to one given spacetime topology, just as in Chern--Simons theory. On the torus, this procedure of course breaks modular invariance at the onset, since the choice of bulk manifold selects which cycle is contractible and which is not. As we will see, this will be reflected in the calculation of the amplitude. Of course, the model can be enhanced via some prescription to sum over different topologies or by restoring more directly modular invariance of the boundary theory. Summing over topologies can be done either ``by hand'' or e.g. via a Group Field Theory approach, which uses the technology of matrix and tensor models to generate discretized manifolds any spacetime dimensions. We restrain from investigating these possibilities here. 
 
As a consequence of the topological nature of 3d quantum gravity, the bulk variables can be exactly integrated out, possibly---as we just emphasized---modulo some global combinations thereof, e.g. monodromies, hence leaving us with some purely two-dimensional object.
It is this object, which we will interpret as the dual boundary theory. 
Manifestly, it will directly depend on the chosen boundary conditions encoded in a boundary state that has to be chosen for the calculation of the quasi-local amplitude.

Again, this has to be contrasted with the contemporary AdS/CFT philosophy, where different boundary CFTs defined at asymptotic infinity are interpreted as giving by duality different definitions of bulk quantum gravity theories. In our case, the quantum gravity theory is instead considered to be known, and the dual theory---not necessarily a CFT---emerges at a finite boundary as a reflection of given boundary condition.\footnote{Notice that at asymptotic infinity ``natural'' boundary conditions emerge.} This situation is, indeed, completely analogous to the way Wess--Zumino--Novikov--Witten and Liouville theories emerge on the boundary of a Chern--Simons theory.

Thus, this series of paper is dedicated to exploring different boundary states and the induced dual theories. 
The Ponzano--Regge model, being inherently discrete, requires two types of choices when defining a boundary state: the choice of an underlying discretization, or graph, and that of a family of states supported on these graphs. 
For what concerns the first choice, so far, the underlying discretization will be fixed to that of a homogeneous quadrangulation of the 2d torus, and leave to future investigations the study of alternative choice and of their local perturbations (dislocations, addition of links, etc).
Our first papers will focus on the second type of choice and on the ensuing interplay between the geometrical meaningfulness of the boundary states and the desirability of the dual theory it induces.

In particular, in this paper we will present the case of boundary states describing a boundary geometry with the smallest possible discretization scale. Such states can be defined using so-called spin networks states specialized to spin labels all taking the smallest possible value $j=1/2$. This corresponds to the length $\ell_\text{min} = 4\sqrt{3}\pi G_\text{N}\hbar$. As we will see this choice still leaves a considerable freedom in completely specifying the states. This freedom is related to the choice of four-valent intertwiners, that we will suppose  to be homogeneous. We will prove that the resulting states map onto a 2d  six-vertex (or ``ice-type'') statistical model, which in the case at hand is integrable. The deep implications of this fact---e.g. the geometrical meaning of boundary phase transitions---will be investigated in a later paper of the series.

We will also compute in detail the case of a homogeneous lattice characterized by two spins and a specifically degenerate intertwiner. Beside computability, these states have the curious property of reproducing---in some appropriate sense---some characteristic divergences of the partition function of three dimensional gravity.%

The second paper of the series\footnote{ We will refer to this second paper \cite{Part2} in the series as Part II trough out this text.}, which is made available at the same time as the present one, will focus instead on coherent spin-network states in the large-spin limit. Hence, in a sense, the second paper focuses on the very opposite regime of the one considered here in the $j=1/2$ calculation. The large-spin limit is arguably the most well-studied regime of spinfoams and spin-networks, both in Mathematics and Physics.  It describes the  semiclassical limit of the quantum gravitational model at large discretization scale. 
 We will derive the dual boundary theory also for these states. This time it will be accurately described by the semiclassical regime of a (peculiar) $\SU(2)$ non-linear sigma model. The accuracy of this continuous description can be understood as the result of the large number of (magnetic-index) degrees of freedom that get involved in these boundary states.

Finally, as a way to cross-check our results,  we compare in all cases the partition functions we obtain on the solid torus with the same quantity calculated in other approaches, ranging from direct path integral calculations in metric General Relativity (in both AdS$_3$ and Minkowski$_3$ spaces), to holographic duals in terms of characters of representations of the relevant boundary symmetry groups (i.e. Virasoro and BMS$_3$, respectively), as well as to the perturbative discrete approach known as linearized quantum Regge calculus.

Since the way these results match is quite subtle and highly non-trivial, we dedicate a long section of this first paper to the review of all these other methods. 
We invite the reader to skip whatever they are familiar with, since the second part of the paper is essentially independent of this introduction, although the conclusions will---of course---rely on it. 
Similarly, Part II can be read independently from Part I, although, again, reference to the review part will be constantly made throughout the series whenever necessary.\\

Here is a roadmap of the present paper.

In section \ref{sec_birdeye}, we start with a bird's eye review of 3d quantum gravity on the torus, focusing on different techniques to compute its partition function, while ignoring pretty much everything else. We discuss both the AdS$_3$ case as well as its flat-space limit. A good deal of attention will be given to the role played by different boundary conditions. At the end of this section, we will dedicate quite some room to the discussion of the perturbative quantum Regge calculus approach. This is not only because we deem this approach might be the least familiar to the reader, but also because it is the only other discrete approach to the present problem, beside our work, and comparison between the two will lead to possibly unexpected conclusions. And furthermore, some of the discussion about Regge calculus and its dual theory is actually original.

In section \ref{sec_PRreview}, the reader can find a short overview of the Ponzano--Regge model. We invite the reader familiar with it to skip this section altogether.

In section \ref{sec_bdrystate}, we introduce the spin-network boundary states for the Ponzano--Regge model, and we will discuss their interpretation in terms of which boundary conditions they impose to the bulk gravitational state sum. We will also focus on the geometry encoded specifically by a 4-valent spin-network dual to a quadrangulation.  Readers familiar with spin-network states and their geometric interpretation in terms of piecewise flat (2d) geometries, can safely skip the first part of this section, too.

Sections \ref{sec_PRreview} and \ref{sec_bdrystate} were written to make this series of papers as self-contained as possible, keeping specifically in mind those readers who might be unfamiliar with spin-network and Ponzano--Regge techniques.
 In fact, we suspect that readers familiar with the material reviewed in section \ref{sec_birdeye}, will not be as familiar with the material reviewed in sections \ref{sec_PRreview} and \ref{sec_bdrystate}. And vice versa.

Section \ref{sec_dualtheoriesgeneralities} is a very short section discussing in  general terms what sort of dual boundary theories one can expect to arise from the Ponzano--Regge model. 

The core of the paper is section  \ref{sectionQu5}. There, we start by deriving the general formula for  Ponzano--Regge amplitudes on the  quadrangulated torus, we then move on to the analysis of the spin-network characterized by two spins and a peculiar intertwiner (spin $0$ in $s$-channel), and finally conclude with the correspondence of general spin $1/2$ spin-network states and a 2d statistical model.

The paper ends with a discussion and outlook section, that is section \ref{sec_conclusions}.

As a complement, there are also five appendices. These are: appendix \ref{app_onshellactions}, dealing with the geometries of thermal AdS$_3$ and of the BTZ (Ba\~nados--Teitelboim--Zanelli) black hole, as well as a discussion of the on-shell value of their gravitational actions, for different boundary conditions; appendix \ref{app_gaugefixing}, where a detailed example of (bulk) gauge fixing of the Ponzano--Regge partition function on the torus is worked out; appendix \ref{app_spinintertwiner}, on the recoupling theory of spin 1/2 and spin 1 underlying the space of four-valent intertwiners; appendix \ref{app_partitionfunction} detailing the exact computation of the partition function for the $0$-spin intertiwiner in the $s$-channel and appendix \ref{App_onehalf} explaining the combinatorics involved in evaluation of the partition function for the spin $1/2$ boundary state.

\section{A bird's-eye review of three dimensional quantum gravity on the torus\label{sec_birdeye}}

Three dimensional gravity has been one of theorists' favourite practising terrains for developing and testing new ideas about quantum gravity \cite{Carlip_book}.
A certain number of approaches to its quantization have been developed over the last thirty years and---despite the remarkable variety of techniques---it is fair to claim a remarkable overall convergence and agreements of results.
The fact that there are several successful approaches to three dimensional quantum gravity,  and their convergence, is for sure a consequence of the lack of local degrees of freedom in this theory. It nevertheless remains that all these approaches provide us with invaluable insights in the possible structure of four-dimensional quantum gravity and geometry.
A first model of three-dimensional quantum gravity already appeared in 1968, by Ponzano and Regge, formulated as a discrete state-sum model \cite{PR1968}.
But the first successful formulation of three dimensional quantum gravity as a quantum field theory only appeared much later, in work by Witten \cite{Witten1988}. He reformulated first-order gravity in terms of the Chern--Simons TQFT (acronym of Topological Quantum Field Theory, \cite{WittenTQFT1988,AtiyahTQFT1988}) and showed how to extract expectation values of Wilson loop observables from a formal path integral formulation of that theory.
Thereafter, a long series of works put Witten's findings on firmer mathematical ground and set the foundations of other approaches.

In \cite{ReshetikhinTuraev1991}, Reshetikhin and Turaev provided a mathematically rigorous formulation of Witten's Chern--Simons TQFT (which actually dispensed all-together with the path integral picture).
At the same time, Turaev and Viro introduced a new way of constructing ``quantum'' invariants of three manifolds \cite{TuraevViro1992}. Known as the Turaev--Viro model, this invariant is formulated in terms of a state-sum model, i.e. as the sum over certain representation--theoretical weights one can associate to a given simplicial decomposition of the manifold under study.
These two ways of building invariants of three manifolds have been shown to be essentially equivalent first by Walker \cite{Walker1991}, and then more straightforwardly by Roberts \cite{Roberts1995} using skein-theory techniques developed by Kauffman and Lins, and Lickorish.

In a spirit similar to Witten's, Horowitz \cite{Horowitz1989} discussed the quantization of another class of TQFTs , ever since known as $BF$ theories \cite{BlauThompson1989}, to which three-dimensional first-order gravity belongs. Such theories provide today the basis of the spinfoam approach to quantum gravity (e.g. \cite{Freidel:1998pt,Oriti:2003wf,Livine:2010zx,Perez:2012wv} for reviews) initiated as a covariant version of Loop Quantum Gravity (LQG) by Reisenberger and Rovelli, and Baez \cite{Reisenberger:1996pu,Baez:1997zt}.
In three space-time dimensions,  first order graivity {\it is} a $BF$ theory, and the relevant spinfoam model is the Ponzano-Regge model \cite{Rovelli:1993kc}. Its relation to Hamiltonian LQG was  rigorously shown in \cite{Noui:2004iy}, and also its relation to Chern--Simons theory was shown to hold both at the covariant level---where the Ponzano--Regge model\footnotemark~is formally a (mathematically degenerate) limit of the Turaev--Viro model which admits a Reshetikhin--Turaev-like formulation too \cite{FreidelLouapre2004b}---and at the canonical one---where its relation to combinatorial quantization was proven even in presence of particles (defects) \cite{FreidelLouapre2004,NouiPerez2005,Noui2006,MeusburgerNoui2010,Meusburger2016,Delcamp:2016yix}.
\footnotetext{
Proposals for a viable definition of the Ponzano--Regge model {\it per se} have been put forward and subsequently refined by a series of authors  \cite{FreidelLouapre2004,Freidel:2005bb,BarrettNaishgusman2008,BonzomSmerlak2010,BonzomSmerlak2012}.
}

The Ponzano-Regge proposal as a model for 3d quantum gravity was somewhat fortuitous: while studying the asymptotics of Wigner $3nj$-symbols from the recoupling theory of $\SU(2)$ representations, Ponzano and Regge realized that in the limit of large spins, the $6j$-symbol reproduced the complex exponential of a discrete version of the (boundary) action of General Relativity for a (flat) tetrahedron with edge lengths quantized in Planck units. This action had been proposed only a few years before by Regge himself \cite{Regge1961}, in the context of a discrete formulation of general relativity, now commonly known as Regge calculus \cite{HartleSorkin1981}.

Soon after its appearance, Mizoguchi and Tada \cite{MizoguchiTada1991} suggested that---in an analogous asymptotic limit---the Turaev--Viro model was related to a version of the Regge action involving a cosmological term proportional to the tetrahedron's volume \cite{BahrDittrichNewRegge}.
This was rigorously proven by Taylor and Woodward, who showed that the asymptotics of the Turaev--Viro model involves homogeneously curved tetrahedra \cite{TaylorWoodward2003}.
In presence of a cosmological constant, the status of the network of correspondences delineated above is still work in progress \cite{Freidel:1998ua,Noui:2011im,Pranzetti:2014xva,Dupuis:2013lka,Bonzom:2014wva,Bonzom:2014bua,Dupuis:2014fya,Livine:2016vhl,Dittrich:2016typ}.

During this condensed survey of approaches to 3d quantum gravity, we have mostly stressed those which are either first-order (Palatini), such as $BF$ theories and loop gravity, or based solely on connection variables, such as Chern--Simons theory\footnote{Notice, however, that the Chern--Simons connection is {\it not} the spin-connection, but a linear combination of the spin-connection and the conjugated dreibein field.}. 
This is arguably because the Wheeler-DeWitt equation of three dimensional gravity with vanishing cosmological constant is nothing but a flatness constraint, which is most easily dealt with in terms of connection variables.
Notice, however, that the discrete and combinatorial models by Turaev and Viro, and Ponzano and Regge (PR), are best interpreted in terms of a (discrete) metric formulation: indeed, the representation-theoretic labels one is summing over, i.e. representations of the $\SU(2)$ Lie group or ``quantum'' deformations thereof, can be understood as (discrete) lengths variables according to the semiclassical correspondence $l \sim( j + \f12)\,l_\mathrm{Planck}$ with $j$ the spin labeling the $\SU(2)$ irreducible representations. In this sense, the PR state-sum model is nothing but a discretized version of Hawking's integral over spacetime metrics.

Such discrete models, however, are of course not the only treatments of three dimensional quantum gravity in metric variables. See e.g. \cite{IizukaTanakaTerashima2015,HondaIizukaTanakaTerashima2015,BarnichGonzalezMaloneyOblak2015}. We will come back on some of these results in a few paragraphs.

But what ``induces'', already at the classical level, a specific choice of variables?
This is an important question, since different choices of variables naturally lead to different preferred quantization schemes. From a mathematical perspective, the answer is clear and is ``{boundary conditions}''.
\footnote{To be precise, when passing from a metric to a viel-bein formulation, changing boundary conditions is not enough: the two have different field spaces and technically are quite different theories, see e.g. \cite{Matschull1999} for a discussion of this point.} 
We should ask, which quantity is kept fixed on the boundary when varying the action, or which boundary fields are kept fixed when performing the path integral on a manifold with boundary, or which polarization---or basis of observables---is chosen in the quantum mechanical formulation. All this viewpoints are intimately related to one another.

\subsection{Boundary conditions \label{sec_bdrycond}}

In the metric formalism, the natural quantity to be kept constant at the boundary is the induced metric. The corresponding boundary term is the Gibbons--Hawking--York (GHY) term \cite{GibbonsHawking1977,York1986}  possibly augmented with local counterterms depending on the induced metric only. E.g. in $\text{AdS}_3$ the ``natural'' action is \cite{Carlip2005}
\be
S_{\text{GHY}} = \frac{1}{2\ell_\text{Pl}} \left[ \int \sqrt{g} \left(R + \frac{2}{\ell_\text{cc}^2}\right) + 2 \oint \sqrt{h} \left(K - \frac{1}{\ell_\text{cc}}\right) \right] ,
\label{eq_GHY}
\ee
where $\ellpl = 8 \pi G_\text{N}$ is the Planck length in three dimensions ($\hbar =1$) and $\Lambda = - 1/\ellcc^2$ the cosmolgoical constant, $g_{\mu\nu}$ is the three dimensional metric, $g$ its determinant and $R$ its Ricci scalar curvature; $h_{\mu\nu}$ is the pullback of $g_{\mu\nu}$ on the boundary surface and $K$ the (trace of) the extrinsic curvature thereof. 

Conversely, in a first order $BF$-like formulation, no boundary term is needed if it is the pullback of the connection to be kept fixed at the boundary,
\be
S_{BF|\omega} = \frac{1}{2\ell_\text{Pl}} \int e_a\wedge F^a[\omega] + \frac{1}{3\ell_c^2} \epsilon_{abc} e^a \wedge e^b \wedge e^c,
\ee 
while a term numerically equivalent to the GHY one---provided an appropriate gauge is chosen
\footnote{I.e. a gauge such that $\pp_\mu n^a = 0$, where $n^a = e^a_\mu n^\mu$ and $n^\mu$ is the tangent-space unit vector orthogonal to the boundary. See e.g. \cite{Corichi:2016zac}.}
---is needed if it is the pullback of the dreibein to be kept fixed,
\be
S_{BF|e} = \frac{1}{2\ell_\text{Pl}} \left[ \int e_a\wedge F^a[\omega] + \frac{1}{3\ell_c^2} \epsilon_{abc} e^a \wedge e^b \wedge e^c -  \oint e_a\wedge \omega^a \right]
 \;\hat = \; S_{\text{GHY}}.
\ee
Here, $e^a = e^a_\mu\d x^\mu$,  $\omega^a = \tfrac12 \epsilon^a{}_{bc}\omega^{bc}_\mu \d x^\mu$ and $F^a[\omega] = \d\omega^a + \tfrac12 \epsilon^{a}_{bc}\omega^b\wedge \omega^c$ are the dreibein one form, the spin-connection one-form, and the curvature two-form, respectively. The index $a$ is a tangent space index to the spacetime manifold, in agreement with the standard relation between the dreibein and the metric, i.e. $g_{\mu\nu}=\eta_{ab}e^a_\mu e^b_\nu$, where $\eta_{ab}$ is the flat metric of signature equal to that of $g_{\mu\nu}$.

Finally, in the Chern--Simons formulation, the action reads
\be
S_{\text{CS}} =  \I W[A] - \I W[\bar A]
\qquad\text{with}\qquad
A = \omega^a + \I \frac{e^a}{\ell_c},
\ee
where $A$ is the Chern--Simons one-form and $W[A]$ the Chern--Simons action functional,
\be
W[A] = \frac{k_\text{CS}}{4\pi} \int A_a \wedge \d A^a + \frac23 \epsilon_{abc} A^a\wedge A^b \wedge A^c
\qquad\text{where}\qquad
k_\text{CS} = 2\pi\frac{\ell_\text{cc}}{\ell_\text{Pl}}.
\ee
Translated in dreibein and connection variables, the action $S_\text{CS}$ features a boundary term which is {\it one-half} of the one appearing in $S_{BF|e}$.
The boundary conditions this action implies are more subtle, and naturally lead to a holomorphic (or antiholomorphic) Fock--Bargman-like representation.
In \cite{BanadosMendez1998}, Ba\~nados and Mendez argued that this set of boundary conditions corresponds to a {\it canonical ensemble} for the three-dimensional BTZ black hole in AdS$_3$ \cite{BanadosTeitelboimZanelli1992,Banados:1992gq}, where it is its inverse temperature and angular velocity to be kept fixed instead of some induced boundary metric.
They hence define the action
\be
\; S_{\frac12\text{GHY}} 
= \frac{1}{2\ell_\text{Pl}}\left[ \int \sqrt{g} \left(R + \frac{2}{\ell_\text{Pl}^2} \right) +  \oint\sqrt{h} \; K \right] \;\hat=\; S_{\text{CS}}.
\ee

Both $S_\text{GHY}$ and $S_{\frac12\text{GHY}}$ are finite when evaluated on the (Euclidean) BTZ spacetime and their on-shell values do, curiously, coincide.
The same is observed on the thermal $\text{AdS}_3$ space considered in the following sections.

Interestingly, it turns out that the on-shell value of $S_\text{GHY}$ is sensitive to the order in which the large radius and flat spacetime limit $\ellcc\to\infty$ are taken.
\footnote{This is due to the parametric disappearance of the counterterm proportional to the boundary area in $S_\text{GHY}$. }
Not so $S_{\frac12\text{GHY}}$.
For this reason in \ref{sec_flatlimit} we will refer to the action $S_{\frac12\text{GHY}}$ only. 
As we will see, it is this version of the gravitational action---also leading to finite results in the context of twisted thermal%
\footnote{This is implicitly proved in \cite{Carlip2005}.}
$\text{AdS}_3$---which is consistent with boundary CFT-theoretical constructions.
In appendix \ref{app_onshellactions} we provide elementary calculations of these on-shell actions.

\subsection{Twisted thermal $\text{AdS}_3$}

The status of dual boundary theories of three dimensional gravity is most thoroughly developed in the case of negative cosmological constant and asymptotic boundaries, where it can also be seen as a very particular case of the AdS/CFT duality \cite{Witten1998}.
In such a case, the emerging dual field theory is a specific conformal field theory (CFT), known as Liouville theory \cite{Carlip2005Review}.
The first evidence for the emergence of a conformal field theory at the asymptotic boundary of $\text{AdS}_3$ was given back in 1986 by Brown and Henneaux \cite{BrownHenneaux1986}. More specifically, they built on work by Regge and Teitelboim \cite{ReggeTeitelboim1974}, to show how the asymptotic symmetries respect a Virasoro algebra with central charge $c = 12 \pi \ell_\text{cc}/\ell_\text{Pl}$, with $\Lambda=-1/\ell_\text{cc}^2$ the negative cosmological constant.
A concrete construction of the Liouville field in terms of ``broken radial diffeomorphisms'' at the asymptotic boundary was provided much later by Carlip \cite{Carlip2005} .

On a completely general ground, however, it is possible to show that Chern--Simons theory's boundary dual  theory is a non-linear sigma model conformal field theory, known as the Wess--Zumino--Novikov--Witten model \cite{Wess:1971yu,Witten:1983tw,Knizhnik:1984nr,Gawedzki:1999bq}. This, in turn, reduces to a Liouville field theory once its field content is appropriately constrained so to encode Brown--Henneaux asymptotic conditions (see the review \cite{Carlip2005Review}, as well as \cite{Banados1999} for a brief pedagogical discussion). 

Thus, there is therefore a convergence of results between the metric and Chern--Simons approaches for what concerns the dual field theory at the boundary of $\text{AdS}_3$, up to certain details which can possibly be traced  back to the use of different boundary conditions (i.e. the two theories lead to different Liouville potential, see \cite{Carlip2005, Carlip2005Review}).
 
So far, we have discussed a purely classical convergence of results.
There is, however, at least one important piece of evidence that this convergence holds beyond the classical regime. 

\subsubsection{Metric variables}

In \cite{GiombiMaloneyYin2008}, the one-loop perturbative partition function was computed for metric (Einstein--Hilbert) three dimensional quantum gravity with a negative cosmological constant. 
More precisely, this was done in the setting of the thermal partition function---at inverse temperature $\beta$---of Euclidean three dimensional quantum general relativity. In particular, they allowed the $\beta$-periodic boundary condition to involve an angular twist $\gamma\in[0,2\pi)$ applied before the identification of the Cauchy surfaces at Euclidean (Killing) times\footnotemark~$t$ and $t+\beta/ \ell_\text{cc}$.
The space we just described is called ``Twisted Thermal AdS$_3$'', or TTAdS$_3$ for short, and its relation to the BTZ black hole is detailed in appendix \ref{app_TAdS_BTZ}.%
\footnotetext{Notice that in our conventions all coordinates are dimensionless. On the other hand, $\beta$ is dimensionful, and has dimensions of inverse energy or, equivalently, of a length.} 

Put in other words, the authors of \cite{GiombiMaloneyYin2008}  computed the partition function of gravity on a solid torus of modulus
\be
\tau = \frac{1}{2\pi}\left(\gamma +\I \frac{\beta}{\ell_\text{cc}} \right),
\ee
obtained as the quotient of the three-dimensional hyperbolid
\be
\d s^2 = \ell_\text{cc}^2 \left(\; \cosh^2 r\; \d t^2 +  \sinh^2 r\; \d \phi^2 + \d r^2 \;\right),
\ee
with $t\in \mathbb{R}$, $r \in \mathbb R^+$ and $\phi\in[0,2\pi]$, under the identifications 
\be
(r, \phi, t) \sim (r, \phi + \gamma, t +  \beta/\ell_\text{cc}).
\label{eq_identification}
\ee

The result of their calculation is remarkable. 
It will be at the center of the investigations of the present series of papers.
Thus, we are now going to present it in some detail.

The 1-loop partition function---sometimes claimed to be perturbatively exact\footnotemark~ \cite{GiombiMaloneyYin2008} ---involves a classical contribution together with a combination of functional determinants, among which there is a combination of Faddeev--Popov determinants for the scalar and vector gauge modes. The ensuing result, after some non-trivial algebraic simplifications between scalar, vector and tensor (graviton) mode contributions, is found to be 
\be
Z_\text{TTAdS}(\tau,\bar\tau) = \E^{ -S_\text{TTAdS}} Z^\text{1-loop}_\text{TTAdS}(\tau,\bar\tau)
\qquad\text{with}\qquad
Z^\text{1-loop}_\text{TTAdS}(\tau,\bar\tau) = \prod_{p=2}^\infty \f1{\left|1 - \E^{2\pi\I \tau \cdot p} \right|^{2}}.
\label{eq_AdS_1loop}
\ee
Here,  $S_\text{TTAdS}$ is the on-shell evaluation of Einstein--Hilbert action on $\text{TTAdS}_3$. In appendix \ref{app_onshellactions}, it is shown that in the appropriate ``infinite radius'' limit, the on-shell values of the actions $S_\text{GHY}$ and $S_{\frac12\text{GHY}}$ are both finite and equal.
\footnotetext{ Of course, metric three-dimensional general relativity is perturbatively non-renormalizable. Therefore, this claim should be carefully understood. In this paper we do not dwell with this fact any further, since this series of works is precisely concerned with non-perturbative evaluation of this partition function through the Ponzano--Regge model. }
Their value is
\be
S_\text{TTAdS}(\tau,\bar\tau) =  -2  \pi^2 \Im(\tau)  \frac{ \ell_\text{cc}}{\ell_\text{Pl} }  =  -   \frac{\pi \beta}{\ell_\text{Pl} } .
\ee

Recall, however, that the two actions encode two different asymptotic boundary conditions at asymptotic infinity. 
Whereas $S_\text{GHY}$ fixes the induced metric on the boundary, and therefore is arguably better suited for a finite spacetime calculation, the action $S_{\frac12\text{GHY}}$ corresponds to the choice of a canonical ensemble for the BTZ black hole formulation of $\text{TTAds}_3$ \cite{BanadosMendez1998} (see appendix \ref{app_TAdS_BTZ} and references therein). 
 The calculation of \cite{GiombiMaloneyYin2008}, however, does not make reference to any specific boundary condition, the gravitational theory being defined on the whole infinite 3-hyperboloid periodically identified as above.

One crucial, and non-trivial, fact about the result of equation \eqref{eq_AdS_1loop} is that the product over $p$ starts at $p=2$. We will comment in a second what is the CFT-theoretical meaning of this fact. 
First, though, let us recall the {\it interpretation} of $p\in \mathbb Z$ in the calculation of \cite{GiombiMaloneyYin2008}, since this very same parameter will acquire (very) different interpretations in different realizations of the above calculation. 
In the present case, the periodic identification of the twisted thermal $\text{AdS}_3$ spacetime is obtained as the quotient of the three-dimensional hyperboloid by the transformation of equation \eqref{eq_identification}, and $p$ labels the copies of this spacetime in $\mathbb H_3$  (method of images).

\subsubsection{Dual CFT}

From the metric perspective, the reason why the product starts at $p=2$ is quite mysterious, and so is the fact that $Z(\tau,\bar \tau)$ factorizes in a holomorphic and an antiholomorphic contribution.
Both facts become, however, transparent once the result is viewed from the vantage point of the boundary \cite{BrownHenneaux1986,MaloneyWitten2007}.

From the boundary perspective, one has two uncoupled CFTs, composed of right- and left-movers respectively, with Hamiltonians $L_0$ and $\tl L_0$, and Brown--Henneaux central charges\footnotemark~ $c = \tl c = 12 \pi \ell_\text{cc}/\ell_\text{Pl}$.
To the AdS/CFT-oriented reader, the units in which these central charges are expressed will look quite odd: recall that we use geometrized units, where $\ellpl=8\pi G_\text{N}\hbar$ (and $\hbar =1$), as it is most natural from the (quantum) gravitational perspective.

Then, the CFT's total Hamiltonian and momentum operators are
\be
H =\frac{1}{\ell_\text{cc}} ( L_0 + \tl L_0 )
\qquad\text{and}\qquad
P =\frac{1}{\ell_\text{cc}} ( L_0 - \tl L_0 ).
\ee
The partition function $Z(\tau,\bar \tau)$ can hence be expressed in the dual theory as
\be
Z(\tau,\bar\tau) = \Tr\left( \E^{ -\I \gamma P} \E^{ -\beta H } \right) = \Tr_L\left( \E^{2\pi \I \tau  L_0}\right) \Tr_R\left(   \E^{-2\pi\I \bar \tau \tl L_0}\right),
\ee
where in the last equality we have highlighted the factorization in the left- and right-movers' theories. 
This encodes a sum over all CFT states which have been first evolved for an Euclidean time $\beta$, then translated by an amount $\gamma$ to the left, and finally re-identified with themselves. 

Following Maloney and Witten \cite{MaloneyWitten2007}, one can argue that the trace can be calculated as the sum of the contribution of the CFT fundamental state $|\Omega\rangle$ and of its descendants, the latter being obtained via the action of the left- and right-moving Virasoro modes $L_{-p}$ and $\tl L_{-p}$ on $|\Omega\rangle$. Now, the CFT fundamental state $|\Omega\rangle$ has vanishing momentum and energy $E_\Omega = - (c + \tl c)/24\ell_\text{cc}^{-1} = - \pi \ellpl^{-1}$, which gives precisely the classical contribution to $Z$:
\be
\langle \Omega| \E^{ -\I \gamma P} \E^{ -\beta H } |\Omega \rangle  = \E^{ \frac{\pi \beta}{\ell_\text{Pl}}} \equiv \E^{-S_\text{TTAdS}}.
\ee
The remaining contributions start at modes $p=2$ given that $L_{-1}$ and $\tl L_{-1}$ annihilate $|\Omega\rangle$.
Therefore, from the dual CFT viewpoint, the index $p$ in equation \eqref{eq_AdS_1loop} labels the contributions from each vacuum descendant.

An interesting generalization of this result consists in considering characters of the operators $\E^{2 \pi i \tau L_0}$ and $\E^{-2\pi i \bar \tau \tl L_0}$ in representations with highest weights $h$ and $\tl h$ respectively, different from the vacuum one. This would lead to
\be
Z_{h,\tl h} (\tau, \bar \tau) = \frac{\E^{2 \pi\I \tau (h - \tfrac{c}{24})} \E^{- 2\pi \I \bar \tau (\tl h - \tfrac{\tl c}{24})} }{\prod_{p=1}^\infty \left|1 - \E^{2\pi \I\tau \cdot p}\right|^2}.
\ee 
Notice that in this case the product starts at $p=1$.
 This is, up to a phase, the inverse of the Dedekind $\eta$-function, which is a typical example of a modular form.  Modular forms are holomorphic functions defined on the upper-half part of the complex plane, which have extremely simple transformation properties under modular transformations. They cannot be modular invariant unless they develop poles.

\subsection{Flat space and the limit $\ell_\text{cc}\to \infty$\label{sec_flatlimit}}

From these results on $\text{TTAdS}_3$, one can hope to extract meaningful predictions for flat (Euclidean) three-dimensional space, understood as the limit of $\text{TTAdS}_3$ in which $\ell_\text{cc} \to \infty$.

Let us start from the on-shell action. Its value depends only on the dimensionful quantity $\beta$, which is untouched by the limit we are considering, and on the Planck length.
For this reason one expects the value of the on-shell action to be preserved by the limiting procedure $\ellcc\to\infty$.
This is easily shown to be the case if the on-shell action is $S_{\frac12\text{GHY}}$, since it is immediate to check the consistency of this result with a direct evaluation of $S_{\frac12\text{GHY}}$ on flat spacetime.\footnotemark~
On the other hand, the evaluation of $S_\text{GHY}$ in the flat limit is more subtle, and the result depends on whether the limit $\ellcc\to\infty$ is take before or after the infinite radius limit. 
It is easy to see that the results will differ by a factor of one-half.
For this reason, in the flat case, we will use the two notations $S^{cl}_\text{GHY}$ and $S^{cl}_{\frac12\text{GHY}}$ for the respective actions' on-shell evaluations.
\footnotetext{Here is the brief computation. The bulk contribution vanishes trivially, since the spacetime is on-shell Ricci- (and actually Riemann-)flat. On the other hand, the extrinsic curvature of the boundary, at some fixed radius $r=a$, is constant and equal to $K = a^{-1}$ (since in flat space there is no cosmological scale to use as a length unit, we reintroduced dimensionful radial and time coordinates). Hence the total action equals $K$ times the cylinder's area $\text{Area}(a) = 2\pi a \beta $ divided by $2\ell_\text{Pl}$. The result is $S_{\frac12\text{GHY}}^{cl}|_\text{flat} = \pi \beta/ \ell_\text{Pl}$, a value independent from $a$, and therefore valid as-well in the limit $a\to \infty$. Clearly, as long as the cylinder is a right cylinder, the value of the boundary action does not depend on the shape of its section, thanks to the Gau\ss-Bonnet theorem.}

Moreover, further subtleties arise also for the limit of the one-loop (or vacuum descendants, from the CFT perspective) contribution.
In the flat limit, the torus modulus $\tau$ becomes effectively real, and the convergence properties of the partition function \eqref{eq_AdS_1loop} get spoiled. For this reason, it is convenient to keep track of a positive infinitesimal regulator $\epsilon^+$, as proposed in \cite{Oblak2015}\cite{BarnichGonzalezMaloneyOblak2015}. Hence,
\be
\lim_{\ell_\text{cc}\to \infty} \tau = \frac{1}{2\pi}( \gamma + \I \epsilon^+).
\label{eq_tau_reg}
\ee
 Notice how this regularization keeps $\tau$ slightly within the upper-half part of the complex plane, where modular forms are defined.

In \cite{BarnichGonzalezMaloneyOblak2015}, it is shown---using techniques analogous to those used for the AdS case \cite{GiombiMaloneyYin2008}---that the twisted thermal partition function of three dimensional flat gravity naturally matches the limit $\ell_\text{cc}\to\infty$ discussed above. 

Moreover, in \cite{Oblak2015}, certain induced representations are studied of the (centrally extended) Bondi--Metzner--Sachs group in three spacetime dimensions ($\text{BMS}_3$)---i.e. the group of asymptotic symmetries of three dimensional Lorentzian gravity with vanishing cosmological constant---and their characters are computed. These are encoded in formulas analogous to those obtained above for the conformal group at the asymptotic boundary of  (``Lorentzian''!) $\text{AdS}_3$. Let us briefly discuss these characters.

The $\text{BMS}_3$ group  \cite{Ashtekar:1996cd} is an infinite dimensional group, with the following semidirect-product structure
\be
\text{BMS}_3 = \mathrm{Diff}^+(S_1) \ltimes_\mathrm{Ad} \mathrm{Vect}(S_1),
\ee
where $ \mathrm{Diff}^+(S_1)$ denotes the group of orientation-preserving diffeomorphisms of the circle, $\mathrm{Vect}(S_1)\cong \mathrm{Lie}(\mathrm{Diff}^+(S_1))$ the Abelian additive group of vector fields on the circle, and $ \ltimes_\mathrm{Ad}$ the semidirect product of these two groups, with the first acting on elements of the second via the adjoint action.

Physically, $ \mathrm{Diff}^+(S_1)$ corresponds to the group of ``super-rotations'' i.e. to time-independent diffeomorphisms of the ``equal advanced-time'' cuts of null asymptotic infinity; $ \mathrm{Vect}(S_1)$, instead, corresponds to the group of ``super-translations'', i.e. translations of asymptotic null infinity in the ``advanced time'' direction which are constant in time but have arbitrary space-dependence. E.g. a constant vector $\alpha_\beta = \beta\pp_\phi$ field corresponds to a rigid advanced-time translation by $\beta$, while the diffeomorphism of $S_1$, $f_\gamma(\phi)  = \phi + \gamma$, corresponds to a rigid rotation by $\gamma$. 

From a generic element $(f,\alpha)\in\text{BMS}_3$, one can extract its rigid super-translational part as the zeroth Fourier mode $\beta$ of $\alpha(\phi)$ in $\alpha = \alpha(\phi)\partial_\phi$, and its rotation angle via the formula
\be
\gamma = \lim_{n\to\infty}\frac{f^n(\phi) - \phi}{n}.
\ee

The BMS$_3$ characters of interest turn out to depend only on these two properties of $(f,\alpha)\in\mathrm{BMS}_3$.
These are the characters obtained by studying certain induced representations of the {\it centrally extended} BMS$_3$ group \cite{Oblak2015}.
These representations are labeled by two real parameters, $(m,j)$. 
The parameter $m\geq0$ represents the spacetime mass. 
It can be seen as the constant representative in the BMS$_3$ orbit of the supermomentum\footnotemark~ $P(\phi) = m(\phi) - \pi \ellpl^{-1}$, with $m(\phi)$ the Bondi mass aspect of the spacetime \cite{Barnich:2014kra,Barnich:2015uva}.
The parameter $j$, on the other hand, is a representation label of the $\mathrm{U}(1)$ stabilizer group of the above constant momentum representative. As such it corresponds to the spacetime spin.
\footnotetext{Recall that $-\pi\ellpl^{-1}=-\frac{1}{24}(c+\tl c)=E_\Omega$ in the Virasoro CFT at the boundary of AdS$_3$. This shift in $P$ is due to the central extension of the BMS$_3$ group, via a central charge $c_2=c+\tl c$.}

The character for a non-vanishing value of the mass, regularized as in equation \eqref{eq_tau_reg} via $\tau\equiv\frac{1}{2\pi}(\gamma + \I \epsilon^+)$, is then
\be
\label{BMScharacter}
\chi^{m,j}( (f,\alpha) ) = \frac{ \E^{\I j \gamma} \E^{\I \beta (m- \pi \ell_\text{Pl}^{-1})}  }{\prod_{p=1}^\infty \left| 1 - \E^{2\pi \I \tau \cdot p}\right|^2} \qquad \text{if} \qquad m\neq0\,.
\ee

If the mass vanishes, on the other hand, the stabilizer group of the super-momentum is the whole of $\mathrm{PSL}(2,\mathbb R)$ and no spin label is present (since the orbit of the vacuum is restricted to one single point). The following character formula ensues
\be
\chi^\text{vac}( (f,\alpha) ) = \frac{ \E^{- \I \frac{\pi \beta}{ \ell_\text{Pl}}}  }{\prod_{p=2}^\infty \left| 1 - \E^{2\pi \I \tau \cdot p}\right|^2} \qquad \text{if} \qquad m=0,
\ee
with the product starting at $p=2$ for reasons analogous to the Virasoro case. 
Indeed, after an analytic continuation of time, these character formulas can be fully understood as the $\ell_\text{cc}\to\infty$ limits of those for the Virasoro case discussed above, where
\be
m = \lim_{\ell_\text{cc}\to\infty} \frac{h + \tl h}{\ell_\text{cc}}
\qquad\text{and}\qquad
j = \lim_{\ell_\text{cc}\to\infty} (h - \tl h) - \frac{c-\tl c}{24}.
\ee
(the last term vanishes in the gravitational case $c=\tl c= 12\pi \ellcc/\ellpl$.
 
Following \cite{Oblak2015} and \cite{BarnichGonzalezMaloneyOblak2015}, the interpretation attached to the $p$ label is that of higher Fourier modes in the super-momenta (Bondi mass aspect) associated to the various elements in the (coadjoint) orbit of the constant super-momentum $p$. These are in turn closely related to the Fourier modes of the diffeomorphisms $f$.

Recently, progress in the identification of the field theory dual to flat three dimensional gravity---that is of the analogue of the Liouville CFT at the boundary of $\text{AdS}_3$---have been made \cite{BarnichGonzalez2013,BarnichGomberoffGonzalez2013,Carlip:2016lnw}, and interesting hints also came from the semiclassical discrete approach of \cite{BonzomDittrich2016}.
Carlip's recent derivation \cite{Carlip:2016lnw} follows the same logic as his previous derivation of Liouville theory from broken radial diffeomorphisms in the context of $\text{AdS}_3$ gravity.
In this way he manages to identify the dual degrees of freedom with broken super-rotations.
The ensuing theory has close ties with Liouville theory and the so-called Schwarzian action (also discussed in \cite{BarnichGonzalez2013,BarnichGomberoffGonzalez2013}). However, it has the quite puzzling feature of lacking a kinetic term \cite{Carlip:2016lnw}: the time dependence of its field is frozen. Notice that the conformal boundary of asymptotically flat space is null, hence no natural ``time'' exists on it in the firsts place. This, however, only makes it more intriguing that the absence of a ``kinetic'' term is also observed in the completely different finite-space Euclidean approach of Bonzom and Dittrich \cite{BonzomDittrich2016} (see next section).  Notice, however, that in the latter case the missing ``kinetic'' term is the one corresponding to the would-be angular direction. This could possibly be explained through the fact that an exchange of cycles is involved in the mapping between TTAdS$_3$ and of the imaginary-time BTZ black-hole. For other attempts in the direction of defining a boundary CFT with symmetry group given by BMS$_3$, see \cite{BarnichGomberoffGonzalez2013}.
 
Also, another rather mysterious fact in Carlip's derivation is that the background metric for the Liouville-like field theory is not the induced metric on the boundary, as it was the case in the $\text{AdS}_3$ derivation, but rather an auxiliary one. 

We conclude this section noticing that, beyond convergence issues due to the infinite product, seen as a function of $\tau$ the one-loop determinant in AdS$_3$ {\it formally} has poles at all real ``rational'' points, which in the limit $\ell_\text{cc}\to\infty$ are hit---at any temperature $\beta$---by all rational values of the twist angle, i.e. $\gamma\in 2\pi (\mathbb Q\cap [0,1])$.
This pole structure is deeply connected to the theory of modular forms, which are in turn a crucial ingredient of the AdS/CFT approach to quantum gravity insofar they are fundamental building blocks of 2d conformal field theories \cite{MaloneyWitten2007,Witten:2007kt}. We invite the reader to keep this remark in mind, because we will provide quite a different viewpoint on this pole structure later in this paper.

\subsection{Flat space perturbative quantum Regge calculus \label{sec_qRegge}}

We continue our review with the discussion of the above duality in the context of perturbative quantum Regge calculus performed by Bonzom and  one of the authors \cite{BonzomDittrich2016}. This is best understood as an intermediate step between the perturbative metric calculations discussed so far and the full non-perturbative calculations within the Ponzano--Regge model presented in this paper, and for this reason we will dedicate to it more room.  Parts of the following discussion, moreover, are original.

Regge calculus \cite{Regge1961} is a discrete approach to general relativity, based on a piecewise flat simplicial decomposition of the underlying spacetime manifold $M$.
In three dimensions and with vanishing cosmological constant, the Regge action augmented by the Hartle--Sorkin boundary term \cite{HartleSorkin1981} is%
\footnote{Generalization to arbitrary dimensions is straightforward: the $e$ {\it on the rhs} of this formula should be understood as a codimension 2 sub-simplex, $l_e$ its volume, and $\theta_e^\sigma$ (see below) the internal hyper-dihedral angle at $e$. All these quantities must be understood as functions of the simplex' {\it edge} lengths, in any dimension.}
\be
S_\text{R-HS}[l_e] = \frac{1}{\ell_\text{Pl}}\left[ \sum_{e \in \mathrm{int}(M)} l_e \epsilon_e + \sum_{e \in \partial M} l_e \psi_e \right],
\label{eq_ReggeAction}
\ee
where $l_e$ is the length of the edge $e$ of the triangulation, $\epsilon_e$ is the deficit dihedral angle at the edge $e$, while $\psi_e$ is the angle between the normal to the two boundary tetrahedra\footnote{I.e. tetrahedra having at least one face being part of the boundary triangulation.} hinging (possibly among other tetrahedra) around the boundary edge $e$. Both angles must be understood as functions of the triangulation's edge lengths. 
In formulas, by introducing the internal dihedral angle at the edge $e$ within the tetrahedron $\sigma$, $\theta_e^\sigma$, 
\begin{subequations}
\begin{align}
\epsilon_e & = 2\pi - \sum_{\sigma \supset e} \theta_e^\sigma 
\qquad\text{for}\qquad e\in\mathrm{int}(M),\\
\psi_e &= \pi - \sum_{\sigma \supset e} \theta_e^\sigma
\;\;\qquad\text{for}\qquad e\in\partial M.
\end{align}
\end{subequations}
The Regge--Hartle--Sorkin action is the proper discretization of the Einstein--Hilbert--Gibbons--Hawking--York action, has the correct composition properties under gluing of manifolds, and most importantly  implements boundary conditions for the Regge equations of motion where the induced metric (i.e. the boundary edge-lengths) are kept fixed.

Regge calculus admits a compelling generalization to the cosmological case, where $\Lambda\neq0$. In its most elegant version, one not only ads an obvious cosmological term $\Lambda\sum_{\sigma} V_\sigma$ to the action, but also makes use of homogeneously curved simplices of constant curvature ($R=\tfrac{2d}{d-2} \Lambda$, in $d\geq3$ spacetime dimensions) rather than flat ones \cite{TaylorWoodward2003,BahrDittrichNewRegge}. The main motivation for this modification is to obtain homogeneously curved solution through a vanishing deficit-angle condition. However, the main reason why this works surprisingly well is the fundamental interplay between the Regge equations of motion and the generalization of the Schlaefli identities to curved simplices, see e.g. \cite{BahrDittrichNewRegge, Haggard:2014gca} for classical applications of this idea, and \cite{HaggardHanKaminskiRiello2015} for a quantum geometrical one involving Chern--Simons theory.

Quantum Regge calculus on a manifold $M$ can be defined via the following finite-dimensional ``path'' integral \cite{Regge:2000wu,Hamber:1985qj}
\be
Z_\text{R} = \int \mathcal D \mu (l) \;\E^{- S_{\text{R-HS}}(l)}.
\ee
In three dimensions, the problem of fixing the quantum measure can be elegantly solved by requiring that invariance under changes of the bulk triangulation holds at least at the linearized level around some background solution\footnotemark~ $\{l_e^0\}_e$ \cite{DittrichSteinhausMeas}.
The resulting measure coincides with the measure one would deduce from asymptotic limit of the Ponzano--Regge model (see later sections) \cite{BarrettNaishgusman2008,BarrettFoxon1993,Roberts1999}.
\footnotetext{Interestingly, no measure with this property exists in four dimensions \cite{DittrichKaminskiSteinhaus2014}.}

Using this result, a perturbative theory of three dimensional quantum gravity can be defined which is at 1-loop ``diffeomorphim invariant'', i.e. invariant under displacements of the triangulation's bulk vertices \cite{Rocek:1982fr,Dittrich:2008pw, Dittrich:2012qb}.
This theory is formally defined via the path integral 
\be
Z_\text{pert-R} = \E^{-S^{cl}_\text{R-HS}(l_e^0)}\int \mathcal D\mu_{l_e^0}(\lambda)\; \E^{- \frac{1}{2\ellpl} \sum_{\sigma,e,e'} H^{\sigma}_{e e'} \lambda_e \lambda_{e'}},
\label{eq_pertR}
\ee
where
\be
H^\sigma_{e e'} = \left.\frac{\partial \theta_e^\sigma}{\partial l_{e'}}\right|_{l_e = l^0_e}.
\ee
and $\lambda_e\ll l^0_e$ are the small edge-length perturbations, $l_e = l_e^0 + \lambda_e$.

We say ``{\it formally} defined'' because the above formula hides two difficulties.
The first one is related to diffeomorphism invariance and can be dealt with by ``gauge-fixing'' the position of the internal vertices of the triangulation.\footnotemark~
The second problem is instead related to an unbounded-from-below mode which is analogous to the conformal mode of continuum gravity. This can be dealt with by analytic continuation. 
For more details, see  \cite{DittrichSteinhausMeas,BonzomDittrich2016} and references therein.
\footnotetext{Here we refer in particular to invariance under 4-1 Pachner moves. Invariance under 3-2 moves is on the other hand exact and not problematic. The same happens in the full Ponzano--Regge model. See \cite{Barrett:1996gd}.} 

Once these issues have been dealt with and triangulation invariance has been established, the partition function $Z_\text{pert-R}$ can be calculated by choosing the most convenient bulk triangulation.
Importantly, the result will still depend on the {\it boundary} triangulation, whose edge lengths are kept fixed in the process consistently with the chosen action principle.

This means that, although this discrete theory in a sense captures all the symmetries of the continuum regime for what concerns the bulk of the spacetime, its boundary is {\it discrete and finite}.
The physical and conceptual role of dealing with finite boundaries can be physically justified in terms of the {\it general boundary} framework \cite{Oeckl2003,Oeckl2016}, which focuses on the realistic operational structure of any intrinsically localized measurement. In the case of a single boundary, one can think the resulting partition function as a generalized version of Hartle--Hawking state \cite{Hartle:1983ai}.

The discreteness of the boundary is also less severe than it looks at first sight (recall that in 3d gravity, bulk discreteness is irrelevant thanks to the triangulation invariance of the model). At the classical level, boundary discreteness can in fact be understood as the imposition of peculiar, i.e. piecewise linear, boundary conditions within the {\it continuum} theory.\footnotemark~
At the quantum level, this is reflected by the fact that spin-network states can be embedded into a continuum Hilbert space. This is indeed a key achievement of loop quantum gravity \cite{Ashtekar:2004eh,Thiemann:2007zz}. 
\footnotetext{Alternatively, discreteness can be seen as a proxy for a finite resolution measurement, which probes only a finite number of degrees of freedom.}  
There is, however, a caveat to this statement: a priori there are different possible embeddings leading to inequivalent Hilbert spaces. Accordingly, the quantum geometries encoded in the spin-network states are completed to continuum quantum geometries in very different manners. 
Indeed, a choice of embedding into a continuum Hilbert space assigns to all degrees of freedom finer than the spin-network scale a natural geometric vacuum state. In the case of 3d gravity this is the $BF$ vacuum state. See \cite{DittrichGeiller2014,BahrDittrichGeiller2015} for detailed discussions of these subtle points. Notice that due to the presence of local degrees of freedom, the identification of a suitable vacuum state for 4d gravity is a key open issue. See \cite{Dittrich:2012jq,Dittrich:2013xwa,Dittrich:2014ala} for a framework to address this problem.\footnotemark
\footnotetext{This is the so-called {\it consistent boundary} framework. It can be seen as an extension of the general boundary framework designed to address the relation between discrete and continuum boundary states while at the same time incorporating a background independent notion of renormalization.}
%

\subsubsection{Partition function of twisted thermal flat space} \label{partition_function_twisted_thermal_flat_space}

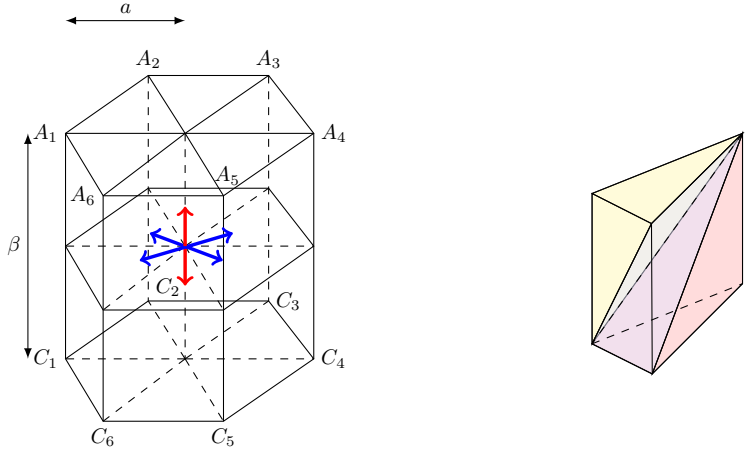
\begin{figure}[h!]
	\begin{center}
		\begin{tikzpicture}[scale=1]
			\coordinate (OA) at (1.59,0);
			\coordinate (A1) at (0,0);
			\coordinate (A2) at (1.1,0.77);
			\coordinate (A3) at (2.7,0.77);
			\coordinate (A4) at (3.3,0);
			\coordinate (A5) at (2.1,-0.83);
			\coordinate (A6) at (0.5,-0.83);
			
			\coordinate (OB) at (1.59,-1.5);
			\coordinate (B1) at (0,-1.5);
			\coordinate (B2) at (1.1,-0.73);
			\coordinate (B3) at (2.7,-0.73);
			\coordinate (B4) at (3.3,-1.5);
			\coordinate (B5) at (2.1,-2.35);
			\coordinate (B6) at (0.5,-2.35);
			
			\coordinate (OC) at (1.59,-3);
			\coordinate (C1) at (0,-3);
			\coordinate (C2) at (1.1,-2.23);
			\coordinate (C3) at (2.7,-2.23);
			\coordinate (C4) at (3.3,-3);
			\coordinate (C5) at (2.1,-3.83);
			\coordinate (C6) at (0.5,-3.83);
			 
			\draw (A1) -- (A2) -- (A3) -- (A4) -- (A5) -- (A6) -- cycle; 
			\draw (OA) -- (A1) ; \draw (OA) -- (A2); \draw (OA) -- (A3); \draw (OA)--(A4); \draw (OA) --(A5); \draw (OA)-- (A6); 
			\draw (A1) node[scale=0.8,left] {$A_1$}; \draw (A2) node[scale=0.8,above] {$A_2$}; \draw (A3) node[scale=0.8,above] {$A_3$}; \draw (A4) node[scale=0.8,right] {$A_4$}; \draw (2.15,-0.78) node[scale=0.8,above]{$A_5$}; \draw (A6) node[scale=0.8,left]{$A_6$};
			 
			\draw (B1) -- (B2) -- (B3) -- (B4) -- (B5) -- (B6) -- cycle;
			\draw [dashed] (OB) -- (B1); \draw [dashed] (OB) -- (B2); \draw [dashed] (OB) -- (B3); \draw [dashed] (OB)--(B4); \draw [dashed] (OB) --(B5); \draw [dashed] (OB)-- (B6);
			
			\draw (C1) -- (C2) -- (C3) -- (C4) -- (C5) -- (C6) -- cycle;
			\draw [dashed] (OC) -- (C1); \draw [dashed] (OC) -- (C2); \draw [dashed] (OC) -- (C3); \draw [dashed] (OC)--(C4); \draw [dashed] (OC) --(C5); \draw [dashed] (OC)-- (C6);
			\draw (C1) node[scale=0.8,left] {$C_1$}; \draw (1.1,-2.05) node[scale=0.8,right] {$C_2$}; \draw (C3) node[scale=0.8,right] {$C_3$}; \draw (C4) node[scale=0.8,right] {$C_4$}; \draw (C5) node[scale=0.8,below]{$C_5$}; \draw (C6) node[scale=0.8,below]{$C_6$};
			
			\draw (A1)--(B1)--(C1); \draw [dashed] (A2)--(B2)--(C2); \draw [dashed] (A3)--(B3)--(C3); \draw (A4)--(B4)--(C4); \draw (A5)--(B5)--(C5); \draw (A6)--(B6)--(C6);
			\draw [dashed] (OA)--(OB)--(OC); 
			
			\draw [<->,>=latex] (-0.5,0) -- (-0.5,-3); \draw (-0.5,-1.5) node[scale=0.8,left]{$\beta$};
			\draw [<->,>=latex] (0,1.5) -- (1.59,1.5); \draw (0.8,1.5) node[scale=0.8,above]{$a$};
			
			\coordinate (T1u) at (9,0);
			\coordinate (T1d) at (9,-2);
			\coordinate (T2d) at (7.8,-3.2);
			\coordinate (T3d) at (7,-2.8);
			\coordinate (T2u) at (7.79,-1.2);
			\coordinate (T3u) at (7,-0.8);

			\draw[fill=yellow!20,opacity=.9] (T3d)--(T1u)--(T3u)--cycle;			
			\draw[fill=red!15,opacity=.9] (T3d)--(T1u)--(T1d)--(T2d)--cycle;
			\draw[fill=blue!10,opacity=.5] (T3d)--(T2u)--(T1u)--(T2d)--cycle;
			\draw[black] (T3u)--(T2u)--(T3d)--cycle; \draw[black] (T3u)--(T1u)--(T2u)--cycle; \draw[black] (T3d)--(T2u)--(T1u);
			
			\draw[black] (T3d)--(T2d)--(T1u); \draw[black] (T1u)--(T2d)--(T1d)--cycle;

			\draw[dashed] (T3d)--(T1d);  \draw[dashed] (T1u)--(T3d); \draw (T2u)--(T2d);

	\draw [->,red, line width = 1.3 pt] (OB) -- ( $ (OB)!.35!(OC) $ ); 	
	\draw [<->,blue, line width = 1.3 pt] ( $ (OB)!.45!($(B3)!.5!(B4)$)$ ) -- ( $ (OB)!.45!($(B1)!.5!(B6)$) $ ); 	
	\draw [<->,blue, line width = 1.3 pt] ( $ (OB)!.45!($(B1)!.5!(B2)$)$ ) -- ( $ (OB)!.45!($(B4)!.5!(B5)$) $ ); 	
	\draw [->,red, line width = 1.3 pt] (OB) -- ( $ (OB)!.35!(OA) $ );		
		\end{tikzpicture}
	\end{center}
	\caption{Example of the background triangulation with $N_x=6$ and $N_t=2$. The effect of the twist $N_\gamma$ appear when we identified $A_i$ and $C_i$ through $A_{i}=C_{i+N_\gamma}$. Each prism is triangulated with tree tetrahedra, that can be construct by considering a diagonal per vertical faces of the prism. In the right panel we draw a prism triangulated with three tetrahedra, draw in red, blue and white.}
	\label{fig_slicedcyl}
\end{figure}

In \cite{BonzomDittrich2016}, perturbative quantum Regge calculus is applied to the calculation of the twisted thermal partition function of flat gravity. 
There, Bonzom and one of the authors, considered as the background triangulation a solid flat right cylinder of height (time extension)\footnotemark~ $\beta$, and radius $a$, regularly divided into $N_t$ time slices each in turn subdivided into $N_x$ cake-slice-like prisms, see figure \ref{fig_slicedcyl}.
This discretization is turned into a triangulation by subsequently dividing each cake-slice prism into three tetrahedra.
The twist is introduced by identifying the bottom and the top of the cylinder after a twist of $N_\gamma$ cake-slice steps.
Consequently, the boundary triangulation consists of a regular rectangular lattice subdivided into triangles along the rectangles' diagonals.
The lengths of the background edges are then fixed in terms of $a$ and $\beta$ by the flatness requirement, i.e. by the embeddability of the cylinder in $\mathbb R^3$.
\footnotetext{Our notation differs from that in the original reference.}

Once the background is fixed, the partition function can be calculated.
The classical contribution is readily found to be
\be
S^{cl}_\text{R-HS} = \frac{2\pi \beta}{\ellpl},
\ee
in agreement with the continuum (infinite space) result for the GHY boundary conditions, $S^{cl}_\text{GHY}$.
This action is twice the one obtained in the sections above. 
We remind the reader why this had to be expected: in TTAdS$_3$, $S^{cl}_\text{GHY}\,\hat=\,S^{cl}_{\frac12\text{GHY}}$, however the $\ellcc\to\infty$ limit of the on-shell GHY action does not commute with its $a\to\infty$ limit (here $2\pi a$ is meant to be the {\it physical} circumference of a Cauchy slice of constant Killing time). Changing the action to the would-be analogue of $S_{\frac12\text{GHY}}$ in the above computation, however, would be a non-trivial task, since it corresponds to some quite non-trivial boundary conditions for the Regge path integral, which would reflect itself in some non-standard choice of polarization of the resulting wave function(al).

The 1-loop contribution is finally obtained by performing the Gaussian integral for the fluctuating {\it bulk} edges using formula \eqref{eq_pertR} at fixed boundary edges. For $N_x$ odd, the result of this computation is
\be
Z_\text{pert-R}(\beta, \gamma) = \mathcal{N}\,\E^{-\frac{2\pi\beta}{\ellpl}} \;\prod_{p=2}^{\tfrac12({N_x-1})} \frac{1}{\left| 1 - \E^{ i \gamma \cdot p} \right|^2} ,
\label{eq_R1loop}
\ee
where
\be
\gamma := 2\pi \frac{N_\gamma}{N_x}. 
\ee
and $\mathcal N = \mathcal N(\beta, a, N_x, N_t) $ is a complicated normalization factor. Contrary to the $S^{cl}_\text{R-HS}$ contribution, however, it does not contain any exponential dependence on neither $\beta$ nor $a$, the cylinder's Euclidean time span and radius respectively. Importantly, it also features {\it no} dependence on the twist $N_\gamma$. Such dependence is limited to the familiar product of \eqref{eq_R1loop}.

The result above shows that the perturbative Regge calculation displays a regularized version of the flat limit of the $\text{TTAdS}_3$ result \eqref{eq_AdS_1loop}. The regularization appearing here is, however, different from the one obtained in the continuum field theoretical calculation: there the product (has to) extend up to infinity and the regularization through an infinitesimal mass \cite{BarnichGonzalezMaloneyOblak2015} turn equivalent to that discussed in section \ref{sec_flatlimit}, i.e. $2\pi \tau \to (\gamma + i \epsilon^+)$.
Here, on the other hand, it is the discrete lattice used to describe the finite-resolution boundary state which naturally provides a cut-off for the product. 

The only cut-off appearing in relation to the twist-dependence is that in the ``spatial'' direction, i.e. the one along the bulk-contractible cycle of the torus. This is the result of a non-trivial resummation of the modes propagating along the time direction, that can be physically understood as the fact that the background geometry is (extrinsically) flat along the time-direction and might therefore be insensitive to the number of steps taken to discretize it.

Once again, this product starts at $p=2$, where $p$ labels the Fourier modes in the spatial direction for the fluctuations of the bulk radial edges. 
Let us explain how this fact is related to diffeomorphism invariance also within perturbative quantum Regge calculus.
To do this, we look at the geometrical meaning of the missing $p=0$ and $p=\pm1$ modes. 
On each constant-time hypersurface, these modes correspond to rigid translations of the unique internal vertex in directions orthogonal and parallel  to the constant-time hypersurface, respectively. In figure \ref{fig_slicedcyl}, these are marked by blue and red arrows respectively.
These modes therefore represent the residual diffeomorphism symmetries of the action and as a consequence should be dropped from its mode expansion.
All other modes, would involve a change of the boundary's shape.
This geometrical interpretation of the 1-loop contribution is in beautiful agreement with Carlip's picture of boundary modes as would-be normal-to-the boundary diffeomorphisms whose action is broken by the presence of the boundary itself \cite{Carlip2005} \acite{Carlip2016}.
The dual boundary theory for the Regge theory is discussed in the next section.

Another interesting feature of the above result, is that it diverges whenever there is a $p\in\{0,\dots,\frac12(N_x-1)\}$ such that $\gamma p \in 2\pi \mathbb Z$. 
This is the case if and only if
\be
K:=\mathrm{GCD}(N_\gamma, N_x) > 1,
\ee
which can be seen as replacing the $\gamma \in 2\pi \mathbb Q$ condition of the continuum.

The geometric origin of this fact is  analogous to what happens in the Ponzano--Regge case with coherent state boundary conditions, and as such is thoroughly discussed in the second paper, Part II, of this series (in Section III.C.5). 
For the moment, it should be enough to say, that if $K>1$, the homogeneous boundary structure  is not ``rigid enough'' to provide a unique solution for the lengths of the bulk edges of the {\it linearized} equation of motions. These ambiguities show up as null modes of the (bulk) Hessian, leading to poles  for the inverse of its determinant, which determines the one--loop correction. On the other hand one finds that for (a certain class of) inhomogeneous perturbations of the boundary data one does not find any solution to the linearized equations of motion. Therefore this situation rather   
describes the emergence of an accidental  symmetry of the linearized theory due to the (homogeneous) boundary conditions, rather than the emergence of a new {\it gauge} symmetry. 
This indicates a breakdown of the linear approximation and hence the cases with $K>1$ need in principle a more refined analysis. However the Regge result reproduces in the continuum limit  the  one-loop calculation of \cite{BarnichGonzalezMaloneyOblak2015}. To regulate the one-loop partition function in   \cite{BarnichGonzalezMaloneyOblak2015} $\gamma$ had to be complexified  to $\gamma + i \epsilon^+$. For this reason, we will understand these poles  as a property of the ``analytic continuation'' of the computed partition function seen as a function of $\gamma$ only. With this subtlety understood, we will  speak in the following of ``the pole structure''

 The crucial point is that the pole structure of the one--loop partition function does almost uniquely determine its dependence on the twist angle $\gamma$. We will use this fact to compare between the various quantum gravity models and approximations.
Interestingly, it will turn out that---despite the fact that the Ponzano--Regge partition functions take {\it finite} values by construction---effective poles will emerge in different limits, e.g. the continuum limit $N_x\to \infty$ (like in this paper, where boundary edge-lengths are kept fixed and Planckian), or the large edge-length limit (like in Part II).

\subsubsection{Dual theory}\label{Bdual}

Consider a field theory defined on the boundary and coupled to the boundary metric, seen as a classical object. 
The partition function of this field theory defines a functional of the boundary metric.
If this functional is the same as the one obtained via the (bulk) gravitational path integral at fixed boundary metric, then we say that the field theory is ``dual'' to the gravitational theory (so far we ignore the introduction of perturbations associated to operators other than the boundary-theory stress-energy tensor).

In the Ponzano--Regge case analyzed in this paper (or in Part II), the homogeneity of the boundary metric implies that we will be able to reliably extract the non-trivial dependence of the gravitational amplitude only on a couple of global measures of the boundary metric, i.e.  the circumferences $a$ and $\beta$ and the twist angle $\gamma$.
The analysis could, and should, be extended to include at least perturbations of the boundary metric, and we plan to do it in future works.
This perturbative analysis has been, however, performed for the perturbative quantum Regge calculus partition function \cite{BonzomDittrich2016}.
Before coming to their result, let us notice that there is another---poorly explored---type of perturbations to be studied in the discrete setting, which has to do with changes of the discretization (or of the spin-network graph) itself. This would be an ideal setting for this study. Again, we leave this to future works.

Coming  back to the dual theory of \cite{BonzomDittrich2016} it was shown that the Regge calculus partition function for a perturbed boundary metric coincides with that obtained by considering a Liouville-like scalar field theory on the boundary. By Liouville-like, we mean a scalar field which couples to the (linearized) curvature $^{(2)}R$ of the perturbed boundary metric as the Liouville field, but has a modified kinetic term. As already mentioned, the kinetic term which is found in \cite{BonzomDittrich2016} lacks derivatives in the spacelike direction, and reduces to a Laplacian in the (Euclidean-)time direction, i.e. along the non-contractible cycle of the solid torus.

Most interestingly,  this theory can be given a geometric interpretation.
In fact, integrating out the variables associated to the bulk edges which  connect two different constant-time surfaces (i.e. both those parallel to cylinder's axis and those which develop ``diagonally''), one automatically obtains a boundary theory where the field is given by the length variations of the radial bulk edges. This is a trivial statement. The non-trivial fact is that this theory is sufficiently local and well behaved to admit a continuum limit. 
This theory corresponds to the Liouville-like theory described above, hence mimicking the mechanism of would-be diffeomorphism turned boundary degrees of freedom envisioned by Carlip \cite{Carlip2005,Carlip:2016lnw}.
See the discussion at the end of section \ref{sec_flatlimit}.

 Note that the process of constructing a dual boundary theory by integrating out a suitable subset of bulk variables from the original partition function guarantees that  (gauge) symmetries of the full model that affect left--over bulk variables are inherited by the boundary theory. On the other hand, gauge symmetries which are `fully' integrated out with the above set of bulk variables will affect the resulting measure for the boundary theory. Hence, the boundary theory will indeed be equivalent to the original model. 
 
To conclude this section, we notice that Regge calculus automatically associates a local boundary theory to any convex spacetime region with the topology of a 3-ball. Indeed, by choosing the coarsest bulk triangulation, i.e. the one with a single bulk vertex, one obtains that the radial edge-length immediately provides a local boundary theory from the onset. A similar result is found for the Ponzano--Regge model, where the dual boundary theories for a 3-ball region arise from a so-called spin network evaluation. The latter process defines a partition function for a local 2d model, corresponding in some case to well--known statistical models. On the other hand, we will show that the presence of (bulk-)non-contractible cycles, as for the solid torus, does actually translate into the insertion of a non-local operator associated to the opposite cycle. In the torus case, this corresponds to the insertion of a Haar intertwiner where the boundary cylinder is glued back to form a torus.

This ends our review of approaches to 3d gravity on the solid torus. We will now concentrate on the Ponzano--Regge model, which allows us to systematically study dual boundary theories arising from a fully non-perturbative model at finite boundaries.  Previous work on the general theme of holographic properties and boundary theories in various non-perturbative approaches to quantum gravity include  the  forayers \cite{Smolin:1998qp,Arcioni:2001ds,Arcioni:2003xx,Arcioni:2003td,Freidel:2008sh} and the more recent works \cite{Geiller:2013pya,Dittrich:2013jxa,BonzomCostantinoLivine2015,Smolin:2016edy,Han:2016xmb,Chirco:2017vhs}.

\section{Review of the Ponzano-Regge model}\label{sec_PRreview}

 The goal of this series of papers is to analyze the Ponzano--Regge amplitude of the twisted torus with various boundary conditions.
To get there, we first review the Ponzano--Regge model in its two formulations on closed manifolds, sections \ref{sec_PRgroup} and  \ref{sec_PRspin}.
Then we move on to the all-important discussion of boundaries and boundary states and their geometric interpretation in  \ref{sec_bdrystate}. In  section \ref{sec_dualtheoriesgeneralities} we will also give a first preview on the type of dual boundary theories expected from the Ponzano--Regge model.
With a definition of boundary states at hand we will  determine the associated PR amplitudes and dual boundary theories in section \ref{sectionQu5}. 

\subsection{Ponzano Regge model as a path integral for first order gravity}\label{sec_PRgroup}

Here we give a short  (heauristic) derivation of the Ponzano Regge mode. Note that our presentation does {\it not} follow the way the Ponzano Regge model  was originally constructed in \cite{PR1968}. The original construction was rather based on the observation that the $6j$ recoupling symbols reproduce in their large-$j$ asymptotics the Regge action for a single tetrahedron \cite{Regge1961}. Following this observation, one can understand the Ponzano--Regge model as a quantization of Regge calculus, alternative to quantum Regge calculus\footnotemark~ \cite{WilliamsTuckey1992,Hamber:1985qj}. 
It was recognized only much later that the Ponzano--Regge model can be derived from a (discretized) path integral of first order general relativity \cite{Rovelli:1993kc,Perez:2003vx}. Therefore, our presentation follows the historical course in reverse. This choice is made for purely pedagogical reasons.
\footnotetext{A key difference between the Ponzano--Regge model and quantum Regge calculus is that the Ponzano--Regge model implements proper quantum mechanical, that is complex amplitudes, and furthermore that the edge length is quantized in terms of spin labels. In contrast quantum Regge calculus uses a formally Wick-rotated action, so that the amplitudes are real $A\sim \exp(-S_\text{Regge})$ and the path integral is over continues length variables.} 

The first order formulation of general relativity, also known as Einstein--Cartan(--Palatini) gravity, is written in terms of a spin connection $\omega^{ab}_\mu\d x^\mu$ and a vielbein $e_\mu^a\d x^\mu$, such that $g_{\mu\nu}=\eta_{ab}e^a_\mu e^b_\nu$ with $\eta_{ab}$ the flat metric of appropriate signature.
This two sets of variables are considered to be independent. The equations of motion, for invertible vielbeins, will imply that $\omega$ is a torsion free Levi-Civita connection and $g_{\mu\nu}$ satisfies Einstein equations.
In 3d, and in absence of a cosmological constant, first order general relativity is just an instantiation of $BF$ theory.
The associated partition function $Z(M)$ for a (closed) 3-manifold $M$ is formally given by
\be
Z(M)=\int \mathcal D e \mathcal D \omega\; \E^{-\I S_{BF}[e,\omega]} 
\qquad\text{where}\qquad
S_{BF} = \frac{1}{2\ellpl}\int_M e_a \wedge F^a[\omega],
\ee
where $F^a=\epsilon^a{}_{bc} F^{bc}$, and $F[\omega] = \d\omega + \omega\wedge\omega$ the curvature of $\omega$.

Using the linearity of the $BF$ action in the triad field $e$, one can formally integrate it out, obtaining
\beq\label{Z-flat1}
Z(M)&=&\int \mathcal D\omega\; \delta(F[\omega]).
\eeq
Thus, we see that $Z(M)$ computes the volume of the moduli space of flat spin connections on the manifold $M$.

This form of the partition function can be easily regularized  through a discretization.%
\footnote{An alternative derivation of the Ponzano--Regge model discretizes the BF action $S_\text{BF}$ together with its dynamical variables. This makes the  formal integration over the triad fields in the now discretized path integral well defined. One then arrives also at (\ref{eq_PRgroupdef}) --  however with $\SO(3)$ delta--functions   instead of $\SU(2)$ delta--functions  \cite{FreidelLouapre2004,Livine:2008sw}. }

Let us first introduce some notation, that will useful in the following too.
Denote by $\Delta=\{v,e,t,\sigma\}$ a simplicial decomposition of the three manifold $M$, where $(v, e, t, \sigma)$ are labels for its vertices, edges, triangles, and tetrahedra, respectively.
Denote then by $\Delta^*$ the Poincar\'e dual cellular complex. 
It is composed by {\it nodes} $n = \sigma^*$ (zero-dimensional dual tetrahedra), {\it links}  $l = t^*$ (dual to triangles), {\it faces}  $f=e^*$ (dual to triangulation edges), and ``{\it bubbles}'' $b = v^* $ (dual to triangulation vertices). 
Faces and edges need to be considered as (arbitrarily) oriented objects.  

The introduction of $\Delta$ allows us to replace the continuum spin connection $\omega$ in (\ref{Z-flat1}) by $\SU(2)$ holonomies $g_l$ associated to the dual links $l$.
Thus the discretization of the partition function (\ref{Z-flat1}) is given by
\be
Z_\text{PR-group} = \left[\prod_{l}\int_{\SU(2)} \d g_{l}\right] \prod_{f} \delta\left( \;{\overleftarrow\prod}_{l\ni f}\, g_{l}^{\epsilon(l,f)} \right),
\label{eq_PRgroupdef}
\ee
where $\epsilon(l,f)=\pm1$ is the relative orientation of the link $l$ and the face $f$, $\d g$ is the Haar measure on $\SU(2)$, and $\delta(\cdot)$ the corresponding group delta-function, $\int \d g \delta(h^{-1}g) f(g) = f(h)$. Notice that $Z_\text{PR-group}$  computes also the  (Haar) volume of the moduli space of flat $\SU(2)$ connections, but now on the discretized manifold. 

As we will explain this form of the partition function gives already the group (or holonomy) representation of the Ponzano--Regge model, hence the label `PR-group'. We will arrive at the original form of the Ponzano--Regge model, or the so--called spin representation, after a group Fourier transform. Before explaining this we make a couple of remarks about the group representation.

First of all, in order to write the amplitude in the group representation, any cellular decomposition of the manifold would work.
The condition of $\Delta$ being a {\it simplicial} decomposition of $M$ can hence be dropped. 
This generalized form of the partition function can also translated into the spin picture, thus yielding a generalization of the original PR model, which is conversely based on a simplicial formulation. In this case, the $6j$-symbols appearing in the original PR model are replaced by higher coupling symbols.
 
Furthermore, in the group formulation, formal invariance under refinement is manifest. Hence, a cellular decomposition can be always refined to a simplicial one.

Of course, we speak only of {\it formal} invariance because the partition function \eqref{eq_PRgroupdef} is in general divergent. This is due to possible redundancies in the delta-functions distributions that appear in \eqref{eq_PRgroupdef}. Thus one can regularize the partition function by simply removing the redundant delta-functions. 
Compellingly, these divergencies can be understood to result from a residual diffeomorphism symmetry and the integration over the non--compact orbits this symmetry implies. The regularization is thus equivalent to a gauge fixing procedure \cite{FreidelLouapre2004, Bonzom:2012mb}.

More precisely these divergences appear 
in the presence of bubbles $b$ \cite{Perez:2000fs}.
Intuitively, a bubble $b$ is a polyhedron in the dual complex $\Delta^*$ which is bounded by a set of dual faces, $f\in b$. 
Each of these faces $f$ is in turn bounded by a set of links $l\in f$ and carries a group $\delta$ function for $h_f = \overleftarrow \prod_{e\in f} g_e^\epsilon$. 
It is hence immediate to see that for each bubble there is a redundant delta function. This description, although intuitive, is however not completely general (especially for $BF$ theories in dimensions higher than three). A more refined study of the homological aspects involved in the appearance of divergences was performed in \cite{BarrettNaishgusman2008,BonzomSmerlak2010,BonzomSmerlak2012}. There, the most general types of divergences have been identified and a relation between $Z_\text{PR-group}$ and the Reidemeister torsion uncovered.

Fortunately, we will have to deal with only the simple bubble divergences which can be easily taken care of by removing redundant delta-functions.
An efficient way to do this is to remove delta-functions associated to faces dual to triangulation edges belonging to a spanning tree of the triangulation 1-skeleton \cite{FreidelLouapre2004b}.  (These ``naive'' divergences were already known to Ponzano and Regge, who proposed a different regularization to the one presented here.)

As mentioned, bubble divergencies can be also understood as resulting from a residual diffeomorphism symmetry, which is on-shell equivalent to the translation symmetry of the BF theory \cite{FreidelLouapre2004}. 
This relationship is more evident in the spin picture of the PR model, where spins represent the triangulation edge lengths (or the norm of the triad fields) that are transformed by the residual diffeomorphism symmetry. This residual symmetry acts on the vertices of the simplicial discretization simply by translation, and is also present in the Regge action \cite{Dittrich:2008pw,Bahr:2009ku} (see also section \ref{sec_qRegge}). 
To add plausibility to this arguments, notice that bubbles are indeed dual to triangulation vertices.
The gauge fixing procedure sketched in the previous paragraph can be also performed in the spin picture, where it amounts to fix the spins associated to the spanning tree mentioned there.\footnote{``Fixing the spin'' $j_{e}$ means giving it a fixed value, generally zero, and removing the sum over it.} 

As final side note, we mention that in Group Field Theories \cite{Boulatov:1992vp, Freidel:2005qe,Baratin:2011tg,Oriti:2014uga} the amplitudes $Z_\text{PR-group}$ associated to dual complexes $\Delta^*$ are generated by a quantum field theory as a higher categorical analogue of Feynman amplitudes. In these theories, divergencies appear as higher categorical versions of loop divergences, resulting from unbounded summations over spin labels, which in turn are the analogue of momentum labels of Quantum Field Theory.

\subsection{Spin formulation and relation to Regge-calculus\label{sec_PRspin}}

We now sketch the derivation of the spin representation of the PR model, that is the PR model in its original version. For a more detailed derivation see \cite{BarrettNaishgusman2008}. As already said, the spin labels are interpreted as edge lengths, and thus the summation over them as the summation over (simplicial) geometric data in a gravitational path integral. 

The main idea is to apply a group Fourier-decomposition to the group representation \eqref{eq_PRgroupdef} of the PR model. One proceeds as follows:
\begin{itemize}
\item[($i$)] For each edge of the simplicial decomposition, apply the Peter--Weyl theorem to the group delta-function
\be
\delta(h) = \sum_{j\in\frac12\mathbb N} d_j \chi^j(h),
\ee
where $\chi^j(h) = \Tr \,D^j(h)$ is the character of  the spin $j$ representation $V_j$, with  $D^j(h)^{m'}{}_m$ the corresponding Wigner matrices.
As a  result, every dual face $f=e^\ast$ will carry a spin label $j_f$ to be summed over. 
This step can be interpreted as a Fourier-decomposition on the group manifold.
Geometrically, the delta function associated to a dual face imposes flatness of the spin connection holonomy around the corresponding triangulation edge (``zero deficit angle'' in Regge calculus parlance). A Fourier transform trades two conjugated variables, in this case holonomies for spins, i.e. connections for dreibeins, or deficit angles for edge-lengths.
 
\item[($ii$)] Then, expand the characters associated to each face making explicit the dependence on each link's group element:
\beq
 \chi^j(h_f)\,=\, \chi^j \left( \overleftarrow\prod_{l\supset f} g_{l}^{\epsilon(l,f)} \right)\,=\,  { \sum_{\{m_l\}}}'\prod_{l\supset f}  D^{j_l}(g_{l}^{\epsilon(l,f)} ) ^{m'_l}{}_{m_l} .
\eeq
On the right hand side, the primed sum means that magnetic indices $(m'_l,m_l)$ are contracted and summed over according to the connectivity of the links going around the face.
Geometrically, we are here decomposing the holonomy around a triangulation edge in the product of local contribution associated to paths piercing the various triangles hinging on the triangulation edge the delta function is associated with.

Now, recall that in the group representation \eqref{eq_PRgroupdef}, there is an integral over each group variable $g_l$. Each $g_l$ appears precisely three times, one per edge of the triangle $t=l^*$.
With this observation, the group variables $g_l$ can be integrated out by means of a standard identity for the Clebsch--Gordan coefficients $C_{j_1,j_2,j_3}\in V_{j_1} \otimes V_{j_2} \otimes V^*_{j_3}$
\be\label{CGD1}
\int \d g \, D^{j_1}(g) \otimes D^{j_2}(g) \otimes \overline{D^{j_3}(g)} = \frac{1}{d_{j_3}}  C_{j_1 j_2 j_3} \otimes \overline{C_{j_1 j_2 j_3}},
\ee
where we identified $D^j(g): V_j \to V_j$ with an element of $V_j\otimes V_j^*$, and omitted the six magnetic indices associated to each copy of $V_j$ or $V_j^\ast$.

In this way, we are left with two Clebsch--Gordan coefficients for each link, one associated to its source, one to its target. This corresponds to two Clebsch--Gordan coefficients per triangle, i.e. one for each tetrahedron the triangles is shared by.

\item[($iii$)] Finally,  $\Delta^*$ being dual to a simplicial decomposition, each of its nodes is 4-valent. According to the above analysis, to each (half-)link ending at the node, there is Clebsch--Gordan coefficient: one per triangular face of a tetrahedron. These four Clebsch--Gordan coefficients are naturally contracted into a $6j$-symbol by the summation over the common magnetic indices. Hence, one $6j$-symbol gets naturally associated to each tetrahedron. 

A detailed analysis of orientations shows, moreover, that specific signs must be associated to triangles.  
 
This finally gives the original Ponzano--Regge partition function
\be\label{PRpart1}
Z_\text{PR-spin}  = \sum_{\{j_e\}} \prod_{e} v^2_{j_e} \prod_t (-1)^{\sum_{e'\in t} j_e'} \prod_\sigma \Big\{ 6 j_{e''\in\sigma} \Big\}.
\ee
This is a local state sum model: local weights are associated to edges, triangles, and tetrahedra of the triangulation. The (formal) sum extends over all the $\SU(2)$ spin labels $j\in \frac{1}{2}\mathbb N$, one per triangulation edge.
We also introduced the common notations
\be
v^2_j = (-1)^{2j} d_{j} = (-1)^{2j}(2 j +1),
\ee
for the (signed) dimension of the $j$-th representation of $\SU(2)$, and $\{ 6j \}$ for the the $6j$-symbol. As mentioned, this symbol is built out of the contraction of  four Clebsch--Gordan coefficients, and we refer to \cite{BarrettNaishgusman2008} for their precise definition. 

Crucially, for these coefficients not to vanish, the three spins must satisfy triangular inequalities and sum to an integer. The latter condition implies that in the formula above the triangles weight is a $\pm1$ sign. 
\end{itemize}

In \eqref{PRpart1} the PR partition function is written as a sum over the spin labels $j$.  These, together with the magnetic indices $m$ which are summed over in contracting the Clebsch--Gordan coefficients together, encode the  quantum numbers of the dreibein field of the first order gravitational action. This interpretation is confirmed by the way we arrived at \eqref{PRpart1} via a group Fourier transform: this replaces a variable with its conjugated one, in our case the discretized spin connection with spin labels, hence giving a pairing which matches the canonical one provided by the first-order ($BF$) action. 

 More specifically, the spin labels encode the rotation invariant part of the triad, that is the triangulation edge lengths. This interpretation of the spin labels as edge lengths was originally suggested to Ponzano and Regge by the large spin asymptotics of the $6j$-symbol, which in this limit tends to the on-shell Regge action of the associated tetrahedron, cf. \eqref{eq_ReggeAction} \cite{PR1968,Roberts1999}:
\be
\Big\{ 6j_{e\in\sigma} \Big\} \xrightarrow{j\gg1} \frac{1}{\sqrt{12 \pi V_\sigma}} \cos\left(\sum_{e} (j_e + \tfrac12) \psi_e + \frac{\pi}{4}\right)
\qquad\text{if}\qquad
V^2_\sigma(j_{e\in\sigma} )>0,
\label{eq_PRspindef}
\ee
and is exponentially suppressed in the spin scale otherwise (for an interpretation of these ``negative square volume'' configurations as Lorentzian geometries, see \cite{BarrettFoxon1993}

In the above formula, the quantity $\psi_e$ is the (unoriented) external dihedral angle associated to the edge $e$ of the unique (unoriented) Euclidean tetrahedron of side lengths%
\footnote{Notice that $j + \tfrac12 = \sqrt{j(j+1)} + \mathrm O(j^{-1})$, where the term under the square root is an eigenvalue of the $\SU(2)$ Casimir operator.} 
$\{l_e = \ellpl (j_e + \tfrac12)\}$.
This tetrahedron might of course not exists: this happens if and only if the relevant Caley--Menger determinant\footnote{Cayley--Menger determinants are the generalization of Heron's formula to arbitrary dimensions.} $-6V^2_\sigma(j_{e\in\sigma} )$ is positive. When negative, $V^2_\sigma$ is the square of the volume of the relevant tetrahedron. 

The cosine is best understood as due to the two possible orientations the tetrahedron can have. 
This observation is reinforced by the presence of the $\pi/4$ phase shift, which can in turn be understood as due to the two possible signs in $\pm\sqrt{V^2_\sigma}$.

Thus, ignoring for a moment the presence of a cosine rather than of a complex exponential, we see that \eqref{PRpart1} defines a version of the quantum Regge calculus discussed in section \ref{sec_qRegge}, but featuring fully quantum mechanical amplitudes.
In this quantum amplitudes, the tetrahedral volume factor appearing in the asymptotics \eqref{eq_PRspindef}, as well as the weights associated to edges and triangles, are readily interpreted as the ``quantum measure'' of the path integral. As we discussed, similar measure factors have to be used for (linearized) Regge calculus to achieve invariance under triangulation changes \cite{DittrichSteinhausMeas}, see section \ref{sec_qRegge}. 
Back to the cosine, it can be understood as resulting from considering a path integral for a first order formulation of gravity, where one integrates over both positively and negatively oriented triads. We will see in a moment why its presences is also related to $BF$ divergences.

Speaking of divergences, let us observe that the spin representation \eqref{PRpart1} resulted from a variable transformation of the formally triangulation invariant group representation of the PR model. Thus one expects \eqref{PRpart1} to be both formally invariant  under changes of the triangulation and generally divergent. These two expectations can be studied in detail and are related. 

In the spin formulation the divergencies can be heuristically explained as follows: whenever a triangulation vertex (whose dual has a ball topology) is present in the bulk, the spins associated to the edges sharing this vertex are unconstrained. The sum diverges as the amplitude is constant in directions for which the spin labels represent a flat geometry. To understand this divergence in these terms it is crucial to notice that the vertex in question can also be ``moved' outside the region made up of the tetrahedra sharing this vertex, and that the amplitude remains constant in this case. This is due to the sum over orientations  that is implemented in the PR model \cite{Christodoulou:2012af}. 
The fact that the amplitude is constant for deformations of the edge lengths that preserve flatness shows that there is a residual diffeomorphism symmetry. A symmetry we already discussed in section \ref{sec_qRegge} on Regge calculus. The existence of this symmetry also explains the invariance under triangulation changes of the partition function
\cite{Bahr:2011uj, Dittrich:2012qb}.

Changes of triangulations can be implemented by Pachner moves. Now, while the $1-4$ move is unsurprisingly plagued with the ``bubble'' divergences we just discussed (in this move, one replaces one tetrahedron with four sharing a common internal vertex), the $3-2$ move holds exactly\footnote{The $3-2$ move is also known as the Biedenharn--Elliott identity.} and is in fact directly related to the action of the quantum Hamiltonian of 3d General Relativity \cite{Barrett:1996gd}.

Apart from gauge fixing this symmetry, another way to make the partition function (\ref{PRpart1}) finite is to consider a so-called $q$-deformation of the model, known as the Turaev--Viro model \cite{TuraevViro1992}.
In the Turaev--Viro (TV) model the representation theoretical elements of the PR formula are substituted with their quantum-group counterparts at deformation parameter $q=\E^{i\pi/r}$ a root of unity. 
The resulting formula is naturally cut-off at the spin $j=\tfrac12 (r-2)$ and is therefore {\it always} finite. 
Furthermore, the TV model is {\it exactly} triangulation invariant and thus provides a local formula for a three manifold invariant \cite{TuraevViro1992}.
Each simplicial weight, in the limit $q\to1$ (or $r\to\infty$) converges to the corresponding PR weight. 
The convergence of the partition function, on the other hand, is a much more complicated issue we are not going to discuss. 
In the large spin asymptotic, where $j,r\gg1$ uniformly, the quantum $6j$-symbol has  the same asymptotic relation to the cosmological Regge action%
\footnote{The cosmological Regge action \cite{BahrDittrichNewRegge} features spherical simplices rather than flat ones, as well as the cosmological term $\sim\Lambda V_\sigma$ with the right prefactor accounting for both the Ricci and the cosmological term contributions to the bulk action.
}
as the PR model has to the flat-space one \cite{MizoguchiTada1991, TaylorWoodward2003}.

Unfortunately, the TV model is based on a spin description, which is quite cumbersome to work with.\footnote{The TV model can also be defined as sum over class angles conjugated to the spins \cite{Barrett:2002vi, Barrett:2004im,Freidel:2006qv}. But these class angles are---due to the quantum deformation of the group at root of unity---quantized and represented again by spin labels (of the representation category of $\SU(2)_q$).  See also \cite{Dittrich:2016typ,Dittrich:2017nmq} for a definition of a basis dual to the $q$-deformed spin network basis in 3d and 4d respectively, \cite{Riello:2017iti} for its classical counterpart in 4d, as well as \cite{Dupuis:2014fya,Bonzom:2014wva} for the analogous case in 3d with negative $\Lambda$.}
For this reason we are focusing here on the PR case, which admits a dual, group-based, formulation.


\section{Boundaries, boundary states and boundary dual field theories \label{sec_bdrystate}}

In presence of boundaries, the partition function becomes a (Schr\"odinger) function(al) of those variables which are kept fixed there.
It is then crucial to use an action which is compatible with the chosen boundary conditions, that is with the chosen polarization. 
For example, $S_{BF}$ is the correct action for an $\omega$ polarization:
\be
Z_{BF}[\varpi] = \int_{\underleftarrow\omega|_\pp = \varpi} \mathcal D \omega \mathcal D e \; \E^{-\I S_{BF}}.
\ee
This is because the boundary contribution to $\delta S_{BF}$ is $e_a \wedge \delta \omega^a$. 
Similarly, the GHY action \eqref{eq_GHY} for metric gravity is adapted to wave functionals of the induced boundary metric. 

Similarly $Z_\text{PR-spin}$ and $Z_\text{PR-group}$ are naturally and automatically adapted to different boundary conditions, which are fixed boundary spins (i.e. metric) and fixed boundary connection, respectively.
In the quantum theory, we can however consider any kind of boundary states representing superpositions of the classical boundary conditions. These boundary states are contracted with the `bulk' amplitude to define transition or Hartle--Hawking amplitudes \cite{Oeckl2003,Oeckl2016}.

As we can express boundary states in any polarization we can use either $Z_\text{PR-spin}$ or $Z_\text{PR-group}$ to compute such transition amplitudes.
For its computational simplicity, we will work from now on in the spin-connection (group) representation.\footnote{This is also a much simpler choice in the continuum. See \cite{DupuisFreidelGirelli2017} for a preliminary discussion of the $e$-polarization.}
In the continuum, a boundary state $\Psi$ and its amplitude $\la {BF} | \Psi\ra$ formally read $\Psi[\varpi]$ and
\be
\la {BF} | \Psi \ra = \int \mathcal D \varpi \;  \overline{Z_{BF}[\varpi]}\; \Psi[\varpi].
\ee
Starting in this paper, we will provide various different---physically relevant---boundary states for the PR model. To evaluate their amplitude it will be easiest to express them in the group representation.

\begin{figure}[h!]
	\begin{center}
		\begin{tikzpicture}[scale=1.5]
		\coordinate (OA) at (1.59,0);
		\coordinate (A1) at (0,0);
		\coordinate (A2) at (1.1,0.77);
		\coordinate (A3) at (2.7,0.77);
		\coordinate (A4) at (3.3,0);
		\coordinate (A5) at (2.1,-0.83);
		\coordinate (A6) at (0.5,-0.83);
		
		\coordinate (OB) at (1.59,-1.5);
		\coordinate (B1) at (0,-1.5);
		\coordinate (B2) at (1.1,-0.73);
		\coordinate (B3) at (2.7,-0.73);
		\coordinate (B4) at (3.3,-1.5);
		\coordinate (B5) at (2.1,-2.35);
		\coordinate (B6) at (0.5,-2.35);
		
		\coordinate (OC) at (1.59,-3);
		\coordinate (C1) at (0,-3);
		\coordinate (C2) at (1.1,-2.23);
		\coordinate (C3) at (2.7,-2.23);
		\coordinate (C4) at (3.3,-3);
		\coordinate (C5) at (2.1,-3.83);
		\coordinate (C6) at (0.5,-3.83);
		
		\draw[black,fill=red!30] (A1)--(B1)--(B6)--(A6);
		\draw[black,fill=red!10] (OA) -- (A1) -- (A6) --cycle;
		\draw[fill=red!10] (OA)--(OB)--(B6) -- (A6);
		
		\draw (A1) -- (A2) -- (A3) -- (A4) -- (A5) -- (A6); 
		\draw (OA) -- (A2); \draw (OA) -- (A3); \draw (OA)--(A4); \draw (OA) --(A5);
		
		\draw (B1) -- (B2) -- (B3) -- (B4) -- (B5) -- (B6);
		\draw [dashed] (OB) -- (B1); \draw [dashed] (OB) -- (B2); \draw [dashed] (OB) -- (B3); \draw [dashed] (OB)--(B4); \draw [dashed] (OB) --(B5); \draw [dashed] (OB)-- (B6);
		
		\draw (C1) -- (C2) -- (C3) -- (C4) -- (C5) -- (C6) -- cycle;
		\draw [dashed] (OC) -- (C1); \draw [dashed] (OC) -- (C2); \draw [dashed] (OC) -- (C3); \draw [dashed] (OC)--(C4); \draw [dashed] (OC) --(C5); \draw [dashed] (OC)-- (C6);
		
		\draw (B1)--(C1); \draw [dashed] (A2)--(B2)--(C2); \draw [dashed] (A3)--(B3)--(C3); \draw (A4)--(B4)--(C4); \draw (A5)--(B5)--(C5); \draw (B6)--(C6);
		\draw [dashed] (OA)--(OB)--(OC); 
		
		\coordinate (Dint1) at (0.8,-1.02);
		\coordinate (Dbdry1) at (0.25,-1.17);
		\coordinate (Dint2) at (0.8,-2.52);
		\coordinate (Dbdry2) at (0.25,-2.72);
		
		\draw (B1)--(B6) node[pos=0.4,above,sloped]{$e$}; 
		
		\draw[dashed,BrickRed,line width=0.5mm] (Dint1)--(Dbdry1);
		\draw[BrickRed,line width=0.5mm] (Dbdry1)--(Dbdry2);
		\draw[dashed] (Dint2)--(Dbdry2);
			
		\draw (Dint1) node {$\bullet$}; \draw (Dint1) node[scale=0.8, below right] {$O_1$};
		\draw (Dint2) node {$\bullet$}; \draw (Dint2) node[scale=0.8,below right] {$O_2$};
		\draw (Dbdry1) node {$\bullet$}; \draw (Dbdry1) node[scale=0.8,above] {$B_1$};
		\draw (Dbdry2) node {$\bullet$}; \draw (Dbdry2) node[scale=0.8, left] {$B_2$};

		\end{tikzpicture}
	\end{center}
	\caption{ We have depicted in red  a boundary 3-cell and the associated boundary 2-cell is in deep red. The set of boundary 2-cells  provides a cellular decomposition of the boundary manifold. The dual of the boundary 3-cell is denoted by $O_1$, and the dual of the boundary 2-cell is denoted by $B_1$. The dual edge of a boundary 2-cell emanating from the node $O_1$, dual to the red boundary 3-cell, is depicted as a dashed dark red line. The dual of the boundary edge $e$ in solid dark red line. }
	\label{fig:boundary_discretization_convention}
\end{figure}
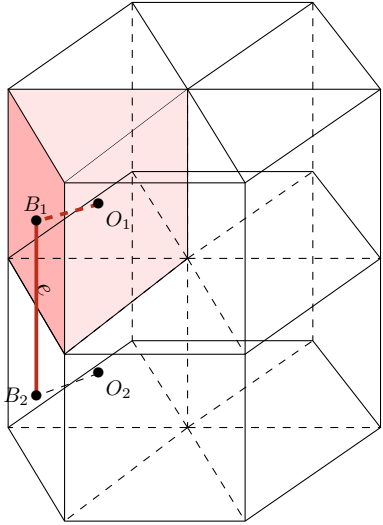

Before moving to the construction of the first boundary states, however, we need to fix conventions and notations for the discretization of a manifold with boundaries.
In particular we require the boundary discretization to be induced by the bulk one in the following way: some bulk codimension 0 cells will have codimension 1 faces belonging to the boundary. 
Since we are working in 3 dimensions, let us call these boundary 3-cells and boundary 2-cells, respectively. The set of boundary 2-cells provides a cellular decomposition of the boundary manifold.
When dualizing the {\it bulk} of the manifolds, one associates nodes to codimension 0 cells,%
\footnote{We keep using the symbols introduced in the case of simplicial decomposition.}
$n = \sigma^*$, links to codimension 1 cells, $l = t^*$, and faces to codimensions 2 cells, $f = e^*$. 
Now, this means that a link dual to a boundary 2-cell $t$ emanating from a node dual to a boundary 3-cell, will be {\it a priori} an open link intersecting the boundary at a point within $t$. Similarly a face dual to a boundary 1-cell (a boundary edge) $e$, will intersect the boundary on a 1 dimensional line transverse to $e$. This situation is pictured in figure \ref{fig:boundary_discretization_convention}.
To deal with this situation, one introduces boundary links and boundary nodes (possibly denoted $l_\pp$ and $n_\pp$ when it is necessary to distinguish them from the bulk ones) which are respectively dual to boundary 1- and 2-cells.%
\footnote{Of course, for completeness one should deal similarly with ``boundary'' bubbles, but we will not need to do that.}

The set of boundary links and nodes constitutes a graph $\Gamma$ dual to the boundary triangulation.
For reasons that will become clear soon, $\Gamma$ is---slightly improperly---called a {\it spin-network (graph)}.

In the group representation, the partition function for a discretization with boundary is given by
\be
Z_\text{PR-group}(g_{l_\pp}) = \left[\prod_{l \neq l_\pp}\int_{\SU(2)} \d g_{l}\right] \prod_{f\neq f_\pp} \delta\left( {\overleftarrow\prod}_{l\supset f} g_{l}^{\epsilon(l,f)} \right),
\ee
that is only those group elements associated to bulk links are integrated over, and delta-functions are associated  to bulk faces only. Notice that here we assume that the dual complex arises from the dualization of a simplicial complex of the type described above. In this case, each boundary face is dual to a boundary vertex and no boundary 3-cell contains more than one boundary face. Thus the flatness conditions for the bulk faces of each boundary 3-cell impose automatically the flatness for the boundary face, too.

\subsection{Boundary Hilbert space and spin network states}

As mentioned, we will be working in the group polarization of the PR amplitude. Hence, the relevant (boundary) Hilbert space describes discretized boundary connections, represented by the group elements  associated to the directed links of the boundary graph $\Gamma$. Thus the boundary Hilbert space is
\be
\mathcal H_\G = L^2(\SU(2)^{|\Gamma|})  \, \ni \, \Psi( g_{l})  .
\ee
where $|\Gamma|$ denotes the number of links $l$ in the boundary graph $\Gamma$. 
Actually, the gauge invariance of the Ponzano--Regge model implies that boundary states are automatically projected onto their gauge invariant part. For this reason one restricts to gauge invariant wave functions, i.e. to wave function satisfying
\be
 \Psi(G_{t(l)} g_l G^{-1}_{s(l)}) = \Psi(g_l),
 \label{eq_SNgaugeinvariance}
\ee
for all possible assignments of group elements $G_n$ to the nodes $n\in\Gamma$. Here and in the following, $s(l)$ and $t(l)$ stand for the source and target nodes of $l$, respectively. 

A basis of gauge invariant states is given by the so-called spin-network basis states  \cite{Rovelli:1994ge,Rovelli:1995ac}. This basis can again be related to a group Fourier transform \cite{Thiemann:2007zz}. Each spin-network state is characterized by a set of fixed spins associated to its links, corresponding to fixed edge lengths in the boundary triangulation. For this reason, spin-network  basis states impose sharp induced-metric boundary conditions to $Z_\text{PR}$. 
If $\Gamma$ is trivalent gauge invariance ensures that the spin labels determine the spin-network state uniquely. If the nodes of $\Gamma$ have higher valency, and hence correspond to polygons, a further specification of non-unique node associated tensors $\iota_n$ is needed to completely characterize the spin-network basis state. Geometrically, these tensors encode the shape of the (non-necessarily flat) polygons, at least compatibly with the indeterminacy principle---more on this in the following.

A spin-network basis state thus takes the form 
\be\label{SNWDef}
\Psi_{j,\iota}( g_l) \,=\, \left(\bigotimes_{n\in \Gamma} \iota_n  \right) \bullet_\Gamma \left(\bigotimes_{l \in \Gamma}   \sqrt{d_{j_l}}  D^{j_l}(g_l)\right),
\ee
where $\bullet_\Gamma$ stands for the contraction of all the magnetic indices as prescribed by the graph $\Gamma$, and $\iota_n$ are intertwining  tensors, or {\it intertwiners}, living in the spaces 
\be \label{invVec}
\iota\in V_{j_1} \otimes \dots \otimes V_{j_k}  \otimes V^*_{j_{k+1}} \otimes \dots \otimes V^*_{j_m} .
\ee
They implement the state's gauge-invariance \eqref{eq_SNgaugeinvariance} thanks to their defining property
\be
\Big(D^{j_1}(G)\otimes\dots \otimes D^{j_k}(G) \otimes D^{j_{k+1}}(G^{-1}) \otimes \dots \otimes D^{j_m}(G^{-1})\Big) \triangleright \iota_n = \iota_n \quad \forall G\in\SU(2).
\label{eq_intertwinerinvariance}
\ee
The subspace of the space $ V_{j_1} \otimes \dots \otimes V^*_{j_m}$ in which this invariance property is satisfied, is the $m$-valent intertwiner Hilbert space
\be
\mathrm{Int}\left(j_1,\dots,j_k,j^*_{k+1},\dots,j^*_m \right)  .
\ee

For an (ortho)normal spin-network basis one needs an (ortho)normal basis of $\mathrm{Int}\left(j_1,\dots,j^*_m \right)$. The orthonormality and completeness of the basis  for the gauge invariant subspace of  $L^2(\SU(2)^{|\Gamma|})$ is then a consequence of the Peter--Weyl theorem which implies that to each link one associates the Hilbert space $L^2(\SU(2)) ]\cong \oplus_j (V_j  \otimes V_j^*)$.

Let us  briefly go back to the case of a trivalent graph $\Gamma$, dual to a boundary triangulation.
As we said, in this case the intertwiners are unique. They are in fact proportional to the Clebsch--Gordan coefficients. (Their orthonormality requires the inclusion of a dimensional factor, which together with the inclusion of an appropriate sign, defines the so-called $3jm$ symbols.)
Then, the amplitude associated to such states $\psi_{j_\pp}$ (we omit the intertwiner labels, since they are unique) is\footnote{We omitted the complex conjugation  of the bulk amplitude in the integral as this amplitude is real.}
\be
\la \text{PR} | \Psi_{j_\pp} \ra
= \left[\prod_{l_\pp }\int_{\SU(2)} \d g_{l_\pp}\right] Z_\text{PR-group}(g_{l_\pp}) \Psi_{j_\pp}(g_{l_\pp}) 
= Z_\text{PR-spin}(j_\pp),
\label{eq_SNgroup}
\ee
where $Z_\text{PR-spin}(j_\pp)$ is the obvious generalization of the (spin-representation) PR amplitude to triangulations with boundaries. (This generalization assigns the same weights as before to tetrahedra, bulk triangles and bulk edges and sums only over the spins associated to the bulk edges only, the boundary spins $\{j_\pp\}$ being held fixed. One thus needs only a specification of the weights for the triangles and edges contained in the boundary. Any specification of these can be matched by adjusting  the sign and dimensional factors in the definition of the spin network state.)

\subsection{Geometric interpretation of  higher valent spin-network states\label{sec_IVB}}

As we discussed above the spins associated to the links of a spin network state can be interpreted as the lengths of the edges dual to these links.
This point of view is further reinforced by the fact that the intertwiner associated to a triple of spins exists and is unique if and only if these satisfy the triangular inequalities, and can thus represent a quantum triangle. 

This interpretation  can be generalized to higher valent nodes: an $m$-valent intertwiner can\footnotemark~  be used to define a quantum $m$-sided polygon with fixed edge lengths determined by the spins $(j_1,\ldots,j_m)$ \cite{Livine:2013tsa}. 
\footnotetext{
Spin-networks are also the boundary states of four-dimensional quantum gravity and can thus  describe quantum states of 3d geometry. From this perspective, intertwiners are naturally interpreted as {\it polyhedra} embedded in the flat 3d Euclidean space $\mathbb R^3$ \cite{Barbieri:1997ks,Freidel:2009ck,Freidel:2009nu,Bianchi:2010gc,Freidel:2010tt,Livine:2013tsa}. Quantum deforming $\su(2)$ allows to extend this geometrical interpretation to polyhedra in homogeneous curvature \cite{Dupuis:2013lka,Bonzom:2014wva,Haggard:2015ima}.}
The intertwiner space is in general not unique anymore, but still finite dimensional. This finite dimensionality of  $\mathrm{Int}(j_1,\dots,j_m)$ indicates that there exist a compact phase space of polygons with fixed edge lengths.\footnotemark~
\footnotetext{
Let us underline an issue with this polygonal interpretation of intertwiners, which is due to the possibility of different orderings of the edges around the polygon. As explained in \cite{Livine:2013tsa}, there are two possibilities. On the one hand, if we do not specify any ordering for the legs of the intertwiner, we can reconstruct multiple possible polygons. It is possible to recover a unique convex polygon, at least in the planar case, in which case the intertwiner contains enough data to deduce an ordering.  On the other hand, the problem is automatically cured by considering graphs embedded in a surface, as we do here.
}

\begin{figure}[h!]
	\begin{center}
		\begin{tikzpicture}[scale=0.8]
		
		\coordinate (C) at (0.09,0.21);
		\coordinate (A) at (-1.68,0.32);
		\coordinate (B) at (0.17,1.81);
		\coordinate (D) at (1.88,0.11);
		\coordinate (E) at (-0.14,-1.74);
		
		\draw (-1.91,1.92) -- (1.97,1.72)--(1.75,-1.97)--(-1.41,-1.59)--cycle;
		
		\draw (C) node[above left]{$\iota_4$};
		\draw (C)--(A) node[midway,below]{$j_3$}; \draw (C)--(B) node[midway,left]{$j_1$}; \draw (C)--(D) node[midway,above]{$j_4$}; \draw (C)--(E) node[midway,right]{$j_2$}; 
		
		\draw (3,0) node{$\rightarrow$};
		
		\coordinate (I1) at (5.25,-0.40);
		\coordinate (I2) at (8.7,0.3); 
		\draw (7.75,-1.97)--(4.59,-1.59)--(4.09,1.92)--cycle; \draw (I1) node{$\bullet$}; \draw (5.30,0) node{$\iota^{1}_3$};
		\draw (6.09,1.92) -- (9.97,1.72)--(9.75,-1.97)--cycle; \draw (I2) node{$\bullet$}; \draw (I2) node[below right]{$\iota^{2}_3$};
		
		\draw (I1) -- (4.4,-0.35) node[midway,above]{$j_3$} ; \draw (I1) -- (6,-1.75) node[midway,right]{$j_2$};
		\draw (I1) to[bend left] node[pos=0.4,above]{$j$} node[pos=0.54,below]{$\theta$} (I2);
		\draw (I2)--(8.1,1.82) node[midway, right]{$j_1$}; \draw (I2)--(9.87,0.1) node[midway,above]{$j_4$};
		
		\end{tikzpicture}
	\end{center}
	\caption{{ Decomposition of a 4-valent intertwiner into two 3-valent ones along one of the diagonals. The recoupling spin associated to the length of the diagonal is denoted by $j$.  $\theta$ is the dihedral angle between the two triangles.} }
	\label{fig:four_intertwiners_to_two_three}
\end{figure}
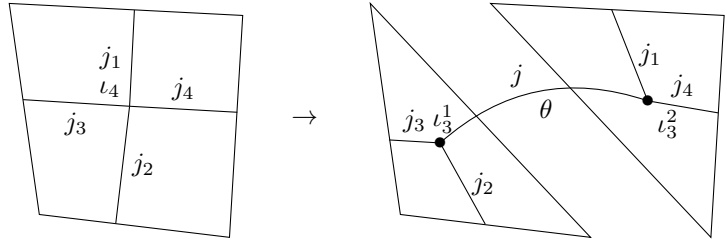

The non-uniqueness of the higher-valent intertwiners fits nicely with the fact that the geometry of polygons with fixed edge length is also not unique. The intertwiner space describes also a possible bending of polygons: that is if we introduce diagonals in the polygon, there might be non-trivial dihedral angle hinging on such diagonals, see figure\ref{fig:four_intertwiners_to_two_three}.

This space of (possibly bent) polygons admits a canonical symplectic structure named after Kapovich and Millson \cite{kapovich1995,kapovich1996} which allows its quantization \cite{Livine:2013tsa,Conrady:2009px}. 
The resulting Hilbert space is indeed $\mathrm{Int}(j_1,\dots,j_m)$  which as mentioned is finite dimensional as a consequence of the compact nature of the space of polygons with fixed side lengths.
 The Kapovich--Millson symplectic structure sets the length of each diagonal and the corresponding dihedral angle to be canonically conjugate variables. Thus, these two variables cannot be determined at the same time by a given intertwiner. This is compatible with the fact that in a boundary polygon the dihedral angle associated to a diagonal encodes some extrinsic curvature of the manifold, dual to the intrinsic metric determined by the length of the diagonal itself.\footnotemark~	
\footnotetext{
The extension of the Kapovich-Milson phase space to the case of hyperbolic and spherical polygons  was introduced in \cite{Treloar1} and \cite{Treloar2}. For the recent work on the quantization of these spaces, their relation to quantum group deformation and their application to quantum gravity with a non-vanishing cosmological constant, the interested reader will find details in \cite{Dupuis:2013lka,Bonzom:2014wva,Haggard:2015ima,Dittrich:2016typ}.
}

Thus we see that higher-valent nodes can encode quantum geometry, which features non-commutative aspects \cite{Freidel:2005bb,Freidel:2005me,Baratin:2010nn}.
In this paper, as well as in Part 2, we will consider boundary state with four-valent nodes. The richness of the possible boundary states---even with fixed spins for the edges of the discretizations---reflects the non-commutative geometry encoded in the four-valent intertwiner spaces. 

Let us discuss in more detail the four-valent intertwiner and its associated geometry.
A 4-valent intertwiner can be decomposed into two 3-valent one glued by a recoupling spin. 
Interpreting the 4-valent intertwiner as a quadrilateral, the decomposion into 3-valent intertwiners  corresponds to cutting the quadrilateral into two triangles along one of its diagonals, see figure \ref{fig:four_intertwiners_to_two_three}. 
The recoupling spin associated to this diagonal corresponds to the length of the diagonal (i.e. the interior edge shared by the two triangles). The conjugate variable to this length according to the Kapovich--Millson symplectic structure is the dihedral angle between the two triangles hinged by the diagonal itself, as illustrated in figure \ref{fig:four_intertwiners_to_two_three}.  We refer to this angle as  the ``extrinsic curvature'' of the quadrilateral. Since the length and the angle are canonically conjugate variables,  a 4-valent intertwiner with fixed recoupling spin, and thus fixed diagonal length, is totally spread in the dihedral angle, i.e. in the extrinsic curvature.

Once the edge lengths of the quadrialteral are fixed, notice that the length of one diagonal together with the related dihedral angle, determine the length of the other diagonal. From the discussion above, it follows that the two diagonals do have a non-trivial commutation relation, both in phase space and as quantum observables.

Thus the  question  arises  whether it is possible to define ``coherent intertwiners'' which are peaked on polygons of a given shape, presenting only minimal fluctuations in the length of their diagonal and in their extrinsic curvature.
This is provided by the coherent intertwiners introduced by Speziale and one of the authors in \cite{Livine:2007vk}. They are usually referred to as the LS (coherent) intertwiners, and were further studied in \cite{Freidel:2009ck,Freidel:2009nu,Freidel:2010tt,Dupuis:2010iq,Bianchi:2010gc}. These intertwiners were originally introduced in a three-dimensional context for polyhedra, but can be readily reinterpreted as coherent quantum polygons in our two-dimensional framework.

We will not use this coherent intertwiner technology in the present paper, but it will be at the heart of the second of our series, where we will focus on building suitable semi-classical boundary state peaked around the correct asymptotic classical geometry.

\section{Dual boundary theories: generalities \label{sec_dualtheoriesgeneralities}}

Recall that the PR amplitude imposes flatness of the bulk connection, and hence the pairing of a boundary state $\Psi_{j,\iota}$ with the PR amplitude will be  schematically given by  
\be
\la \text{PR} | \Psi_{j,\iota} \ra \,=\, \left[\prod_{l_\pp }\int_{\SU(2)} \d g_{l_\pp}\right] Z_\text{PR-group}(g_{l_\pp}) \Psi_{j,\iota}(g_{l_\pp})  \,= \,
\left[\prod_{l_\pp }\int_{\SU(2)} \d g_{l_\pp}\right]_{|\text{bulk flat}}  \Psi_{j,\iota}(g_{l_\pp}),
\label{eq_SNgroup2}
\ee
where we indicated with ``$|\text{bulk flat}$'' the restriction of the integration to those boundary connections compatible with a connections which is flat in the specified bulk. As the bulk amplitude is given by a product over delta functions such constraints for the boundary connection are easily determined. E.g. if we consider a three-dimensional ball bounded by the sphere the boundary connection is constrained to be locally and globally flat. 

Later we will in particular consider the example of a solid torus with a connected boundary given by a two-torus. The boundary connection is again required to be locally flat, that is every boundary-contractible cycle has to be flat. We do have, however, also two non-contractible cycles, an equatorial and a meridian one, which can a priori carry non-trivial (commuting) holonomies. Having fixed the bulk to be a specific solid torus, however, requires---say---the meridian cycle to be trivial.

In general, using the flatness conditions as well as the gauge-invariance of the boundary state, we will be able to fix most of the boundary holonomies to the identity. Eg. if we consider a ball with a spherical boundary we can set in this way all the boundary holonomies to the identity, without loosing any information. In this case the PR amplitude associated to a (spin network) boundary state,
\be
\la \text{PR}(\text{3-ball}) | \Psi_{j,\iota} \ra = \Psi_{j,\iota}(g_{l_\pp} = \mathbb{I} ),
\ee
amounts to a so-called spin-network evaluation. Going back to the definition of the spin-network states (\ref{SNWDef}) we see that this evaluation takes the explicit form
\be\label{Btheory1}
\la \text{PR}(\text{3-ball}) | \Psi_{j,\iota} \ra =    \sum_{m_l,m'_l}   \,\, \left(\prod_{n \in \Gamma}  (\iota_n)_{m'_{l:s(l)=n}}^{m_{l:t(l)=n}} \right)    \left(\prod_{l \in \Gamma}   \sqrt{d_{j_l}}    \delta^{m'_l}_{m_l} \right).
\ee
In other words, apart from dimensional factors, the spin-network evaluation consists just in a full contraction of the intertwiners $\iota_n$ associated to the boundary nodes, as dictated by the connectivity of $\Gamma$. For a regular boundary graph, a homogeneous spin assignement and a homogeneous association of boundary intertwiners defines a specific vertex- (or spin-chain) model.   
In these statistical models, defined by \eqref{Btheory1}, the dynamical variables are given by the magnetic indices $m_l=m'_l$, which are all located on the two-dimensional boundary of the manifold, while their interactions are are encoded in the specific (homogeneous) choice of intertwiners that has been made.

The above description of degrees of freedom and interaction is ``quantum mechanical'', in the sense that it is expressed as a sum over elements of given Hilbert spaces weighted by some interaction matrix elements (the intertwiner components). 
Some choices of intertwiners might allow a sleek rewriting in terms of a (continuum) classical system. The accuracy of the classical description is expected to hold when there are many degrees of freedom. This happens  for large spins --- which is also the limit in which spinfoams are  expected to turn ``semiclassical''. The prototypical example of states admitting a neat semiclassical interpretation in the large spin limit is that of LS coherent states. The corresponding contiuum classical system is described by a peculiar (quite uncoventional) non-linear sigma-model with $\SU(2)$ (or $\SO(3)$) as a target manifold associated to each node of $\Gamma$, and the variables specifying the boundary metric as coupling constants. For more details see Part II.

The choice of a finite boundary allows also to consider more general boundary states that describe superpositions of intrinsic boundary geometry data, and in particular a superposition of the spin labels associated to the boundary edges. This leads to more general boundary theories, e.g. so--called intertwiner models  \cite{Dittrich:2013aia,Dittrich:2013voa},  or theories involving fermions, e.g. \cite{BonzomCostantinoLivine2015}.  Some of these boundary states  will be considered in Part III of this series \cite{Part3}.

We see that there is a large space of possible boundary theories. E.g. for a regular four-valent spin-network with fixed homogeneous spin labels $j$ (and a homogeneous choice of intertwiners), the boundary theories are parametrized by the possible choices of four-valent intertwiners, which in turn constitute a $(2j+1)$-dimensional vector space. 
We will see in section \ref{spinhalf}  that such boundary theories include integrable models (at small spins, where the intertwiner space is small), but so far nothing guarantees that non-integrable models will not arise, too. 
In addition, when considering superposition of spins or inhomogeneities in the spins and/or in the graph, far more possibilities arise. 
The challenge is here to identify particularly interesting classes of boundary states, either for the boundary theories they give, or for their geometrical interpretation. Possibly for both. A concrete quest consists in looking for those boundary states encoding {\it asymptotic} boundary conditions.
A further challenge is to understand the symmetries of these boundary theories, and---at least in the case just mentioned---their relation to the BMS group.
In this and the next paper of the series, we start the explorations of these questions.

In the next section, we will start investigating the specifics of the toroidal case, with a quadrangulated boundary.
Contrary to the 3-ball case mentioned above, the toroidal case sees the introduction of (bulk) non-contractible cycles. 
The presence of such non-trivial topological features requires a modification to the spin-network evaluation formula described above. 
It results in the emergence of a non-trivial monodromy around the non-contractible cycle, which has to be integrated over against the boundary state in order to obtain the sought partition function, which is an integral over {\it all} bulk-flat connections. (Of course, the boundary state can be specifically chosen so that it peaks (possibly sharply) around a specific value of this monodromy.)
From the boundary theory perspective, the non-trivial monodromy appears as a new non-local variable which is integrated over. This procedure---introduction and integration of a new non-local variable---can be equivalently recast as the insertion of a non-local operator, in fact a Haar intertwiner, winding around the opposite cycle.

Thus we see that---at least a priori---the amplitude will break modular invariance of the boundary torus, since a specific cycle is selected by the presence of a given bulk ``filling''.
This dependence can be removed by hand by simply replacing the ``bulk flat'' condition in \eqref{eq_SNgroup2} with the less restrictive ``boundary flat'' condition.
This might be possibly understood as an amplitude of a theory allowing for (superpositions of) different bulk topologies and bulk defects -- but might be better defined than the formal sum over all bulk topologies. For now, we are not investigating these possibilities here.

\section{Quadrangulations and PR amplitudes of the solid torus}\label{sectionQu5}

In this final section, we will compute the Ponzanno-Regge amplitude explicitly for a simple class of boundary spin network states. We will illustrate two points. First we show that  the amplitude depends on the choice of boundary states: even if  states seem to describe the same discretized geometry, the details of the amplitude depends on the details of the quantum state. Second, we will exhibit the basic features of the Ponzano-Regge amplitude for the twisted torus boundary and discuss the extent to which we recover the standard BMS character formula and the partition function of three-dimensional flat-space gravity.

The spin-network states on the boundary will be based on a regular square lattice $Q$, with all links labeled by the same spin $j$  (unless otherwise stated). This boundary discretization, as a square lattice, naturally arises when considering the cellular decomposition described earlier in section \ref{partition_function_twisted_thermal_flat_space}.

\subsection{Spin network evaluation on the torus: the general formula}

Let us put aside for a moment the details of the spin and intertwiner labels, and first focus on the geometrical features  of the cellular decomposition and its square lattice boundary $Q$.
The cellular decomposition is denoted $\Delta$ and is composed by $N_t \times N_x$ triangular prisms organized in cylinders split into ``cake slices'' and stacked onto each other as ``cake layers''.\footnotemark~
\footnotetext{French readers might want to think of this structure in terms of stacked and cut cheese wheels. } 
We will denote by $\Delta_d$ the set of $d$-dimensional subcells, $\Delta_1$ being e.g. the set of its edges, and by $\Delta^*$ its dual. 

We denote the vertices of the quadrangulation $Q$ by coordinates $(t,x)\in\mathbb Z^2$. We demand $Q$ to be trivially periodical in the horizontal direction and periodic up to a Dehn-twist of $N_\gamma$ units in the vertical direction:
\be
(t, x+N_x) \sim (t,x) \sim (t+N_t , x +N_\gamma).
\ee
This has to be compared with equation \eqref{eq_identifcyl}: having in mind a thermal Minkowski space, we think of the horizontal direction labeled by $x$ as the space direction and of the vertical direction labeled by $t$ as the ``Euclidean time'' direction.  The twist angle is here given by
\be
\gamma:=2\pi\frac{N_\gamma}{N_x}.
\ee

The spin-network graph $\Gamma$ is dual to the quadrangulation $Q$. It is also a square lattice whose vertical  and horizontal links, labeled by subscripts $v$ and $h$,  are dual to the {\it space} and {\it time} edges of $Q$, respectively (notice the ``inverted'' relation!).

\begin{figure}[h!]
	\begin{center}
		\begin{tikzpicture}[scale=1.3]
		\foreach \i in {0,...,6}{
			\foreach \j in {-1,...,3}{
				\draw (\i,\j) node {$\bullet$};
				\draw[<-] (\i,\j-.5) --(\i,\j+.5);
				\draw[->] (\i-.5,\j) --(\i+.5,\j);
			}
		}
		
		\foreach \i in {0,...,4}{
			\draw[rounded corners=3 pt,->] (\i,4.5+0.3) --(\i,4+0.1+0.3)-- (\i+2,4-0.1+0.3)--(\i+2,3.5+0.3)   ;
		}
		
		\draw (0,-1.7) node{$0$};
		\draw (2,-1.72) node{$N_\gamma$};
		\draw (3,-1.7) node{$x$};
		\draw (6,-1.7) node{$N_x$-1};
		
		\draw(-.9,-1) node{$N_t$-1};
		\draw(-0.7,1) node{$t$};
		\draw(-0.7,3) node{$0$};
		
		\draw(3.5,1) node[above]{$g^{h}_{t,x}$}; \draw (2.5,1) node[above]{$j$};
		\draw(3,0.5) node[left]{$g^{v}_{t,x}$}; \draw (3,1.5) node[right]{$j$};
		
		\end{tikzpicture}
	\end{center}
	\caption{Oriented square lattice on the twisted torus. The twist angle is $\gamma=2\pi \f{N_{\gamma}}{N_{x}}$. Starting from the vertex $(t,x)$, the edge on the right is associated to $g^{h}_{t,x}$ and the edge below to $g^{v}_{t,x}$. All the spins are fixed to $j$. The horizontal periodic condition is without twist.}
	\label{fig:lattice_Q}  
\end{figure}
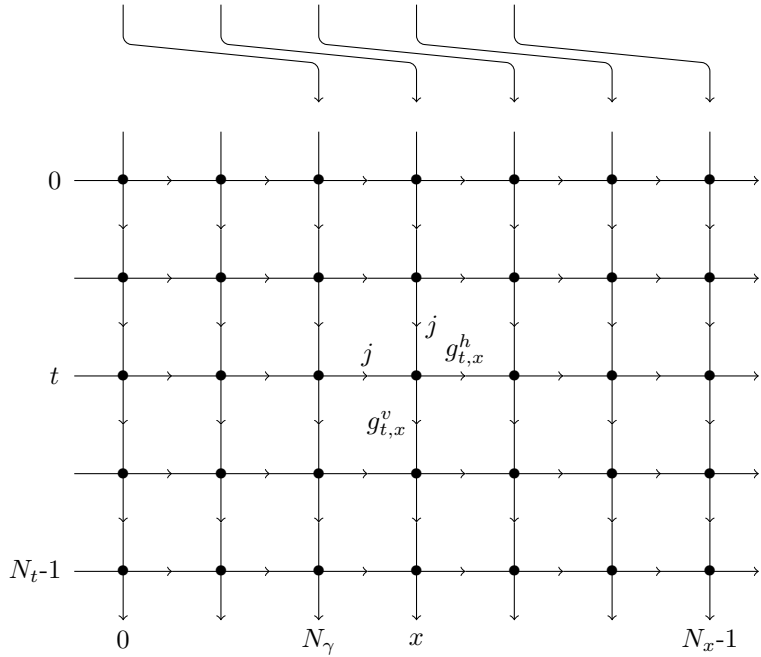

Drawing $Q = \mathbb Z^2/\sim$ on a plane as in figure \ref{fig:lattice_Q}, we take its bottom-left vertex and bottom-left square (corresponding to a node in $\Gamma$) as the origin, $(t,x)=(0,0)$, of $Q$ and $\Gamma$ respectively. 
Vertical links between nodes  $(t,x)$ and $(t+1,x)$ are labeled by $(t,x)$ and similarly for the horizontal ones. 
Vertical and horizontal links are oriented in the direction of growing time and space coordinates, respectively.

A generic spin-network state on the above quadrangulation will be denoted, in the connection representation, by 
\be
\Phi = \Phi( g^h_{t,x}, g^v_{t,x} ).
\ee
We will assume that this state is gauge invariant. (If it weren't, the PR amplitude kernel would anywhoe project it onto its gauge invariant component.)

From this cellular decomposition, it is straightforward to write the formal PR amplitude in the group representation. 
 
 As discussed above, the ``naive'' PR amplitude needs to be regularized. To do so, it is most convenient to choose a maximal spanning tree $T\subset \Delta_1$ of the bulk vertices which starts at one boundary vertex of the triangulation---the root---say $(t,x)=(0,0)$, then moves into the bulk via a radial edge, and hence follows $(N_t-1)$  of the central edges along the core non-contractible circle of the solid-torus discretization described above. The regularization proceeds by removing, for every edge $e\in T$, the corresponding delta-function associated to $f=e^*\in \Delta^*_2$.
In the case at hand, this procedure produces a finite amplitude.

To further simplify the expression of the amplitude, a gauge fixing of the boundary state can be performed.
A maximal gauge fixing is determined by the choice of a spanning tree $T_\pp^*\in \Gamma$, this time made of links of the  spin-network graph  dual to the boundary cellular decomposition.
Performing an appropriate gauge transformation at each node of this tree (root excluded) we can always gauge fix the holonomies $(g_l)_{ l \in T^*_\pp }$ to the identity.

At this point, one can solve for most of the remaining $g_l$'s by using the relations imposed by the remaining (bulk) delta-functions.
If the tree $T_\pp$ has been chosen conscientiously, all the horizontal holonomies have been trivialized and all the vertical ones are set to the one same value $g_t$ throughout each time slice. 
The amplitude then reads:
\be
\la \text{PR} | \Phi  \ra = \left[ \prod_{t=0}^{N_t-1} \int_{\SU(2)} \d g_t \right] \, \Phi(g^h_{t,x} = \id, g^v_{t,x} = g_t ).
\ee

This expression can be further simplified by performing gauge transformations which are constant throughout time slices, so that the gauge condition above $g^h_{t,x}=\id$ stays untouched.
Choosing appropriate gauge parameters the amplitude can be brought to the form
\be\label{PartTorus1}
\la \text{PR} | \Phi \ra =  \int_{\SU(2)} \d g \,\,\, \Phi(g^h_{t,x} = \id, g^v_{t\neq N_t,x} = \id,  g^v_{N_t, x} = g).
\ee

The result of this tedious procedure (exemplified for completeness in appendix \ref{app_gaugefixing} on a toy example) could have been guessed on the basis of triangulation invariance as well as from the fact that the flatness of the model requires all contractible loops to be flat, and integrates over all possible values for the non-contractible cycle. 

Note that we are left only with one integration as there is only one independent cycle which is non-contractible in the solid torus. Notice, also, that there is a residual global symmetry in the left-over holonomy $g$ given by the adjoint action $g \rightarrow G g G^{-1}$. Hence we can gauge fix this global symmetry and restrict the integration in \eqref{PartTorus1} to the class angle of $g$:
\be\label{PartTorus_classangle}
\la \text{PR} | \Phi \ra =  \f2\pi\int_0^{\pi} \d \theta  \; \sin^2\left(\theta\right)\, \Phi(g^h_{t,x} = \id, g^v_{t\neq N_t,x} = \id,  g^v_{N_t, x} = \E^{2\theta \tau_z}).
\ee
(In Part II it will be more convenient to use, instead of $\theta$ the variable $\varphi=2\theta$.)

These formulas are valid for any  state supported on the quadrangulation dual $\Gamma$. As explained above the twist $\gamma$ is imposed in the way the quadrangulation is periodically identified to a torus.

\subsection{Exact Formulas for the Partition Function}

Here, we will study the evaluation of the boundary spin networks for the standard intertwiners in the spin basis. For a special choice---the spin-0 intertwiner in the s-channel---we will be able to compute the Ponzano-Regge amplitude exactly. Even though we will not recover the BMS character formula (since we are not working with semi-classical boundary states), this will allow us to illustrate the dependence of the asymptotic partition function on the twist angle $\gamma$, and especially the distinction that arises between its rational vs. irrational values: in the appropriate limit, we will get poles for all rational angles. We will further give the explicit formulas for the spin-network evaluation with lowest spins $j=\f12$ and arbitrary intertwiner states, hence showing that it maps onto the 6-vertex (or ``ice-type'') model, well-known in statistical physics. The spin $j=1$ case will also be briefly discussed in similar terms.

\subsubsection{Intertwiner basis and spin network evaluation}

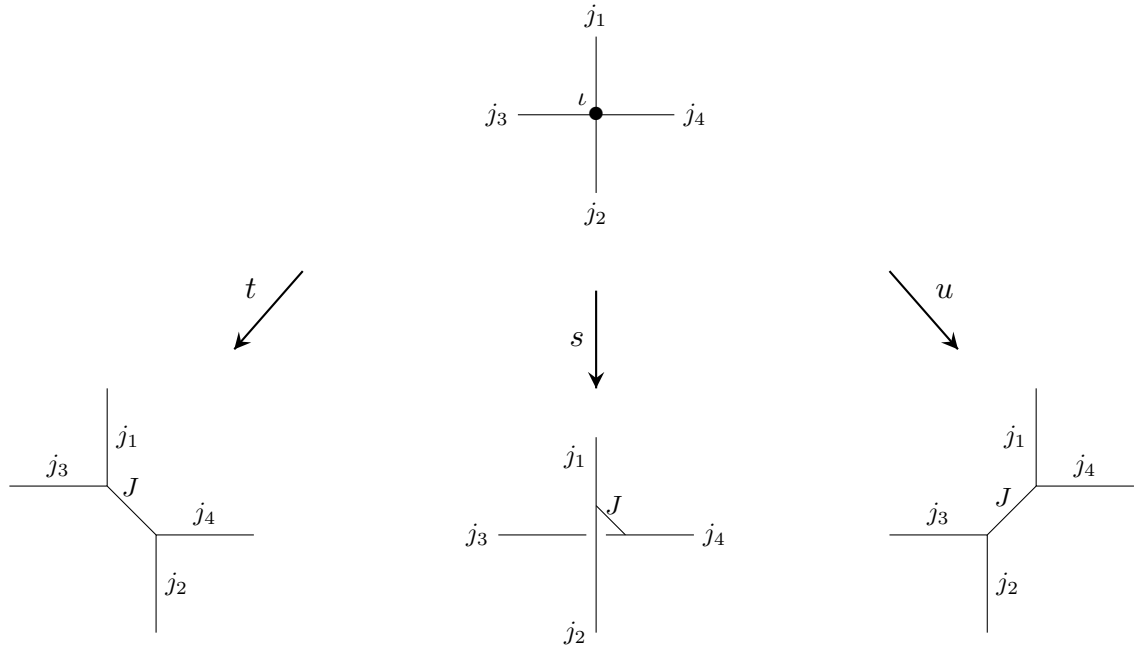
\begin{figure}[t]
	\begin{center}
		\begin{tikzpicture}[scale=1.3]

\draw (-.8,0) -- node[pos=0, left]{$j_3$} node[pos=1, right]{$j_4$} (.8,0) ; \draw (0,-.8)-- node[pos=0, below]{$j_2$} node[pos=1, above]{$j_1$}(0,.8); \draw (0,0) node[scale=1.3]{$\bullet$} node[above left]{$\iota$}; 

\draw[thick,decoration={markings,mark=at position 1 with {\arrow[scale=1.5,>=stealth]{>}}},postaction={decorate}]
(0,-2+0.2) -- (0,-3+0.2) node[midway,left,scale=1.2]{$s$};
\draw[thick,decoration={markings,mark=at position 1 with {\arrow[scale=1.5,>=stealth]{>}}},postaction={decorate}]
(-3,-2+0.4) -- (-3.7,-2.8+0.4) node[midway,above left,scale=1.2]{$t$};
\draw[thick,decoration={markings,mark=at position 1 with {\arrow[scale=1.5,>=stealth]{>}}},postaction={decorate}]
(3,-2+0.4) -- (3.7,-2.8+0.4) node[midway,above right,scale=1.2]{$u$};
			
\draw (-6,-4+0.2)--node[pos=0.5, above]{$j_3$} (-5,-4+0.2)-- node[pos=0.5, right]{$j_1$} (-5,-3+0.2); \draw (-5,-4+0.2) -- node[pos=0.4, above]{\;$J$} (-4.5,-4.5+0.2); \draw (-4.5,-5.5+0.2) -- node[pos=0.5, right]{$j_2$} (-4.5,-4.5+0.2) -- node[pos=0.5, above]{$j_4$} (-3.5,-4.5+0.2);
			
\draw (-1,-4.5+0.2)--node[pos=0, left]{$j_3$}(-0.1,-4.5+0.2); \draw (0.1,-4.5+0.2)--node[pos=1, right]{$j_4$}(1,-4.5+0.2); \draw (0,-5.5+0.2)--node[pos=0, left]{$j_2$} node[pos=0.9, left]{$j_1$} (0,-3.5+0.2); \draw (0,-4.2+0.2)--node[pos=0.6, above]{$J$} (0.3,-4.5+0.2);
			
\draw (3,-5+0.7)--node[pos=0.5, above]{$j_3$} (4,-5+0.7)-- node[pos=0.5, right]{$j_2$} (4,-6+0.7); \draw (4,-5+0.7)-- node[pos=0.4, above]{$J$\;} (4.5,-4.5+0.7); \draw (4.5,-3.5+0.7)-- node[pos=0.5, left]{$j_1$} (4.5,-4.5+0.7)-- node[pos=0.5, above]{$j_4$} (5.5,-4.5+0.7);
			
		\end{tikzpicture}
	\end{center}
	\caption{The three channels for splitting a 4-valent intertwiner into two 3-valents ones linked by an intermediate link carrying a spin $J$.}
	\label{fig_split_intertwiner}
\end{figure}

Let us now describe in detail a choice  of boundary spin network state on the square lattice.
 First we fix the spins on all the lattice links. 
Then we need to choose a 4-valent intertwiner.
As sketched above and reviewed in details in appendix  \ref{app_spinintertwiner}, 4-valent intertwiners are not unique, but a basis of the intertwiner Hilbert (vector) space can be easily constructed. 
For this, we choose a pairing of the spins,  $(12)-(34)$, or $(13)-(24)$, or $(14)-(23)$, and we split the 4-valent intertwiner into two 3-valents ones linked by an intermediate edge. 
Basis states are then defined by the spin $J$ carried by that intermediate link, as shown on figure\ref{fig_split_intertwiner}. Similarly to the terminology used in particle scattering, we refer to the three possible pairings as the channels $s$, $t$ or $u$.

The $t$ and $u$ choices correspond geometrically to fixing the lenghts of a diagonal in the quadrileteral picture we discussed in section \ref{sec_IVB}.

Let us  explicitly  consider the $s$-channel, corresponding to the $(12)-(34)$ pairing.
In this case, the 4-valent intertwiner basis state with intermediate spin $J$ reads 
\be
\label{PRexact1}
|\iota^{s|J}\ra
=
\f1{d_{J}}
\sum_{\{m_{i\}},M} (-1)^{J+M}
|\otimes_{i=1}^4(j_{i}m_{i})\ra \,
\la (j_{1}m_{1})(j_{2}m_{2})| J,M \ra
\la (j_{3}m_{3})(j_{4}m_{4})| J,-M \ra
\,.
\ee
Here, we have taken the two sets of Clebsh-Gordan coefficients, recoupling $j_{1}$ and $j_{2}$ into $J$ on one side and recoupling $j_{3}$ and $j_{4}$ into $J$ on the other, and glued them using the $\su(2)$ structure map $\varsigma$ along the intermediate link.
This map identifies the spin $j$ representation with its conjugate, according to
\be
 D^j(\varsigma ) \, |j,m\ra=(-1)^{(j+m)}\,|j,-m\ra
\nn
\ee
(see appendix \ref{app_spinintertwiner} for further details). 

By convention, the above formula describes an intertwiner for four links {\it outgoing} from a node $n$. Thus for gluing the intertwiners along auxiliary two-valent nodes positioned between two ``half-links'', we need to insert again the $\su(2)$ structure map at each such node, possibly together with the insertion of a group element associated to this link.  

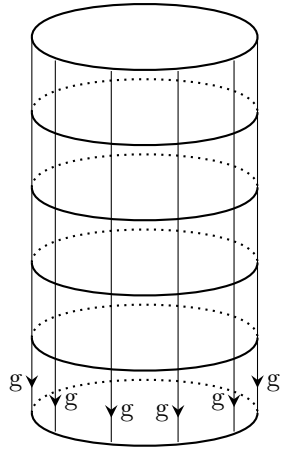
\begin{figure}[h!]
				\begin{tikzpicture}
						\coordinate(ai) at (0,0);
				\coordinate(bi) at (0,5);
				\draw[dotted,thick,in=90,out=90,looseness=.5] (0,0) to (3,0);
				\draw[thick,in=-90,out=-90,looseness=.5] (0,0) to node[pos=0.2,inner sep=0pt](a1){}  node[pos=0.4,inner sep=0pt](a2){}  node[pos=0.6,inner sep=0pt](a3){}  node[pos=0.8,inner sep=0pt](a4){} (3,0);
				\draw[in=90,out=90,looseness=.5,thick] (0,5) to  (3,5);
				\draw[thick,in=-90,out=-90,looseness=.5] (0,5) to node[pos=0.2,inner sep=0pt](b1){}  node[pos=0.4,inner sep=0pt](b2){}  node[pos=0.6,inner sep=0pt](b3){}  node[pos=0.8,inner sep=0pt](b4){} (3,5);
				\coordinate(af) at (3,0);
				\coordinate(bf) at (3,5);
				
				\foreach \i in {1,...,4}{
					\draw[dotted,thick,in=90,out=90,looseness=.5] (0,\i) to (3,\i);
					\draw[thick,in=-90,out=-90,looseness=.5] (0,\i) to (3,\i);
				}
						
				\draw[decoration={markings,mark=at position 0.1 with {\arrow[scale=1.5,>=stealth]{<}}},postaction={decorate}] (ai) -- node[left,pos=0.08]{g}(bi);
				\draw[decoration={markings,mark=at position 0.1 with {\arrow[scale=1.5,>=stealth]{<}}},postaction={decorate}] (a1) --node[right,pos=0.08]{g} (b1);
				\draw[decoration={markings,mark=at position 0.1 with {\arrow[scale=1.5,>=stealth]{<}}},postaction={decorate}] (a2) -- node[right,pos=0.08]{g} (b2);
				\draw[decoration={markings,mark=at position 0.1 with {\arrow[scale=1.5,>=stealth]{<}}},postaction={decorate}] (a3) -- node[left,pos=0.08]{g} (b3);
				\draw[decoration={markings,mark=at position 0.1 with {\arrow[scale=1.5,>=stealth]{<}}},postaction={decorate}] (a4) --node[left,pos=0.08]{g} (b4);
				\draw[decoration={markings,mark=at position 0.1 with {\arrow[scale=1.5,>=stealth]{<}}},postaction={decorate}] (af) -- node[right,pos=0.08]{g} (bf);
				\end{tikzpicture}
\caption{The square lattice on the boundary torus: the boundary spin network state is defined as the assignment of spins on the lattice edges and the choice of a 4-valent intertwiner at each vertex; then the Ponzano-Regge amplitude is defined as the integral over the $\SU(2)$ group element $g$ of the evaluation of the spin network function. All other group elements were gauge-fixed to the identity.}
\label{fig_FIG6}
\end{figure}

Now, that we have fixed the spins $j_{l}$ on all the lattice edges and chosen intertwtiner states $\iota_{n}$ for every node of the lattice, we evaluate the Ponzano-Regge partition function specializing equation \eqref{PartTorus1} to our choice of boundary state. 
In words, we glue and contract the intertwiners together along the lattice links, with a group element insertion along the links of the last slice of the torus: 

\begin{equation}
\label{PRexact2}
\la \text{PR} | \Phi_{j_l,\iota^J_n}  \ra
=
\int_{\SU(2)} d g \,
\bigotimes_{n}\iota_n  \bullet_\Gamma \Big{[}\bigotimes_{l\ne(N_{t},x)} D^{j_{l}}(\varsigma)  \bigotimes_{l=(N_{t},x)} D^{j_{l}}(g\varsigma)\Big{]}
\end{equation}
where $\bullet_\gamma$ stands for a trace over the magnetic indices following the connectivity of $\Gamma$. See figure \ref{fig_FIG6}.

For fixed spins $j_{l}$ and intertwiners $\iota_{n}$, this is the integral of a polynomial over $\SU(2)$. Indeed the Wigner matrix elements $D^j(g)^{m'}{}_m$ are polynomials of degree $2j$ in the $\SU(2)$ group element $g$ (defined as a 2$\times$2 matrix). Hence the Ponzano-Regge amplitude, defined above for a finite lattice, always gives a finite result.

It is, therefore, evident that in order to get an interesting structure, in particular the appearance of poles, we need to take a continuum or asymptotic limit, along three possible lines: 
\begin{itemize}
\item We can take a limit in which the lattice cylinder radius $N_{x}\rightarrow\infty$ goes to infinity, but the spins are kept constant. Divergences might  possibly emerge in such a limit. We explore this scenario below.
\item Alternatively we can consider a limit in which the lattice is kept constant, i.e. $N_{x}=\text{const}$, but the spins are taken large. We will consider such a limit in Part II. 
\item We can also consider superpositions of spins, allowing to peak the lattice edge length on arbitrarily small or large distances. Here we can  get poles as was already shown in \cite{Dittrich:2013jxa,BonzomCostantinoLivine2015}. We will postpone a more detailed study  to Part III \cite{Part3} in this series of papers.
\end{itemize}

In the following, we will focus on a homogeneous state,\footnotemark~ taking all the spins to be equal, $j_{l}=j$, and taking the same intertwiner state at every node of the lattice.
\footnotetext{
It would be interesting to later allow for spin and intertwiner fluctuations, which would amount to allow surface geometry fluctuation on the boundary and would lead to a deeper understanding of the boundary theory at the quantum level.
}
We will then use spin recoupling techniques to express intertwiners in terms of symmetrization and antisymmetrization of spin states, which will in turn allow us to write the partition function as an integral over suitably symmetrized products of $\SU(2)$ characters. This is akin to using the equivalence between the spin-network basis and the (outfashioned) loop-basis of Loop Quantum Gravity \cite{Gambini:1996ik}.
Below, we will apply this method to the $J=0$ $s$-channel intertwiner.
In this case, the loops completely decouple, thereby considerably simplifying the combinatorics and the integral evaluation.

Using a different mapping, we will also show, in section \ref{spinhalf}, that for the spins $j_{l}=\f12$ the partition function for arbitrary intertwiners can be exactly mapped onto a 6-vertex model.

\subsubsection{The $J=0$ s-channel intertwiner}

Consider the spin network basis state on the square lattice defined by a uniform assignation of spins $j_l=j$ to all lattice links\footnotemark{} and a uniform 4-valent intertwiner with spin $J=0$ in the $s$-channel at all lattice nodes.
\footnotetext{
The results of this section are completely straightforwardly generalized to an assignation of spin $j^v=j$ to all vertical edges and of a different spin $j^h=j'$ to all horizontal edges.
}
This choice of intertwiner not only allows to completely decouple the horizontal and vertical lines and hence to explicitly evaluate  the partition function and its asymptotic limit, but it also comes with a clear geometrical interpretation.

\smallskip

{\bf Geometric interpretation of the uniform $J=0$ $s$-channel intertwiner} 
This intertwiner is given by
\begin{align}
\label{I0s}
|\iota^{s|0}\ra
& =
\f1{d_{j}}
\sum_{\{m_a\}_{a=1,\dots,4}} (-1)^{2j+m_1+m_3} \delta_{m_1+m_2,0} \delta_{m_3+m_4,0}
| (j,m_1)(j,m_2)(j,m_3)(j,m_4)\ra\nn\\
&= \f1{d_{j}}
\sum_{m,\tilde{m}} (-1)^{2j+m+\tilde{m}}
| (j,m)(j,-m)\ra_{(12)}\otimes| (j,\tilde{m})(j,-\tilde{m})\ra_{(34)}\,.
\end{align}
Notice that all four edge vectors of the quadrilateral plaquette have equal norm, $\la \vJ_{a}^{2} \ra=j(j+1)$, for $a=1,\dots,4$.
Moreover, the spin-0 on the intermediate link implies that edge 2 is exactly opposite to edge 1, while  edge 3 is exactly opposite to edge 4. 
Indeed, one can check the following expectation values from the intertwiner formula above:
 \be
\la \vJ_{1}\cdot\vJ_{2} \ra=\la \vJ_{3}\cdot\vJ_{4} \ra=-j(j+1)
\,,
\ee
Moreover, one can further compute the expectation values of the angles between the other pairs of vectors:
\be
\la \vJ_{1}\cdot\vJ_{3} \ra=\la \vJ_{1}\cdot\vJ_{4} \ra=\la \vJ_{2}\cdot\vJ_{3} \ra=\la \vJ_{2}\cdot\vJ_{4} \ra=0
\,.
\label{mean}
\ee
This means that, in average, edge 1 (and thus also edge 2) is orthogonal to edges 3 and 4, and therefore the intertwiner seems to be dual to a geometrical square. However, this is only true in average. Computing the variance of the scalar products between ``orthogonal'' edges,
\be
\la (\vJ_{1}\cdot\vJ_{3})^{2} \ra=\f13j^{2}(j+1)^{2}\,,
\label{variance}
\ee
one realizes that it takes its maximal value.

Thus, the $J=0$ $s$-channel intertwiner is the furthest possible from a semi-classical state with a good geometrical interpretation. More precisely, while the spin-0 intertwiner defines a maximal entanglement between the opposite edges (1 and 2 on the one hand, and 3 and 4 on the other), the other pairings are left to be independent random vectors.\footnotemark~ Hence, more correctly, this intertwiner can be said to represent a superposition of all possible parallelograms centered on the rectangular one.
\footnotetext{
In fact, considering two random vectors $\hat{u}$ and $\hat{v}$ on the 2-sphere of radius $r$, $\mathbb S^2_r$, one can easily compute:
$$
\int_{(\mathbb{S}^{2}_{r})^{\times 2}} \f{d^{2}\hat{u}}{4\pi}\f{d^{2}\hat{v}}{4\pi}\,
\big(
\hat{u}\cdot\hat{v}
\big{)}
=
0
\,,\qquad
\int_{(\mathbb{S}^{2}_{r})^{\times 2}} \f{d^{2}\hat{u}}{4\pi}\f{d^{2}\hat{v}}{4\pi}\,
\big(
\hat{u}\cdot\hat{v}
\big{)}^{2}
=
\f13 r^{4}
\,,
$$
which are the corresponding classical calculations for the expectation values \eqref{mean} and for the variances \eqref{variance}.
}

Even if the $J=0$ $s$-channel intertwiner is not fully peaked on the intrinsic geometric data encoded at a node, it nevertheless defines a legitimate quantum state, yielding a well-defined Ponzano-Regge amplitude and offering interesting insight in its structure and potential asymptotic limit. In Part II, we will consider coherent intertwiner states which have semi-classical properties in the intertwiner degree of freedom, thus defining coherent rectangular plaquettes.

\smallskip

\begin{figure}[t]
	
	\begin{tikzpicture}
	\coordinate(ai) at (0,0);
	\coordinate(bi) at (0,5);
	\draw[dotted,thick,in=90,out=90,looseness=.5] (0,0) to (3,0);
	\draw[thick,in=-90,out=-90,looseness=.5] (0,0) to node[pos=0.2,inner sep=0pt](a1){}  node[pos=0.4,inner sep=0pt](a2){}  node[pos=0.6,inner sep=0pt](a3){}  node[pos=0.8,inner sep=0pt](a4){} (3,0);
	\draw[in=90,out=90,looseness=.5,thick] (0,5) to  (3,5);
	\draw[thick,in=-90,out=-90,looseness=.5] (0,5) to node[pos=0.2,inner sep=0pt](b1){}  node[pos=0.4,inner sep=0pt](b2){}  node[pos=0.6,inner sep=0pt](b3){}  node[pos=0.8,inner sep=0pt](b4){} (3,5);
	\coordinate(af) at (3,0);
	\coordinate(bf) at (3,5);
	
	\foreach \i in {1,...,4}{
		\draw[dotted,thick,in=90,out=90,looseness=.5] (0,\i) to (3,\i);
		\draw[thick,in=-90,out=-90,looseness=.5] (0,\i) to (3,\i);
	}
	
	\draw[decoration={markings,mark=at position 0.1 with {\arrow[scale=1.5,>=stealth]{<}}},postaction={decorate}] (ai) -- node[left,pos=0.08]{g}(bi);
	\draw[decoration={markings,mark=at position 0.1 with {\arrow[scale=1.5,>=stealth]{<}}},postaction={decorate}] (a1) --node[right,pos=0.08]{g} (b1);
	\draw[decoration={markings,mark=at position 0.1 with {\arrow[scale=1.5,>=stealth]{<}}},postaction={decorate}] (a2) -- node[right,pos=0.08]{g} (b2);
	\draw[decoration={markings,mark=at position 0.1 with {\arrow[scale=1.5,>=stealth]{<}}},postaction={decorate}] (a3) -- node[left,pos=0.08]{g} (b3);
	\draw[decoration={markings,mark=at position 0.1 with {\arrow[scale=1.5,>=stealth]{<}}},postaction={decorate}] (a4) --node[left,pos=0.08]{g} (b4);
	\draw[decoration={markings,mark=at position 0.1 with {\arrow[scale=1.5,>=stealth]{<}}},postaction={decorate}] (af) -- node[right,pos=0.08]{g} (bf);
	
	\draw[very thick,->,>=stealth] (3.7,2.5) --(5.2,2.5);
	
	\coordinate(ci) at (6,2); 
	\coordinate(di) at (6,3);
	\coordinate(cf) at (9,2);
	\coordinate(df) at (9,3);
	
	\draw[decoration={markings,mark=at position 0.7 with {\arrow[scale=1.5,>=stealth]{>}}},postaction={decorate}, dashed](6,2) to[bend right,in=75,out=-105,looseness=2.5] node[scale=0.9,pos=0, left]{
	} node[scale=0.7,pos=0.75,left]{$+N_\gamma\;$} node[pos=1,above,scale=0.9]{%
	} (7.945,2.55) ;
	
	\draw[opacity=0.2,dotted,thick,in=90,out=90,looseness=.5] (ci) to (cf);
	\draw[opacity=0.2,thick,in=-90,out=-90,looseness=.5] (ci) to node[pos=0.2,inner sep=0pt](c1){}  node[pos=0.4,inner sep=0pt](c2){}  node[pos=0.6,inner sep=0pt](c3){}  node[pos=0.8,inner sep=0pt](c4){} (cf);
	\draw[opacity=0.2,in=90,out=90,looseness=.5,thick] (di) to  (df);
	\draw[opacity=0.2,thick,in=-90,out=-90,looseness=.5] (di) to node[pos=0.2,inner sep=0pt](d1){}  node[pos=0.4,inner sep=0pt](d2){}  node[pos=0.6,inner sep=0pt](d3){}  node[pos=0.8,inner sep=0pt](d4){} (df);
	
	\draw[decoration={markings,mark=at position 0.5 with {\arrow[scale=1.5,>=stealth]{<}}},postaction={decorate}] (ci) -- node[left,pos=0.5]{g}(di);
	\draw[decoration={markings,mark=at position 0.5 with {\arrow[scale=1.5,>=stealth]{<}}},postaction={decorate}] (c1) --node[right,pos=0.5]{g} (d1);
	\draw[decoration={markings,mark=at position 0.5 with {\arrow[scale=1.5,>=stealth]{<}}},postaction={decorate}] (c2) -- node[right,pos=0.5]{g} (d2);
	\draw[decoration={markings,mark=at position 0.5 with {\arrow[scale=1.5,>=stealth]{<}}},postaction={decorate}] (c3) -- node[left,pos=0.5]{g} (d3);
	\draw[decoration={markings,mark=at position 0.5 with {\arrow[scale=1.5,>=stealth]{<}}},postaction={decorate}] (c4) --node[left,pos=0.5]{g} (d4);
	\draw[decoration={markings,mark=at position 0.5 with {\arrow[scale=1.5,>=stealth]{<}}},postaction={decorate}] (cf) -- node[right,pos=0.5]{g} (df);
	
	\node[scale=1.5] at (10.2,2.5){$\times$};
	
	\coordinate(ci) at (11,2.5);
	\coordinate(cf) at (14,2.5);
	\draw[thick,in=90,out=90,looseness=.5] (ci) to (cf);
	\draw[thick,in=-90,out=-90,looseness=.5] (ci) to (cf);
	
	\node[scale=1.2] at (14.2,3){$N_{t}$};		
	
	\end{tikzpicture}
	
	\caption{The figure indicates the loops that arise with the choice of s--channel spin 0 intertwiner. Due to this choice of intertwiner the loops decouple into vertical and $N_t$ horizontal ones. The winding number $\W$  and total number $K$ of the vertical loops on the cylinder is determined by the  shift $N_\gamma$ in the periodic identification of the cylinder and the spatial size $N_x$ of the cylinder.}
	\label{fig:graph_integration}
\end{figure}
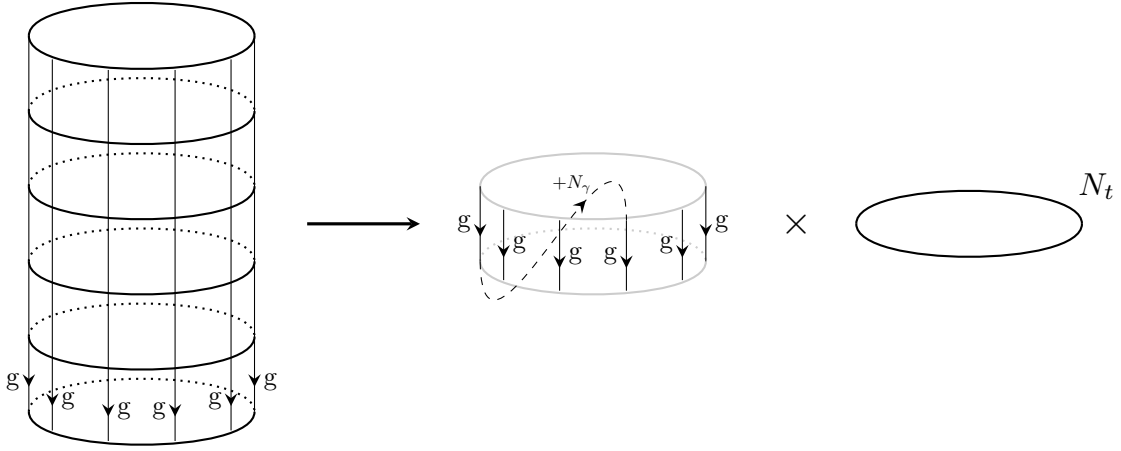

{\bf Evaluation of the partition function}  
As we anticipated, inserting the $J=0$ $s$-channel intertwiner in the Ponzano-Regge amplitude of equation \eqref{PRexact2}, results in the complete decoupling of the horizontal from the vertical links of $\Gamma$.
 Absorbing the $\su(2)$ structure maps $\varsigma$ in the intertwiners, these indeed become
\be
\left( \iota^{s|0} \otimes D^j(\varsigma) \otimes D^j(\varsigma)\right)^{m',\tl m'}{}_{m,\tl m} \,\propto\, \delta^{m'}_m \delta^{\tl m'}_{\tl m} ,
\ee
where $(m',m)$ are the magnetic indices associated to the two vertical links, and $(\tl m',\tl m)$ the magnetic indices associated to the two horizontal ones.  
This formula is evident from the graphical representation of the $s$-channel given in figure \ref{fig_split_intertwiner}: since a 0-spin link is mathematically the same as no link at all, the 0-spin intertwiner means that the two lines go through the vertex without interacting with to each other. 
As illustrated in figure \ref{fig:graph_integration}, this results in $N_{t}$ horizontal loops of spin $j$, completely decoupled from a number of vertical loops carrying the group element $g$ and winding around the torus with twist $N_{\gamma}$.

Whereas horizontal loops simply factor out, with the number of time slices $N_{t}$ only contributing with an overall volume factor, vertical loops acquire a non-trivial structure due to the interplay between the spatial size $N_{x}$ and the shift number $N_{\gamma}$ defining the twist. Putting these two ingredients together, we obtain the Ponzano--Regge amplitude in the form 
\be
\label{PRformula}
\la \text{PR} | \Phi_{j,\iota^{s|0}}  \ra
=
\frac{d_{j}^{N_{t}}}{d_j^{N_t N_x}}
\int_{\SU(2)} dg \; \chi_{j}(g^{\W})^{K}
= 
\frac{1}{d_j^{N_t (N_x-1)}}
\frac{2}{\pi} \int_{0}^{\pi} \text{d}\theta \;\sin^2(\theta) \; \chi_{j}(\W \theta)^K.
\ee
First, $\chi_{j}$ is the character in the spin-$j$ representation, which in terms of the  (half) class angle $\theta$ of $g$ reads (with a slight abuse of notation)
\be
\chi_{j}(g)\equiv\chi_j(\theta) = \f{\sin d_{j}\theta}{\sin\theta}.
\ee
And also, the two integers
\be
K := \mathrm{GCD}(N_x,N_\gamma)\qquad\text{and}\qquad \W := N_x/K.
\ee

In \eqref{PRformula}, the volume factor $d_{j}^{-N_{t}N_{x}}$ comes from the normalization of the intertwiner $\iota^{s|0}$ as given in equation \eqref{I0s}. The factor $d_{j}^{N_{t}}$, on the other hand, comes from the contribution of $\chi_{j}(\id)$ given by each time slice.
Notice that $K = \mathrm{GCD}(N_x,N_\gamma)$ is the number of vertical loops, while $W=N_{x}/K$ is their winding number around the non-trivial cycle of the torus. See figure \ref{fig:graph_integration}. For instance, the case of a vanishing twist $N_{\gamma}=0$ gives $K=N_{x}$ and $\W=1$, while the single increment case $N_{\gamma}=1$ gives $K=1$ and $\W=N_{x}$.

We have two ways to evaluate this integral. We can either express it in terms of random walks and compute it exactly, or we can extract its asymptotic behavior at large $K$'s by a saddle point approximation.
Before proceeding, it is useful to notice that the integral vanishes for odd values of $2jN_x$, since $\chi_j(W(\pi-\theta))^K = \chi_j(W(\theta-\pi))^K =  (-1)^{2j K W} \chi_j(W\theta)^K$ and $N_x \equiv W K$. For the same reason, whenever the integral does not vanish, the integrand is periodic of period $\pi$, and the integration domain can be compactified to a circle $\cong\mathbb S_1$. 

\smallskip
{\bf Exact evaluation}
 First aiming for an exact evaluation, we expand the character into a sum over exponentials by expressing the trace of the group element $g$ in the $|j,m\ra$ basis of the Hilbert space of the spin-$j$ representation,
\be
\chi_{j}(\theta)
=
\f{\sin d_{j}\theta}{\sin\theta}
=\sum_{m=-j}^{+j} \E^{2\I m\theta}
\nn\,,
\ee
from which
\be
\la \text{PR} | \Phi_{j,\iota^{s|0}}  \ra
=
\frac{1}{4\pi d_j^{N_t (N_x-1)}}
\int_{0}^{2\pi} {\d}\theta\,
(2-\E^{2\I \theta}-\E^{-2\I \theta})
\sum_{m_{1},..,m_{K}} \E^{2\I \sum_{k=1}^{K}m_{k}\W\theta}\,.
\ee
This integral has a straightforward combinatorial interpretation: we are counting the number of returns after $K$ steps to either the origin or to the positions $\pm 2$, of a random walk characterized by steps of arbitrary size between $-2\W j$ and $+2\W j$. One must always distinguish the case of a half-integer spin $j\in(\mathbb N+\frac12)$ for which each step is a odd multiple of $\W$, from the case of integer spin $j\in\mathbb N$ for which each step is an even multiple of $\W$.

Due to the measure factor $(2-\E^{2\I \theta}-\E^{-2\I \theta})$, one must also distinguish the special cases $\W=1$ and $\W=2$ from the generic case $\W\ge 3$. 

Curiously, in Part II we will find---in a very different calculation---something analogous, i.e. that the cases of a vanishing $N_\gamma=0$ and of $(N_x,N_\gamma)$ both even have to be treated with special care.

We also notice that the case $\W=1$ corresponds to computing the dimension of the intertwiner space between $K=N_x$ copies of the spin $j$. 

The calculation is simplest to perform for the case that $K=1$, when $N_x$ and $N_\gamma$ are co-prime. (Notice that for a given  $\gamma\in2\pi\mathbb Q$, there is  only one $N_x$ such that the twist $\gamma$ can be implemented exactly and $K:=\mathrm{GCD}(N_x,N_\gamma)=1$.) For $\W=N_{x}\ge 3$, we do not have to take into account the terms in $\E^{\pm2\I \theta}$, and
\be
\label{caseK1}
\la \text{PR} | \Phi_{j,\iota^{s|0}}  \ra_{ K=1 \atop N_x\ge3}
=
\frac{1}{2\pi d_j^{N_t (N_x-1)}}
\int_{0}^{2\pi} {\d}\theta\,
\sum_{m=-j}^{+j} \E^{2\I m N_x\theta}
=
\begin{cases}
1/d_j^{N_t (N_x-1)} \quad &\text{if} \quad j \in \mathbb{N} \\
0 \quad &\text{if} \quad j \in {\N+ \f12}
\end{cases}.
\ee
When $\W=N_x=1$, the integral always vanishes (as soon as the spin is non-zero, $j\ne 0$). When $\W=N_x=2$, one gets half of the volume factor $d_j^{-N_t (N_x-1)}$ when the spin $j$ is an integer and minus half of this volume factor when $j$ is a half-integer. %
The exact expression for arbitrary $K$ as a 
rational function in the spin $j$ is given in appendix \ref{app_partitionfunction} and is related to the Fourier series expansion of the cardinal sine function $\mathrm{sinc}\,\theta\equiv\sin \theta/\theta$ and to the Duflo map coefficients for $\SU(2)$.

\smallskip

\begin{figure}[!]
	\centering
	\begin{minipage}[b]{1\textwidth}
		\centering
		\includegraphics[scale = 0.7]{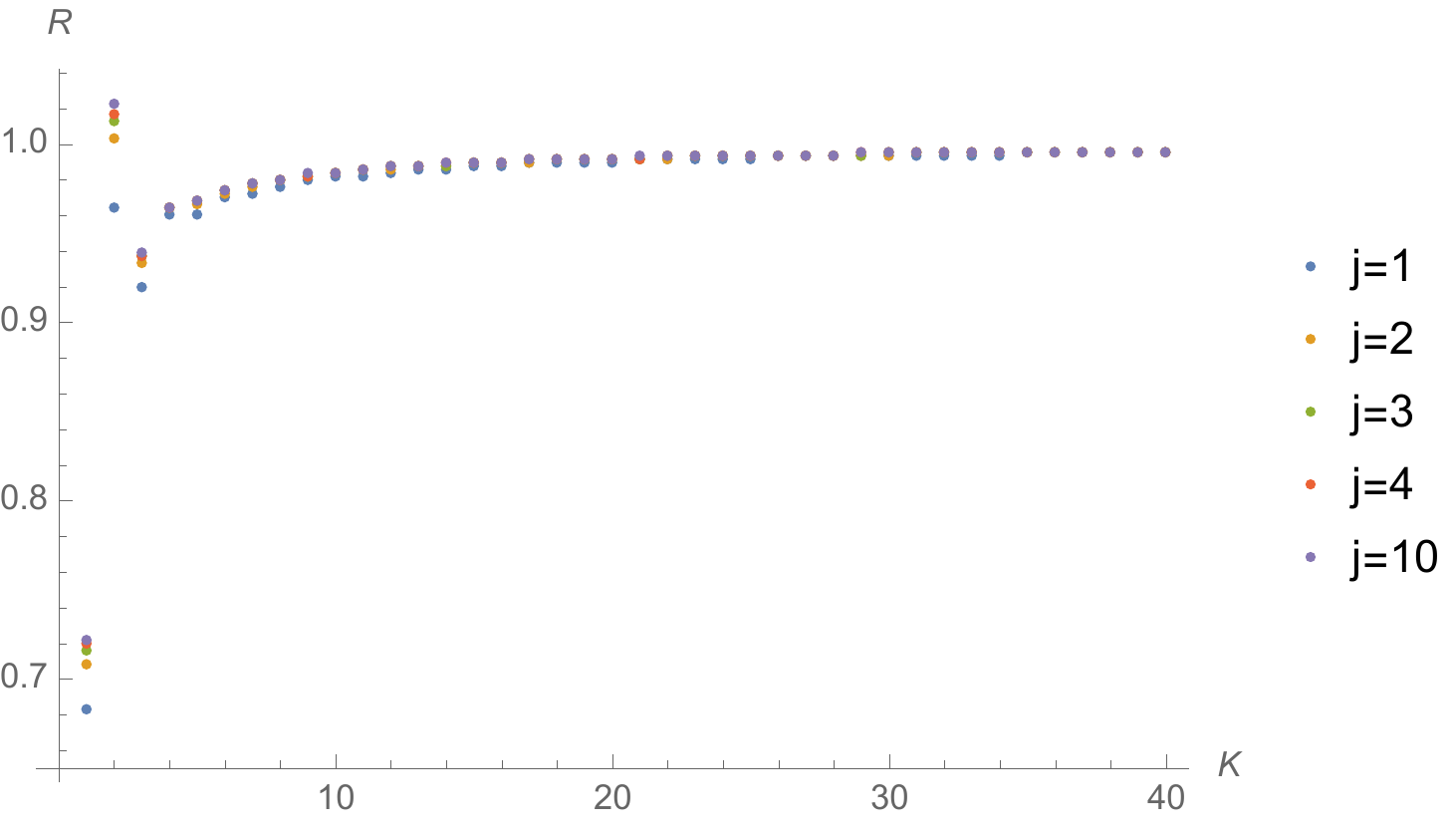}	
	\end{minipage}
	\caption{ This shows a plot of the ratio $R$ of the exact partition function divided by its asymptotic form (\ref{AsymptGen}) as a function of $K$ for several choices of integer spins. The asymptotics gives a good approximation already for $K\sim 10$. The case of half spins shows a similar structure, the only difference is that the partition function vanishes for all odd $K$.)  \label{Compare}}
\end{figure}

{\bf Asymptotic limit} 
To us, the most interesting case is that of an asymptotic limit.
We define it as a double scaling limit where both $N_{x}$ and $N_{\gamma}$ are sent to infinity while their ratio $2 \pi \frac{N_\gamma}{N_x} \rightarrow \gamma \in \mathbb{R}$ is kept finite. 
Although similar in spirit to a lattice refinement limit, this should be more correctly considered as an asymptotic limit: since the spin $j$ is fixed, the spatial size $\sim j N_{x} \ellpl$ diverges.
At this point, it becomes clear that we need to distinguish the cases where $\gamma$ is ``rational'' or not:

\begin{itemize}

\item {\it Irrational twist angle} $\gamma\in 2\pi(\R\setminus\Q)$

In this case we decide to approximate $\gamma$ via a sequence of pairs of integers $\left(N_{\gamma}^{(n)},N_{x}^{(n)}\right)_{n\in\N}$ which are always prime with each other (e.g. taking the continued fraction approximation):
\beq
 2\pi \frac{N_\gamma^{(n)}}{N_x^{(n)}} \underset{n\rightarrow\infty}\longrightarrow {\gamma}
 \,,\quad
 &&N_{\gamma,x}^{(n)}\underset{n\rightarrow\infty}\longrightarrow\infty
 \,,\quad
K^{(n)}:=\mathrm{GCD}\left(N_{\gamma}^{(n)},N_{x}^{(n)}\right)=1
\,,\nn\\
&&\quad
\text{and}\quad
\W^{(n)}=N_x^{(n)}\to\infty.
\eeq
This is exactly the case computed above in equation \eqref{caseK1}. For half-integer spins, the amplitude always vanishes. For integer spins, putting aside the volume factor $d_{j}^{-N_{t}(N_{x}-1)}$, the amplitude  remains finite, always equal to 1.

\item {\it Rational twist angle} $\gamma\in 2\pi \Q$

In this case, the twist angle can be implemented exactly on a whole sequence of discrete lattices, at least provided one chooses $N_x^{(n)}$ appropriately. To do this, one first identifies the corresponding minimal fraction and then  considers multiples of its numerator and denominator:
\be
\gamma={2\pi}\f{P}{Q}\quad\text{with}\quad \mathrm{GCD}(P,Q)=1
\quad\text{and hence}\quad
\left(N_{\gamma}^{(n)},N_{x}^{(n)}\right)=(nP,nQ).
\ee
This case is combinatorially the reverse situation compared to the case of an irrational angle, since the number of loops $K$ grows to infinity while the winding number remains constant:
\be
N_{\gamma,x}^{(n)}\underset{n\rightarrow\infty}\longrightarrow\infty
 \,,\qquad
K^{(n)}:=\mathrm{GCD}\left(N_{\gamma}^{(n)},N_{x}^{(n)}\right)=n\rightarrow\infty
 \,,\quad\text{and}\quad
\W^{(n)}:=\f{N_{x}^{(n)}}{K^{(n)}}\equiv Q
\,.
\ee
In this case, the simplest way to evaluate the Ponzano--Regge partition function is to compute its saddle point approximation at large $n$. We rewrite the integral as 
 \begin{equation}
 	\la \text{PR} | \Phi_{j,\iota^{s|0}}  \ra
	=
	\frac{2}{\pi d_j^{N_t (N_x-1)}} \int_{0}^{\pi} \text{d}\theta \sin^2(\theta) \; \E^{\f{N_x}{\W} \ln(\chi_{j}(\W\theta))} .
 \end{equation}
As noticed above, whenever non-vanishing, the integral can be considered defined on a circle with the points $\theta=0$ and $\theta=\pi$ identified.
On this circle, the exponent $\ln(\chi_{j}(\W\theta))$ reaches its maximal value of $\ln(d_{j})$ exactly $\W$ times at the locations $\theta_{l}= \f{\pi l}{\W}$, with $l=0,..,(\W-1)$. The second derivative at those points is given by the $\SU(2)$ Casimir:
\be
\f12 \left.\f{\pp^{2}\ln\chi_{j}(\W\theta)}{\pp\theta^{2}}\right|_{\theta=\theta_{l}}=-\f16(d_j^2-1)\W^2 = -\f{4}{6}\W^{2}j(j+1)\,.
\nn
\ee
In  the special case $\W=1$, we have a unique stationary point at $\theta=0$, which gives the asymptotics (recall that for $\W=1$, $N_x=K$):
\be
\la \text{PR} | \Phi_{j,\iota^{s|0}}  \ra_{\W=1}
\underset{ N_x\rightarrow\infty}\sim
\sqrt{\f{3}{2\pi}}\,\f{12 d_{j}^{N_x} {N_x}^{-\f32} }{d_j^{N_t (N_x-1)} (d_j^2-1)^{\f32}}
\, 
\ee
Here, the $N_x^{-\f32}$ decrease is due to the measure factor, $\sin^{2}\theta \sim\theta^2$ around the saddle.\footnote{This is analogous to a standard log-correction to the black hole entropy computed from the dimensions of the intertwiner spaces	 \cite{Livine:2005mw,Livine:2012cv,Kaul:2012pf}.}
In the generic case $\W\ge 2$, we sum over all the maxima and get (recall $N_x=\W K$): 
\be\label{AsymptGen}
\sum_{l=1}^{\W-1}\sin^{2}\f{\pi l}{\W}=\f \W2
\quad\Rightarrow\quad
 	\la \text{PR} | \Phi_{j,\iota^{s|0}}  \ra_{\W\geq 2}
\underset{ N_x\rightarrow\infty}\sim
{
=\sqrt{\frac{3 }{2\pi }}\,\f{2 d_{j}^{\f{N_x}{\W}} \left(\f{N_x}{\W}\right)^{-\f12}}{d_j^{\f{N_t (N_x-\W)}{\W}}\left(d_j^2-1\right)^{\f12}} 
}.
\ee
 Notice the usual decrease in $(N_x/W)^{-\f12}\equiv K^{-\f12}$ expected for a random walk, as also the volume factor $d_j^K$ is since it corresponds to $K$ steps with $d_j$ possibilities. 
Figure \ref{Compare} compares this asymptotics to the exact value of the partition function, and shows that a good approximation is already obtained for $K=N_x/\W\sim 10$.

We conclude the discussion of a rational twist angle, by observing that in this case, aside for the volume factor $d_j^{-N_t (N_x-1)}$, the partition function {exponentially {\it diverges} in the asymptotic limit $n\to\infty$, where $N_x^{(n)}=\W K^{(n)}\rightarrow\infty$. }

\end{itemize}

To summarize, we focussed on the asymptotics of the renormalized Ponzano--Regge amplitude $d_j^{N_t (N_x-1)}\,\la \text{PR} | \Phi_{j,\iota^{s|0}}\ra$, where the volume factor $d_j^{N_t (N_x-1)}$ stabilizes the limiting process (similarly to a wave-function renormalization in quantum field theory). In the continuum limit, irrational angles correspond to a trivial renormalized Ponzano-Regge amplitude, always equal to 1, while rational twists lead to divergent amplitudes and thus signify a pole in the asymptotic partition function.

This difference between rational and irrational twist angles is a crucial feature of the BMS character formula for the 3d quantum gravity partition function, see equation \eqref{BMScharacter}.  We do not however obtain the exact formula. This is due to the fact that we are {\it not} working with a semi-classical boundary state (even in the asymptotical limit) since---as previously discussed---the $J=0$ $s$-channel intertwiner is rather as far from classicality as possible. 

Nonetheless, we see that even for such a deeply quantum spin-network state, we obtain the correct pole structure for the {\it asymptotic} partition function. And this happens despite the fact that the partition function is finite for a finite-sized boundary. 

In the follow-up paper of this series, Part II, we will show how to recover both a sensible geometry and the correct pole structure by using coherent intertwiners and by considering a large spin $j$ limit.  This shows, possibly for the first time, a convergence of results between states involving infinitely many links equipped with fixed (small) spins, and states with a fixed number of links in the large spin regime.

\subsubsection{Spin $\f12$: arbitrary intertwiners \& mapping to 6-vertex model}
\label{spinhalf}

In the previous section, we have explicitly computed the Ponzano-Regge partition function for a boundary spin-network state with fixed (but arbitrary) spin $j$ and the special choice of a $J=0$ $s$-channel intertwiner. By considering small spins, however, the involved Hilbert spaces  have low dimensionality and the  characterization of the intertwiners is simple. This suggests that another approach might be possible in this case. In this section, we will therefore consider such opposite regime, where the spin-network state features a uniform choice of smallest possible spin $j=\f12$ and arbitrary intertwiner.

Geometrically, this boundary state corresponds to the finest lattice with all edge lengths set at the shortest allowed distance in Planck units, i.e. $\ell_\text{min}=\f{\sqrt{3}}{2} \ellpl$. In a sense, we are probing the deep quantum regime of the boundary geometry.

For what concerns the dual theory, we will find that this boundary state maps exactly onto the 6-vertex (or ``ice-type'') model of statistical physics, with couplings defined by the choice of intertwiner.\footnote{For details on the 6-vertex model see e.g. \cite{2016arXiv161109909D} and references therein.}
This will provide the archetype of the mapping of spin-network evaluations and quantum gravity amplitudes onto condensed matter models.

\smallskip

Henceforth, we will suppose all spins have been set to $j=\f12$. Then, the  space of 4-valent intertwiners $\mathrm{Int}(j_1=\dots=j_4=\f12)$ has dimension 2. 
Choosing a channel, $s$, $t$ or $u$, an orthonormal basis is provided by the two states with intermediate spins $J=0$ and $J=1$. 
Explicit formulas for those intertwiner states are given in details in appendix \ref{app_spinintertwiner}. 
Instead of working in the intertwiner basis $|0\ra_{s},|1\ra_{s}$, it will prove more convenient to use the over-complete basis of 0-spin intertwiners in the three channels, i.e.  $\{|0\ra_s , |0 \ra_t ,|0 \ra_u\}$, corresponding to the three pairings $(12)-(34)$, $(13)-(24)$ and $(14)-(23)$. 
Indeed, we have already experienced the power of using 0-spin intertwiners, for they decouple the lines meeting at the vertices and hence lead to simpler formulas in terms of products of characters.

Now choosing an arbitrary intertwiner corresponds to choosing its coefficients in the 0-spin intertwiner basis:
\be
|\iota\ra=\lambda|0\ra_s +\mu |0 \ra_t +\rho|0 \ra_u\,.
\ee
Of course, these three intertwiners are not independent and one can always set one of those coefficients to zero.
More generally, we know that $|0\ra_{s}-|0\ra_{t}+|0\ra_{u}=0$, implying that
\be
|\iota\ra=(\lambda+\eta)|0\ra_s +(\mu-\eta) |0 \ra_t +(\rho+\eta)|0 \ra_u
\qquad\forall\eta\in\C
\ee
always defines the same intertwiner.

We will now investigate the structure of the partition function, by setting $\mu=0$, and keeping the two arbitrary couplings $\lambda$ and $\rho$.
The intertwiner $|\iota \ra = \lambda |0\ra_s + \rho |0 \ra_u$ is graphically represented in figure \ref{spinsu}.
\begin{figure}[h!]
	\begin{tikzpicture}
	\draw (-0.6,0) node {$|\iota[\lambda,\rho]\ra$ :=}; \draw (0.5,0) node{$\lambda$};
	
	\draw (1.5,-0.5) -- node[pos=0,scale=0.7,below]{$2$} node[pos=1,scale=0.7,above]{$1$} (1.5,0.5); \draw (1,0)-- node[pos=0,scale=0.7,left]{$3$} (1.4,0); \draw (1.6,0)-- node[pos=1,scale=0.7,right]{$4$} (2,0);
	
	\draw (2.8,0) node{$+ \, \rho$};
	
\draw[rounded corners=5] (4,.5) --node[pos=0,scale=0.7,above]{$1$} (4,0)--(4.5,0)node[pos=1,scale=0.7,right]{$4$} ;
\draw[rounded corners=5] (4,-0.5)-- node[pos=0,scale=0.7,below]{$2$} (4,0)--(3.5,0) node[pos=1,scale=0.7,left]{$3$};	
	
	\end{tikzpicture}
	\caption{An arbitrary 4-valent intertwiner between four spins $\f12$ decomposes onto the non-orthogonal basis of 0-spin intertwiners in the $s$ and $u$ channel,  $|0\ra_s$ and  $|0 \ra_u$, which can be represented as the lines crossing or bending by the vertex  without interacting.}
	\label{spinsu}

\end{figure}

Recall that, geometrically, the $s$-channel intertwiner represents a (maximally fuzzy) square. Mixing it with a $u$-channel intertwiner corresponds to turning the square into (maximally fuzzy) parallelograms with the angle between adjacent edges depending on the ratio $\rho/\lambda$.

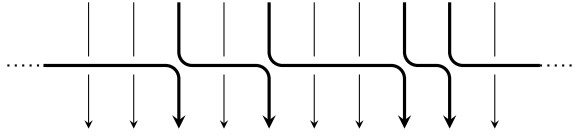
\begin{figure}[h!]
		\begin{tikzpicture}[scale=1.2]

\draw (0,.7)--(0,.1); \draw[->,>=stealth] (0,-.1)--(0,-.7);
\draw (0+.5,.7)--(0+.5,.1); \draw[->,>=stealth] (0+.5,-.1)--(0+.5,-.7);
\draw (0+1.5,.7)--(0+1.5,.1); \draw[->,>=stealth] (0+1.5,-.1)--(0+1.5,-.7);
\draw[very thick,rounded corners=5,->,>=stealth] (0+1,.7)--(0+1,0) -- (0+2,0)--(0+2,-.7);
\draw (0+2.5,.7)--(0+2.5,.1); \draw[->,>=stealth] (0+2.5,-.1)--(0+2.5,-.7);
\draw (0+3,.7)--(0+3,.1); \draw[->,>=stealth] (0+3,-.1)--(0+3,-.7);
\draw[very thick,rounded corners=5,->,>=stealth] (0+2,.7)--(0+2,0) -- (0+3.5,0)--(0+3.5,-.7);
\draw[very thick,rounded corners=5,->,>=stealth] (0+3.5,.7)--(0+3.5,0) -- (0+4,0)--(0+4,-.7);
\draw (0+4.5,.7)--(0+4.5,.1); \draw[->,>=stealth] (0+4.5,-.1)--(0+4.5,-.7);
\draw[very thick,rounded corners=5] (0+4,.7)--(0+4,0) -- (0+5,0);
\draw[very thick,rounded corners=5,->,>=stealth] (0-.5,0)--(0+1,0) -- (0+1,-0.7);

\draw[thick,dotted] (0+5,0)--(0+5.4,0);
\draw[thick,dotted] (-.5-.4,0)--(0-.5,0);
		
		\end{tikzpicture}

	\caption{One time slice: at every node, we insert either an $s$-channel intertwiner, in which case the horizontal and vertical links decouple, or a $u$-channel intertwiner in which case the incoming link bends and propagates along the time slice until it reaches the next $u$-channel intertwiner, where it bends again and propagates to the next time slice. To build the whole partition function, we need to compose such time slices together, insert the group element $g\in\SU(2)$ on all the lines going through the last time slice, take into account the twist when gluing back the final time slice with the initial one, and finally sum over all possible assignements of $s-$ and $u$-channel intertwiners at all the nodes.}
	\label{oneslice}

\end{figure}

Let us focus  on one time slice. We can expand the product of intertwiners over all nodes into a sum of configurations with each node coming equipped with either a $|0\ra_s$ or a $|0 \ra_u$ intertwiner. The $s$-channel goes straight through the time slice, while the $u$ channel bends the line which propagates along the time slice until it reaches the next $u$ channel intertwiner to exit from the time slice. This is illustrated in figure \ref{oneslice}.
To compute the full partition function, we have to stack such time slices together, not forgetting to insert the $\SU(2)$ holonomy along all the lines on the last time slice, and hence to glue back the last time slice with the first slices translated by the twist. At this point, pretty much as in the previous section, we have to follow the lines across the nodes and time slices to see the loops that they form. The computation largely reduces to a purely combinatorial problem.
Notice that, whereas the choice of a purely $J=0$ $s$-channel intertwiner ($\rho=0$) lead to a ``factorization'' of the time slices, this is not anymore the case for an arbitrary intertwiner: the time slices are now non-trivially coupled to each other.

We solve this combinatorial problem in the simple case of a vanishing twist, $N_{\gamma}=0$  and $N_t=1$. First of all, since the amplitude vanishes for an odd number of nodes in the spatial direction, we fix
\be
N_x=2M\qquad  \text{with} \quad M\in \mathbb{N}.
\ee
Then, with a look at figure \ref{oneslice}, the PR partition function is readily decomposed according to the number $k$ of $u$-channel intertwiners on the slice:
\begin{equation}
\la \text{PR} | \Phi_{j = \f12,\iota[\lambda,\rho]}  \ra
=
\sum_{k=0}^{N_x} \binom{N_{x}}{k} \lambda^{N_x-k}\rho^k 
\int_{\SU(2)} dg\,
\chi_{\f12}(g)^{N_{x}-k}\chi_{\f12}(g^{k})
\,.
\end{equation}
As derived in appendix \ref{app_onehalf}, this integral can be exactly computed and gives:
\be
\label{eqn:notwist}
\la \text{PR} | \Phi_{j = \f12,\iota[\lambda,\rho]}  \ra \,\,=\,\,\f2{M+1}\binom{2M}{M}
\lambda^{M-1}(\lambda+\rho)^{M-1}
\bigg{[}\lambda^{2}+\lambda\rho-\f M2\rho^{2}\bigg{]} .
\ee

The case of a non-vanishing twist $N_\gamma \neq 0$ and $N_t>1$, however, leads to a considerable combinatorial problem. 
This is, in our view, is best formalized using transfer matrix techniques.

On one time slice, each set of $u$-channel insertions, at the positions  $0\le x_{1}<\dots<x_{k}\le N_{x}-1$, defines a permutation of $N_{x}$ elements given by the cycle $C_{\{x_{n}\}}\equiv(x_{1},x_{2},..,x_{k})$. The sum over all such cyclic permutations over subsets of nodes defines our transfer matrix.
Now, we have to choose arbitrary $u$-channel insertions on each time slice $\{x_{n}^{(t)}\}$ for $t=0,\dots,(N_{t}-1)$ and compose with a twist, i.e. the cyclic permutation $\cC_{N_{\gamma}}$ sending every position $i$ to $(i+N_{\gamma}) \,\mathrm{mod}\,N_{x}$:
\be
\cC_{N_{\gamma}}\circ C_{\{x_{n}^{(N_{t}-1)}\}}\circ\dots\circ C_{\{x_{n}^{(0)}\}}
\,.
\nn
\ee

The last ingredient before the identification of the first and last time slices, is the introduction of and the integration over the holonomy $g\in\SU(2)$. This sits on all the vertical edges belonging to the final time slice. This means that we have to evaluate the integral
\be
\int_{\SU(2)} dg\,
\prod_{i=0}^{N_{x}-1}D^{\f12}_{a_{i}b_{i}}(g)\,.
\ee
This integral, also known as the ``Haar intertwiner'', is non-vanishing if and only if $N_{x}$ is even, $N_{x}=2M$ as above. It can also be expressed purely in terms of permutations $\omega$'s, which match the incoming magnetic indices $a_{i}$ with permuted outgoing magnetic indices $b_{\om(i)}$. As explained in appendix \ref{app:permutations}, the Haar intertwiner is given in terms of the characters $s_{[M,M]}$ of the symmetric group of $N_{x}$ elements $S_{N_{x}}$ in the representation associated to the partition of the integer $N_{x}=2M$ as $M+M$. That is
\be
\int_{\SU(2)} dg\,
\prod_{i=0}^{2M-1}D^{\f12}_{a_{i}b_{i}}(g)
=
\f{M!}{(2M)!}
\sum_{\om\in S_{2M}} s_{[M,M]}[\om] \prod_{i=0}^{2M-1}\delta_{a_{i}b_{\om(i)}}\,.
\ee
Taking the trace of this expression, we recover the dimension of the intertwiner spaces between $2M$ spins $\f12$, given by the Catalan numbers:
\be
\int_{\SU(2)} dg\,\big{(}\chi_{\f12}(g)\big{)}^{2M}
=
\f1{M+1}\binom{2M}{M}
\,.
\ee 
As explained in appendix \ref{app:permutations}, the character for arbitrary permutations $s_{[M,M]}(\om)$ can be computed using Young tableaux. Putting all these ingredients together, the PR amplitude for $N_{t}$ time slices and a twist $N_{\gamma}$ is expressed as:
\be
\la \text{PR} | \Phi_{j = \f12,\iota[\lambda,\rho]}  \ra
\,=\,
\sum_{\{x_{n}^{(t)}\}}
\lambda^{N_{x}N_{t}-\# x}
\rho^{\# x}
s_{[M,M]}
\Bigg{[}
\Big{(}
\cC_{N_{\gamma}}\circ C_{\{x_{n}^{(N_{t}-1)}\}}\circ\dots\circ C_{\{x_{n}^{(0)}\}}
\Big{)}^{-1}\Bigg{]}
\,,
\label{eqn:permut}
\ee
where $\# x$ is the total number of $u$-channel intertwiner insertions on the whole lattice, i.e. the sum over all time slices of the cardinal of the sets $\{x_{n}^{(t)}\}$.

Studying the statistics of the composition of cycles is definitely a non-trivial combinatorial problem. Since we are mostly interested in the thermodynamical limit $N_{x},N_{t}\rightarrow\infty$, the most efficient approach is to look for a mapping of our spin evaluation onto known statistical models. 
As we have already announced, this spin-network evaluation for spin $\f12$ maps onto the 6-vertex model, which is integrable and for which the transfer matrix is known (and actually expressed and solved in terms of sums over permutations, see e.g. \cite{2016arXiv161109909D}).

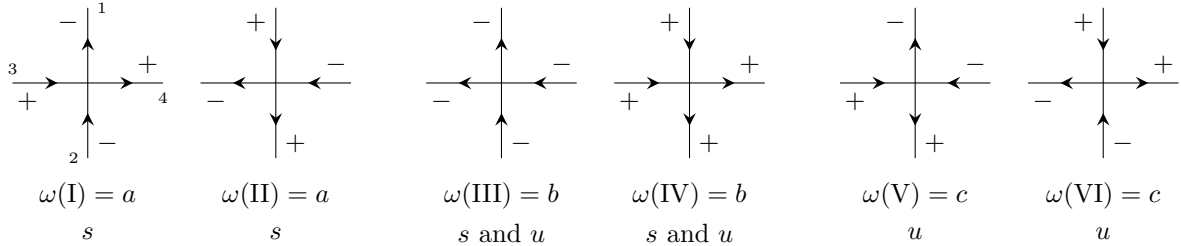
\begin{figure}[t]
	\begin{center}
		\begin{tikzpicture}[scale=1]
			\draw[decoration={markings,mark=at position 0.3 with {\arrow[scale=1.5,>=stealth]{>}}},decoration={markings,mark=at position 0.8 with {\arrow[scale=1.5,>=stealth]{>}}},postaction={decorate}]  (0,-1) node[left ]{\tiny{2}}-- node[pos=0.1,right]{$-$} node[pos=0.9,left]{$-$} (0,1)node[right ]{\tiny{1}}; \draw[decoration={markings,mark=at position 0.3 with {\arrow[scale=1.5,>=stealth]{>}}},decoration={markings,mark=at position 0.8 with {\arrow[scale=1.5,>=stealth]{>}}},postaction={decorate}] (-1,0)node[above ]{\tiny{3}}-- node[pos=0.1,below]{$+$} node[pos=0.9,above]{$+$} (1,0)node[below ]{\tiny{4}};
			\draw (0,-1.5) node[scale=1]{$\omega(\mathrm I) =a$}; \draw (0,-2) node[scale =1]{$s$};
			
			\draw[decoration={markings,mark=at position 0.3 with {\arrow[scale=1.5,>=stealth]{<}}},decoration={markings,mark=at position 0.8 with {\arrow[scale=1.5,>=stealth]{<}}},postaction={decorate}]  (2.5,-1) -- node[pos=0.1,right]{$+$} node[pos=0.9,left]{$+$} (2.5,1); 
			\draw[decoration={markings,mark=at position 0.3 with {\arrow[scale=1.5,>=stealth]{<}}},decoration={markings,mark=at position 0.8 with {\arrow[scale=1.5,>=stealth]{<}}},postaction={decorate}] (1.5,0)-- node[pos=0.1,below]{$-$} node[pos=0.9,above]{$-$} (3.5,0);
			\draw (2.5,-1.5) node[scale=1]{$\omega(\mathrm{II}) =a$}; \draw (2.5,-2) node[scale =1]{$s$};

			\draw[decoration={markings,mark=at position 0.3 with {\arrow[scale=1.5,>=stealth]{>}}},decoration={markings,mark=at position 0.8 with {\arrow[scale=1.5,>=stealth]{>}}},postaction={decorate}]  (5.5,-1) -- node[pos=0.1,right]{$-$} node[pos=0.9,left]{$-$} (5.5,1); 
			\draw[decoration={markings,mark=at position 0.3 with {\arrow[scale=1.5,>=stealth]{<}}},decoration={markings,mark=at position 0.8 with {\arrow[scale=1.5,>=stealth]{<}}},postaction={decorate}] (4.5,0)-- node[pos=0.1,below]{$-$} node[pos=0.9,above]{$-$} (6.5,0);
			\draw (5.5,-1.5) node[scale=1]{$\omega(\mathrm{III}) =b$}; \draw (5.5,-2) node[scale =1]{$s$ and $u$};
			
			\draw[decoration={markings,mark=at position 0.3 with {\arrow[scale=1.5,>=stealth]{<}}},decoration={markings,mark=at position 0.8 with {\arrow[scale=1.5,>=stealth]{<}}},postaction={decorate}]  (8,-1) -- node[pos=0.1,right]{$+$} node[pos=0.9,left]{$+$} (8,1); 
			\draw[decoration={markings,mark=at position 0.3 with {\arrow[scale=1.5,>=stealth]{>}}},decoration={markings,mark=at position 0.8 with {\arrow[scale=1.5,>=stealth]{>}}},postaction={decorate}] (7,0)-- node[pos=0.1,below]{$+$} node[pos=0.9,above]{$+$} (9,0);
			\draw (8,-1.5) node[scale=1]{$\omega(\mathrm{IV}) =b$}; \draw (8,-2) node[scale =1]{$s$ and $u$};
			
			\draw[decoration={markings,mark=at position 0.3 with {\arrow[scale=1.5,>=stealth]{<}}},decoration={markings,mark=at position 0.8 with {\arrow[scale=1.5,>=stealth]{>}}},postaction={decorate}]  (11,-1) -- node[pos=0.1,right]{$+$} node[pos=0.9,left]{$-$} (11,1); 
			\draw[decoration={markings,mark=at position 0.3 with {\arrow[scale=1.5,>=stealth]{>}}},decoration={markings,mark=at position 0.8 with {\arrow[scale=1.5,>=stealth]{<}}},postaction={decorate}] (10,0)-- node[pos=0.1,below]{$+$} node[pos=0.9,above]{$-$} (12,0);
			\draw (11,-1.5) node[scale=1]{$\omega(\mathrm{V}) =c$}; \draw (11,-2) node[scale =1]{$u$};
			
			\draw[decoration={markings,mark=at position 0.3 with {\arrow[scale=1.5,>=stealth]{>}}},decoration={markings,mark=at position 0.8 with {\arrow[scale=1.5,>=stealth]{<}}},postaction={decorate}]  (13.5,-1) -- node[pos=0.1,right]{$-$} node[pos=0.9,left]{$+$} (13.5,1); 
			\draw[decoration={markings,mark=at position 0.3 with {\arrow[scale=1.5,>=stealth]{<}}},decoration={markings,mark=at position 0.8 with {\arrow[scale=1.5,>=stealth]{>}}},postaction={decorate}] (12.5,0)-- node[pos=0.1,below]{$-$} node[pos=0.9,above]{$+$} (14.5,0);
			\draw (13.5,-1.5) node[scale=1]{$\omega(\mathrm{VI}) =c$}; \draw (13.5,-2) node[scale =1]{$u$};
			
		\end{tikzpicture}
	\end{center}
	\caption{The 6 arrow configurations around a node in the 6-vertex model and the 0-spin intertwiner channel that they correspond to, with the magnetic moment $m$ on each link.}
	\label{fig:6_vertex}
\end{figure}

\smallskip
{\bf Mapping onto the 6-vertex model} 
The 6-vertex model is defined on a regular square lattice. The variables of this model  are a sign assigned to each edge of the lattice, which can be thought of as the orientation of the edges as represented on figure \ref{fig:6_vertex}. Each node is required to have the same number of ingoing and outgoing edges. In other words, at each node there are two incoming and two outgoing arrows. The partition function is defined as a sum over all admissible arrow configurations. 
As drawn on figure \ref{fig:6_vertex}, this gives 6 allowed vertex configurations. At each node, the simultaneous reversal of all four arrows is considered to be a symmetry of the model. This leaves us with three pairs of node configurations to which one associates three weights $a$, $b$ and $c$. Finally the partition function is given as
\be
Z_\text{6-vertex} \,=\, \sum_{\text{arrows}} a^{(\#_{\rm I}+\#_{\rm II})} \,b^{(\#_{\rm III}+\#_{\rm IV})}\, c^{(\#_{\rm V}+\#_{\rm VI})}
\ee
where $\#_i$ with $i={\rm I},\ldots,{\rm VI}$ represents the number of vertices in configuration $i$.

Let us compare to the PR partition function.Ignoring for the moment the group element $g$ attached to the last time before, in the PR partition function one sums over magnetic indices $m=\pm\f12$ on each link\footnotemark{} and the weights are determined by the choice of intertwiners. To map the spin-$\f12$ PR model to the 6-vertex model, we match magnetic index configurations with arrow configurations. Specifically, we identify $m=+\f12$ with arrows that point to the right or downwards (along the time direction) and $m=-\f12$ to arrows that point to the left or upwards (opposite to the time direction). 
\footnotetext{Here we consider intertwiners as $\SU(2)$-invariant maps $V^{\f12}\otimes V^{\f12} \rightarrow V^{\f12}\otimes V^{\f12}$ and not as $\SU(2)$ singlet states $(V^{\f12})^{\otimes 4}\rightarrow\C$. And  those maps are always composed in the direction of increasing $t$ and $x$ (in our figures, to the right and downwards).
}

We furthermore choose to expand the general spin $\f12$ intertwiner on the spin-0 intertwiners in the $s$-channel and $u$-channel, leaving the $t$-channel aside, as explained above.\footnotemark{} This general intertwiner can be parametrized in such a way that the PR partition function amounts to contracting the following four-valent tensors associated to the nodes of the graph ( we keep the notation $\iota$, although it already accounts for the presence of $\su(2)$ structure map on each link):
\footnotetext{In order to include the $t$-channel in the mapping by considering a generic intertwiner as:
$$
\iota_{m_1m_2m_2m_4} \,=\,
\lambda \,\delta_{m_1 m_2} \delta_{m_3 m_4}
+ \mu \,\delta_{m_1 m_3} \delta_{m_2 m_4} 
+ \rho \,\delta_{m_1 m_4} \delta_{m_2 m_3}
\,, 
$$
we would need to introduce a new pair of node configurations corresponding to four incoming (or outgoing) arrows. This readily leads to a mapping onto an 8-vertex model. This is however not necessary in our spin-$\f12$ case, but might turn out to be useful for higher spin evaluations.
}
\be
\iota{[\lambda,\rho]}_{m_1m_2m_2m_4} \,=\, \lambda \,\delta_{m_1 m_2} \delta_{m_3 m_4} + \rho \,\delta_{m_1 m_4} \delta_{m_2 m_3} \, . 
\ee
Notice that this tensor only gives non-vanishing weights for the six configurations allowed by the 6-vertex model (after translating arrows into magnetic indices in the way just described). Evaluating this tensor for each of the six allowed node configurations, we obtain the following weights for the 6-vertex models in terms of our intertwiner parametrization
\be
a=\lambda \,,
\quad
b=\lambda+\rho \,,
\quad
c=\rho \, .
\ee
The expression of the $b$-coupling is actually reminiscent of the factors $(\lambda+\rho)$ of the no-twist formula \ref{eqn:notwist} derived earlier.

We can then use all the results obtained for the 6-vertex model to study our spin-$\f12$ Ponzano--Regge amplitude, especially the diagonalization and thermodynamic limit of its transfer matrix \cite{2016arXiv161109909D}.
 Of particular interest will be the gravitational interpretation of phase transitions in the model. However, we defer its study to future work.

Before moving to the last part of this section, let us stress a key point. The Ponzano--Regge partition function on the twisted solid torus does {\it not} get mapped onto the 6-vertex partition function on a twisted torus, for it requires the insertion of the Haar intertwiner on the last time slice.\footnote{Recall that the Haar intertwiner is just another name for the insertion of and integration over the global $\SU(2)$ element $g$ associated to the last time slice before regluing.} This should correspond to the insertion of a specific  (non-local) operator in the 6-vertex model. If we call $\cF$ the 6-vertex transfer matrix, this means that we should not only look at the  6-vertex partition function $\mathrm{Tr}\left( \cF^{N_{t}}\right)$, but rather study $\mathrm{Tr}\left( \cF^{N_{t}}\cT_\gamma\cG \right)$ where $\cT_\gamma$ and $\cG$  are the operators implementing the torus twist and the Haar intertwiner, respectively.
This step is crucial, since it is precisely how the integration over all possible non-trivial (bulk) monodromies is taken into account.

\smallskip
{\bf Mapping onto a loop model (and the $j=1$ case)} 
Finally, another possible way to study the structure of this spin-$\f12$ Ponzano--Regge amplitude.
Instead of focusing on the weights associated to the single nodes, let us go back to the idea of keeping track of the lines that get concatenated into paths on the lattice, once we make an assignement of $J=0$ intertwiners---in the $s$-, $t$-, and $u$-channels---to the nodes.
In this case, following the lines through the various nodes and across the periodic boundary conditions, we see  that they must close into loops. 
These loops form a partition of the set of all links of the lattice and can intersect each other\footnotemark{} (at the vertices carrying a $s$-intertwiner).
\footnotetext{We could avoid loop intersections by considering only the t and u channels, but loops would still touch at the vertices.}
Temporarily forgetting the role of the bulk monodromy $g$, we see that each loop then carries a constant weight of $d_{\f12}=2$  times powers of $\mu^{1/2}$ and $\rho^{1/2}$ counting their left or right bends, and powers of $\lambda^{1/2}$ counting the number of intersections with other loops.
These configurations and weights define a so-called loop model \cite{Fendley2006}. 
Once again, the issue is that the PR partition function on the solid torus involves the introduction of and integration over the bulk monodromy $g$.
This can be dealt with either by assigning a weight $\chi_{\f12}(g^W)$ to each loop winding $W$ times around the torus and hence integrating over $g$,\footnote{This is equivalent to assigning to a loop the weight $\chi_{\f12}(g^W)$ with $W$ its intersection number with the ``last'' time slice.} or by inserting a Haar intertwiner operator (e.g. decomposed in terms of link permutations) which reshuffle the loop structure on the re-glued torus. 

The disadvantage of the loop description is, of course, intrinsic to the involved nature of the weight associated to loops winding around the bulk non-trivial cycle of the torus (a characteristic that is likely to be common to all statistical model we can map onto). 

Conversely, a clear advantage of the loop formulation is that it extends to the analysis of the spin-1 PR partition function.  Indeed, as explained in detail in appendix \ref{app:spinone}, the space of 4-valent intertwiners between four spins 1 is of dimension 3, so that the 0-spin intertwiners in the three channels form a (non-orthogonal) basis of $\mathrm{Int}(j_1=\dots=j_4=1)$. Indeed, an arbitrary intertwiner between four spin-1 representations can be decomposed on the intertwiner basis $\{|0\ra_s , |0 \ra_t ,|0 \ra_u\}$:
\begin{equation}
	| \iota_{1111} \ra = \lambda | 0 \ra_s + \mu | 0 \ra_t + \rho | 0 \ra_u \,.
\end{equation}
Thus we can express the resulting spin-1 PR partition for the choice of a general intertwiner on the boundary lattice in terms of a loops as explained above. It might, thus, be interesting to notice the possibility of rewriting loop models in terms of Grassmannian integrals (e.g. \cite{Samuel1980}). We leave the study of these possibilities to further work.

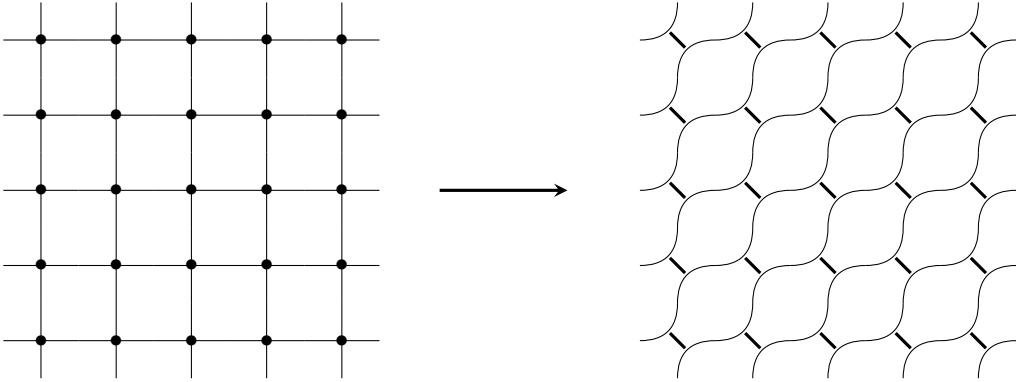
\begin{figure}[t]

		\begin{tikzpicture}[scale=1]

		\foreach \i in {0,...,4}{
			\foreach \j in {0,...,4}{
				\draw (\i,\j) node {$\bullet$};
				\draw (\i,\j-.5) --(\i,\j+.5);
				\draw (\i-.5,\j) --(\i+.5,\j);
			}
		}
		\draw[->,very thick, >=stealth] (5.3,2)--(7,2);
		\end{tikzpicture}
\hspace*{7mm}
		\begin{tikzpicture}[scale=1]

		\foreach \i in {0,...,4}{
			\foreach \j in {0,...,4}{
				\draw[in=180,out=90,looseness=1.2] (\i,\j-.5) to node[midway,inner sep=0pt](a){}  (\i+0.5,\j) ;
				\draw[in=-90,out=0,looseness=1.2] (\i-.5,\j) to node[pos=0.5,inner sep=0pt](b){} (\i,\j+.5) ;
				\draw[very thick] (a)--(b);
			}
		}

		\end{tikzpicture}

	\caption{From the square lattice to the honeycomb lattice: for a homogeneous spin $j=1$ on all the edges, we choose as 4-valent intertwiner at the vertices the  spin-1 intertwiner in the t-channel, thus unfolding the vertices and inserting a little intermediate edge instead. In the end, for this specific choice of intertwiner, we have a regular honeycomb lattice, with a homogeneous spin $j=1$ on all the edges. This configuration can be interesting when studying the relation of spin network evaluations with condensed matter models. }
	\label{honey}
\end{figure}

\smallskip
We conclude this presentation with a last remark. So far, we took advantage of the simplicity of 0-spin intertwiners, which allow to decouple the graph lines meeting at the nodes thus yielding simple expressions involving dimensional factors and $\SU(2)$ characters. Nonetheless, one should not neglect the fact that intertwiner with higher spin in the recoupling channel can also admit interesting geometrical interpretation. For instance, in the case of a spin-network state with $j=1$ on all links, it is possible to choose a state with $J=1$ intertwiner in the $t$- or $u$-channel. This configuration readily maps the square lattice onto a honeycomb one, as shown in figure \ref{honey}. This honeycomb lattice is dual to a triangulation of the boundary torus, of the type considered in the  computation of the  partition function in 3d Regge calculus \cite{BonzomDittrich2016} (there  is, however, an important difference between this two cases: the present spin-network state is dual to an equilateral ``quantum'' triangulation, while the one involved in the cited paper features right triangles coming from a subdivided rectangular lattice). 

Not even mentioning the possibility of spin superpositions, the discussion of the last section can  only give a small impression of the wealth of statistical models that can be encoded into the PR partition function via an appropriate choice of the boundary state.

%
%
%
%

\section{Conclusion \& Outlook}\label{sec_conclusions}

In this paper, the first in a series, we started a systematic investigation of holographic dualities in 3d non-perturbative gravity. 
Our focus is on the properties of 3d quantum gravity as described by the Ponzano--Regge model \cite{PR1968}, and on its relation to other approaches.  The general philosophy is to use the solvability of 3d gravity to derive directly from the non-perturbative partition function a wealth of dual theories defined by the choice of boundary state (that is---more loosely speaking---of boundary conditions). Note that this investigation is performed a priori only for {\it finite} boundaries. Standard asymptotic boundary conditions are expected to arise as a fined tuned choice of boundary states,  whose identification is part of our quest. The use of non-asymptotic boundaries allows us to probe the deep quantum regime of the boundary state. In this context, for the construction, analysis, and interpretation of the bulk and boundary quantum geometries, we make use of a rich set of tools developed over the years within Loop Quantum Gravity. The full power of these geometric techniques is maybe best exemplified in the second paper of the series, Part II \cite{Part2}.  

Here is a short summary of the main results of this first exploration of the above setting in the context of the twisted solid torus manifold.
It is followed by a brief outlook on possible future investigations.

{\bf Amplitude as a function of the twist angle: pole structure} 
As a first test, we considered the question of whether our setting can reproduce the structure of the partition function on the  twisted solid torus as computed by a range of other methods---which we also reviewed in the first part of the paper. We found that already a very simple, and somewhat non-geometric, choice of boundary state can in an appropriate limit reproduce the characteristic pole structure of the one-loop partition function of 3d gravity, seen as a function of the (Dehn) twist angle (notice that, in the case of a vanishing-cosmological-constant, the twisting angle is what is left of the torus modular parameter).
The state involves a specific choice of intertwiner, representing a discretization of the toroidal boundary by ``fuzzy squares'', and of a homogeneous---but otherwise arbitrary---spin.
The appropriate limit, on the other hand, consists of taking an infinitely fine lattice. As a consequence of keeping a fixed value of the spin, this limit corresponds to an asymptotic, infinite radius, limit.
The result is that in the limit, the renormalized amplitude, develops poles at every rational value of the twist angle.
This result is maybe more surprising at the light of the fact that the Ponzano--Regge partition function is finite by construction for all finite boundaries. 
Moreover, it is often declared that the one-loop approximation in the field theoritical description is perturbatively exact. 
The result here shows that for the Ponzano--Regge quantization of three dimensional gravity, this can hold only for asymptotic boundaries. At this purpose, we invite the reader to consult also the more extensive discussion in Part II, where---using states with a neat geometrical interpretation---we recover the pole structure in the ``orthogonal'' asymptotic limit: at fixed lattice and large spin values.
\footnote{Of course, an infinite lattice limit needs also to be eventually considered in order to recover poles at {\it all} rational twist angles.}
This is, at the best of our knowledge, the first time that a regime involving many small (boundary) spins can be matched to a regime involving large (boundary) spins. 

{\bf Boundary with minimal spins and the 6-vertex model} 
We also analyzed in some detail the partition function for a spin-network state featuring only minimal spins $j=\f12$. This corresponds to a lattice with minimal discretization scale equal to $\ell_\text{min}=4\pi\sqrt{3} G_\text{N} \hbar$. In this case we keep the choice of intertwiner completely arbitrary, albeit uniform. What we showed is that the resulting partition function can be mapped to an interesting combinatorial problem, which can also be mapped either onto a version of the 6-vertex (or ``ice-type'') model, or alternatively onto a loop model. 
In both cases, in order to take into account the possibility of monodromies around the single (bulk) non-contractible cycle, both models have to be augmented by the insertion of specific non-local operator, which winds around the opposite cycle of the twisted boundary torus.
This operator is essentially a rewriting of the so-called Haar intertwiner, and can be expressed in terms of combinatorial objects.
A similar mapping onto a loop model can be achieved for spins $j=1$ and arbitrary intertwiner. Larger spins seem to require new techniques. 

{\bf Outlook on further dual boundary theories} 
In the second paper of this series \cite{Part2},  we will consider a large spin asymptotics based on so-called  LS coherent states \cite{Livine:2007vk}. This will lead to a boundary theory readily written in the form of a non-linear $\SU(2)$ (or $\SO(3)$) sigma-model.  In the third paper \cite{Part3}, on the other hand, we will consider states involving superpositions of spins and thus show that a particular choice of superposition allows for an exact solution of the partition function. Such boundary states lead again to a different class of boundary theories, now involving also spin labels as variables, as in \cite{DittrichMartinbenitoSchnetter2013,Dittrich:2013aia}. These theories can be dualized to fermionic systems admitting themselves a loop model formulation \cite{BonzomCostantinoLivine2015}.

{\bf Outlook on general research directions} 
The work presented in this paper hints many directions in which to push and generalize our investigations. For instance, a relevant question is whether and how, for appropriate asymptotic boundary states  (possibly in a Lorentzian version of the PR model \cite{Freidel:2000uq,Girelli:2015ija}), we can recover a dual theory featuring the (centrally extended) BMS$_3$ group as symmetry. 
Or what kind of symmetries characterize dual boundary theories associated to finite boundaries.
This question can probably be clarified by studying a (formal) continuum limit for the boundary theory, or maybe by tuning the boundary state to approach criticality in the dual statistical model description.

Another direction consists to better studying the coupling of the dual theory to the boundary (quantum) geometry encoded in the spin-network state. 
For this, it will be helpful to study (perturbative) deformations of the homogeneous states investigated here and in the following paper. This can happen by perturbing the spin and intertwiner labels of the boundary state (similarly to what is discussed in \cite{BonzomDittrich2016}), or by introducing actual modifications of the boundary graph connectivity. 
 From the geometric viewpoint, a local change in the discretization can be seen as the introduction of a new geometric excitation on the top of a continuum state where the absence of graph features in a region means that there the fields are in their ``vacuum'' state. From the viewpoint of the dual theory the change in the discretization can have drastic consequences: e.g. certain statistical models  can feature frustration on trivalent lattices but do not show such frustration phenomena on fourvalent lattices. In this sense, our framework  opens brand new questions which are unexplored in the quantum gravity literature and that can be approached e.g. perturbatively, by introducing local changes in the graph.

Crucial for the present work has been the fact that 3d gravity is a topological theory described by a 3d $BF$-theory. 
Similarly, we can envisage a study of dualities in 4d $BF$ theories, which feature the same kind of boundary states discussed here.
This would probably turn out to be interesting from a 4d gravitational perspective too, for $BF$ theory serves as the basis for the construction of gravitational spin foam models in four dimensions. 

Going back to the 3d case, it would of course also be interesting to generalize the considerations here firstly to include a Lorentzian signature, and secondly a non-vanishing cosmological constant. The former can be done by replacing the $\SU(2)$ group by $\SL(2,\mathbb{R})$, while the latter can in principle be achieved by introducing various quantum deformations of these groups.

\smallskip
We believe that the current work  not only opens up an exciting interface between non-perturbative gravity approaches and the extensive work on the  AdS/CFT correspondence, but also features interesting intersections with the theory of integrable systems and more generally of 2d (for now) condensed matter and statistical physics models.

\section*{Acknowledgement}
The authors thank FQXi  for the support of the workshop  "Holography in background independent approaches" in Paris, October 2016, where part of this work was initiated.  BD and AR thank Laurent Freidel for discussions, CG acknowledges discussions with Clement Delcamp.  
This work is supported by Perimeter Institute for Theoretical Physics. Research at Perimeter Institute is supported by the Government of Canada through Industry Canada and by the Province of Ontario through the Ministry of Research and Innovation.

\appendix

\section{On-shell actions \label{app_onshellactions}}

\subsection{Geometries of $\text{AdS}_3$ and the BTZ black holes \label{app_TAdS_BTZ}}

This section is adapted from references \cite{Banados:1992gq,CarlipTeitelboim1995,BanadosMendez1998}  to the Euclidean case. We consider the ``Euclideanized'' version of region I  of \cite{Banados:1992gq}, that is the outer region of the Lorentzian BTZ black hole.

Consider $(3+1)$ dimensional Minkowski space
\be
\d s^2 = \ellcc^2\Big( - \d u^2 + \d v^2 + \d x^2 + \d y^2 \Big),
\ee
(we inserted the cosmological scale for later convenience) and in it the upper sheet $u\geq1$ of the hyperboloid 
\be
- u^2 + v^2 + x^2 + y^2 = -1.
\label{eq_hyperboloid}
\ee
Now, identify points on the hyperboloid according to
\be
\Big( u + v, u - v , x + i y \Big) \sim \Big( \E^{\ellcc^{-1}\beta} (u+v), \E^{\ellcc^{-1}\beta} (u-v), \E^{i\gamma} (x + i y) \Big).
\label{eq_identifuvxy}
\ee
The above corresponds to the combination of a boost in the $u$ direction by a velocity $\beta$ and a rotation in the $(xy)$ plane, i.e. around the $u$-axis, by an angle $\gamma$. 

Thermal $\text{AdS}_3$ can be read off by the following parametrization of this space:
\begin{subequations}
\begin{align}
u & = \cosh r \;\cosh t\\
v & = \cosh r \;\sinh t\\
x & = \sinh r \;\cos \phi\\
y & = \sinh r \;\sin \phi
\end{align}
\label{eq_TAdScoords}
\end{subequations}
with the following identifications:
\be
(r, \phi + it + 2\pi )\sim (r, \phi + i t) \sim (r, \phi + i t + 2\pi \tau_\text{TTAdS}) ,
\label{eq_identifcyl}
\ee
where we introduced
\be
\tau_\text{TTAdS} = \frac{1}{2\pi}\Big( \gamma + i \ellcc^{-1}\beta \Big).
\ee
This parameter is readily interpreted as the modulus of the torus defined by $(t,\phi)\in[0,\ellcc^{-1}\beta ]\times[0,2\pi]$ at fixed a radius $r=a$. 
On the hyperboloid \eqref{eq_hyperboloid}, the induced metric then reads
\be
\d s^2 = \ellcc^2\Big( \cosh^2 r \; \d t^2  + \d r^2 + \sinh^2 r \; \d \phi^2 \Big).
\ee

The (exterior region of the) Euclidean BTZ black hole is on the other hand obtained by introducing the following parametrization of the hyperboloid \eqref{eq_hyperboloid}
\begin{subequations}
\begin{align}
u & = \sqrt{A(\tl r)} \;\cosh \tl \Phi\\
v & = \sqrt{A(\tl r)} \; \sinh \tl \Phi \\
x & = \sqrt{B(\tl r)}\;\cos \tl T\\
y & = -\sqrt{B(\tl r)} \;\sin \tl T
\end{align}
\label{eq_BTZcoords1}
\end{subequations}
Here, the functions $A$ and $B$ are defined as
\be
A(\tl r) = \frac{\tl r^2 - \tl r_-^2}{\tl r_+^2 - \tl r_-^2}
\qquad\text{and}\qquad
B(\tl r) = \frac{\tl r^2 - \tl r_+^2}{\tl r_+^2 - \tl r_-^2},
\ee
with $\tl r\geq \tl r_+$, so that $u\geq1$, and $i\tl r_->0$ (the choice of convention will be clear later), while the coordinates $\tl T$ and $\tl \Phi$ are further decomposed into 
\be
\tl \Phi =  \tl r_+ \tl \phi'  + i \tl r_-  \tl t 
\qquad\text{and}\qquad
\tl T =  \tl r_+ \tl t - i\tl  r_- \tl \phi' .
\label{eq_BTZcoords2}
\ee

In these coordinates the metric on the hyperboloid reads
\begin{subequations}
\begin{align}
\d s^2 %
& =\ellcc^2\left(   \frac{\tl r^2 \d \tl r^2}{(\tl r^2 - \tl r_+^2)(\tl r^2 - \tl r_-^2)} + A \d\tl \Phi ^2 + B\d \tl T^2    \right)\\
& = \ellcc^2\left(   %
\frac{\tl r^2 \d \tl r^2}{(\tl r^2 - \tl r_+^2)(\tl r^2 - \tl r_-^2)} %
+ (\tl r^2 - \tl r_+^2 - \tl r_-^2)\d \tl t ^2 %
+ \tl r^2 (\d \tl \phi')^2 %
+ 2 i\tl r_- \tl r_+ \d \tl t \d\tl \phi' %
\right)
\end{align}
\end{subequations}
Define
\be
M =\tl r_+^2 + \tl r_-^2,
\qquad
J = 2 i \tl r_- \tl r_+
\qquad\text{and}\qquad
N^2(\tl r) = \tl r^2  - M - \frac{J^2}{4\tl r^2},
\ee
and notice that $\tl r_\pm^2$ are the positive and negative solution to the equation $N^2=0$, respectively.
Then,
\begin{subequations}
\begin{align}
\d s^2  & = \ellcc^2\left(   %
N^{-2}\d\tl r^2 %
+ \Big(N^2 + \frac{J^2}{2\tl r^2}\Big) \d \tl t ^2 %
+ \tl r^2 (\d \tl \phi')^2 %
+ J  \d \tl t \d\tl \phi' %
\right)\\
& = \ellcc^2\left(   N^{-2}\d\tl r^2 %
+ N^2  \d \tl t ^2 + \tl r^2 \Big( \frac{J}{2\tl r^2} \d \tl t +  \d \tl \phi'\Big)^2 %
\right).
\end{align}
\end{subequations}
This is the Euclidean BTZ black hole metric. 
Schwarzschild-like coordinates (see below) are obtained by letting the angular variable spin appropriately:
\be
\tl \phi' = \tl \phi + \Omega \tl t ,
\qquad\text{where}\qquad
\Omega = -\frac{i \tl r_-}{\tl r_+}.
\ee
Hence,
\be
\d s^2  = \ellcc^2\left(   N^{-2}\d\tl r^2 %
+ N^2  \d \tl t ^2 + \tl r^2 \left( \Big(\frac{J}{2\tl r^2} + \Omega\Big)\d \tl t +  \d \tl \phi\right)^2 %
\right).
\ee
As usual, the periodicity of $\tl t$---coinciding with the black hole temperature---is calculated by invoking the absence of conical singularities at the horizon%
\footnote{Recall this is the minimal value $\tl r$ can take on the chosen component of the hyperboloid, $u\geq1$.}
$\tl r = \tl r_+$ of the $(\tl r, \tl t)$ plane. 
Writing $\tl r = \tl r_+ + \epsilon^2 + \mathrm O(\epsilon^4)$, the metric becomes
\be
\d s^2_{(\tl r,\tl t\,)} = \frac{2 \ellcc^2 \tl r_+}{\tl r_+^2 - \tl r_-^2} \left( \d \epsilon^2  + \Big(\frac{\tl r_+^2 - \tl r_-^2}{ \tl r_+} \Big)^2\epsilon^2 \d \tl t^2  \right) + \mathrm O(\epsilon^4)
\ee
and thus
\be
\tl t \sim \tl t +  \ellcc^{-1}\tl \beta
\qquad\text{where}\qquad
\frac{\tl\beta}{\ellcc} = \frac{2\pi  \tl r_+}{\tl r_+^2 - \tl r_-^2}.
\ee
This coordinates are said to be Schwarzschild-like because $\tl \phi$ is not involved in the above identification. 
Furthermore, $\tl \phi$ is a regular  angular coordinate of period $2\pi$, as . (Notice, however, that circles of constant $(\tl r, \tl t)$ are non-contractible, since the length of their circumference is larger than $2\pi \tl r_+$.) 

In the BTZ (primed) coordinates introduced above, on the other hand, the above identification reads
\be
\left(\tl t + \ellcc^{-1}\tl \beta, \tl \phi' + \tl \gamma \right)
\sim (\tl t , \tl \phi' ) 
\qquad\text{where}\qquad
\tl \gamma = \frac{\tl \beta \Omega}{\ellcc} = -\frac{2\pi i \tl r_-}{\tl r_+^2 - \tl r_-^2}.
\ee
Therefore, we see that the BTZ coordinates describe at fixed radius a torus of modulus
\be
\tau_\text{BTZ} = \frac{1}{2\pi}\Big( \tl \gamma + i \ellcc^{-1}\tl\beta \;\Big)
\ee
since
\be
\left(\tl r,  \tl \phi' + i \tl t +  2\pi \tau_\text{BTZ}\right)
\sim \left(\tl r,  \tl \phi' + i \tl t \right)
\sim \left(\tl r,  \tl \phi' + i \tl t +  2\pi \right) .
\ee
Back in the coordinates $(\tl T, \tl \Phi)$, and using
\be
  \tl T- i \tl \Phi 
=-i ( \tl r_+ +  \tl r_-) (   \tl \phi' + i \tl t) 
= \frac{\tl \phi' + i\tl t }{\tau_\text{BTZ}} ,
\ee
the identification is immediately found to read
\be
\left(  \tl T -i  \tl\Phi  + 2\pi \right)
\sim \left( \tl T -i  \tl\Phi  \right)
\sim \left(\tl T -i   \tl \Phi  + \frac{2\pi }{\tau_\text{BTZ}}\right).
\ee
And comparison with equation \eqref{eq_identifuvxy} gives
\be
\tau_\text{TTAdS} = - \frac{1}{\tau_\text{BTZ}}.
\ee
(Notice that to obtain this equation without a further conjugation, the angular coordinates $\phi$ and $\tl T$ had to be chosen with opposite orientation.)

\subsection{Thermal $\text{AdS}_3$}

Thermal $\text{AdS}_3$ is defined as 
\be
\d s^2 = \ellcc^2\Big( \cosh^2 r \; \d t^2  + \d r^2 + \sinh^2 r \; \d \phi^2 \Big),
\ee
with $r\in \mathbb R^+$, $\phi\in[0,2\pi]$, and $t\in[0,\beta/\ellcc]$. 
The slices at $t=0$ and $t=\beta/\ellcc$ are identified, possibly after the application of a twist $\phi\mapsto\phi+\gamma$. 

The on-shell Einsten--Hilbert action is infinite on thermal $\text{AdS}_3$, because of the infinite radial extension of this space. 
Therefore, we regularize it by inserting a boundary at $r=a$.
This gives
\be
\frac{1}{2\ellpl}\int \sqrt{g}\; \left(R + \frac{2}{\ellcc^2} \right) 
= - 2 \frac{ \text{Vol}(a)}{\ellpl \ellcc^2} 
= - \frac{\pi\beta}{\ellpl}\Big(\cosh(2a) - 1\Big).
\ee

The GHY boundary term is most easily calculated as follows:
\be
\frac{1}{\ellpl} \oint \sqrt{h} \; K = \frac{1}{\ellpl} \pp_n \oint \sqrt{h},
\ee
where $\pp_n$ stands for the normal derivative (this follows from $K_{ab} = \tfrac{1}{2}\pounds_n h_{ab}$). 
In the coordinates above, the area of the boundary is
\be
\text{Area}(a) = \pi \beta \ellcc \sinh(2a),
\ee
and thus
\be
\frac{1}{\ellpl} \oint \sqrt{h} \; K =  \frac{1}{\ellpl \ellcc} \frac{\pp}{\pp r}_{|r=a} \text{Area}(r) = \frac{2\pi \beta}{\ellpl}\cosh(2a).
\ee

Therefore, combining these results and taking the limit $a\to\infty$, one finds
\be
S_\text{GHY} = \frac{\pi \beta}{\ellpl}\Big( 1 - \cosh(2a) + 2 \cosh(2a) - \sinh(2a)\Big) \xrightarrow{a\to\infty} \frac{\pi \beta}{\ellpl},
\ee
and
\be
S_{\frac12\text{GHY}} = \frac{\pi \beta}{\ellpl}\Big( 1 - \cosh(2a) + \cosh(2a)\Big) = \frac{\pi \beta}{\ellpl}.
\ee
Thus, we denote
\be
S_\text{TTAdS} = 2\pi^2 \Im(\tau_\text{TTAdS})\frac{\ellcc}{\ellpl}.
\ee

\subsection{Euclidean BTZ black hole}

In Schwarzschild-like coordinates the Euclidean BTZ black hole \cite{BanadosTeitelboimZanelli1992,Banados:1992gq}  physical mass and angular momentum proportional to%
\footnote{The precise conversion factors will be given at the end of this section}
 $M$ and $J$, and with angular ``velocity'' $ \ellcc^{-1} \Omega$, is given by
\be
\d s^2 = \ellcc^2\Big( N^2 \d \tl t^2 + N^{-2} \d \tl r^2 + \tl r^2 (N_{\tl \phi} \d \tl t + \d \tl \phi )^2 \Big)
\label{eq_BTZmetric}
\ee
with%
\footnote{To pass to the Lorentzian BTZ black hole, not only the time becomes imaginary $t\mapsto it $, but also the angular momentum and velocity do $(J,\Omega) \mapsto (-i J, -i \Omega)$.  Also, for the Lorentzian metric to define a black-hole, one requires $J\leq M$ (as for the 3+1 Kerr metric).}
\be
N^2 = \tl r^2 - M  - \frac{J^2}{4\tl r^2}
\qquad\text{and}\qquad
N_{\tl \phi} = \frac{ J}{2 \tl r^2} +  \Omega.
\ee
Coordinates have the following range: $\tl r\in[\tl r_+, \infty]$, $\tl \phi\in[0,2\pi]$, and $\tl t\in [0,\ellcc^{-1} \tl\beta]$ with $\tl\phi$ of period $2\pi$.
Explicitly, in terms of $M$ and $J$
\be
\frac{\tl\beta}{\ellcc} = \frac{2\pi  \tl r_+}{\sqrt{M^2 + J^2}}
\qquad\text{and}\qquad
\tl r_+ = \left(\frac{\sqrt{M^2 + J^2} + M}{2}\right)^{1/2} 
\ee
Similarly $\Omega$ is fixed to be
\be
\tl\gamma = \frac{\tl\beta\Omega }{\ellcc}= \frac{- 2\pi i \tl r_-}{\sqrt{M^2 + J^2}}
\qquad\text{and}\qquad
\tl r_- = - i \left(\frac{\sqrt{M^2 + J^2} - M}{2}\right)^{1/2} 
\ee
Notice, however, that $\beta$ and $\gamma$ are the (thermodynamically) conjugate variables to $M$ and $J$, respectively. 
In this sense, their expressions in terms of $M$ and $J$ have to be understood as equilibrium conditions, corresponding to minima of the Euclidean action, i.e. of the free energy.

Again, we compute first the Einstein--Hilbert contribution up to $\tl r= \tl a$,
\be
\frac{1}{2\ellpl}\int \sqrt{g}\; \left(R + \frac{2}{\ellcc^2} \right) 
= - 2 \frac{ \text{Vol}(\tl a)}{\ellpl \ellcc^2} 
= -\frac{2 \pi \beta}{\ellpl} \Big( \tl a^2 - \tl r_+^2 \Big),
\ee
then the area of the spacetime boundary defined by $\tl r = \tl a$,
\be
\text{Area}(\tl a) = 2\pi \tl\beta \ellcc N(\tl a) \tl a,
\ee
and finally the GHY boundary term
\be
\frac{1}{\ellpl}\oint \sqrt{h} K = \frac{N}{\ellpl \ellcc} \frac{\pp}{\pp \tl r}_{|\tl r = \tl a} \text{Area}(\tl r) = \frac{2 \pi \tl \beta}{\ellpl} \Big( 2 \tl a^2 - M \Big).
\ee

From this, one finds
\be
S_\text{GHY} =  \frac{2 \pi \tl\beta}{\ellpl} \Big( \tl r_+^2  -  \tl a^2 + 2 \tl a^2 - M - N(\tl a) \tl a \Big) 
 \xrightarrow{a\to\infty} \frac{2\pi \tl\beta}{\ellpl}  \Big(\tl r^2_+   - \frac{1}{2} M \Big) 
=  \frac{2\pi^2 \tl r_+}{\ellpl},
\ee
and
\be
S_{\f12\text{GHY}} =  \frac{2 \pi \tl\beta}{\ellpl} \Big( \tl r_+^2  -  \tl a^2 +  \tl a^2 - \frac12 M \Big)
  = \frac{2\pi^2 \tl r_+}{\ellpl}.
\ee

As a function of $\beta$ and $\gamma$ alone, the above results, denoted $S_\text{BTZ}$, read
\be
S_\text{BTZ} = \frac{\pi \tl\beta}{\ellpl \left| \frac{1}{2\pi} \left(\tl\gamma + i \ellcc^{-1}\tl\beta \right) \right|^2} =  - 2\pi^2 \Im\left(\tau_\text{BTZ}^{-1}\right)\frac{\ellcc}{\ellpl}.
\ee
The fact that 
\be
S_\text{BTZ} = S_\text{TTAdS}
\ee
should not be too surprising, since these are nothing but the values of the gravitational on-shell actions associated to the same hyperboloid sheet.

Rewritten in terms of the physical mass and angular momentum%
\footnote{Being a mass times an angular velocity, the angular momentum $J_\text{phys}$ requires a Planck-scale conversion factor for the mass, like in $M_\text{phys}$, and a cosmological-scale conversion factor for the time, like in $\beta$. The factors of $\pi$ are due to our choice of ``geometrized'' Planck unit $\ellpl=8\pi G$ as well as to our will of keeping the above formulas uncluttered.}
\be
M_\text{phys} = \frac{\pi M}{\ellpl} 
\qquad\text{and}\qquad
J_\text{phys} =  \frac{\pi \ellcc J}{\ellpl} ,
\ee
the on-shell action takes the form of a Bekenstein--Hawking generalized free energy \acite{??}
\be
S_\text{BTZ} = F_\text{BH} = \frac{2\pi}{\ellpl}(2\pi \ellcc \tl r_+) - \tl \beta M_\text{phys} - \tl \gamma J_\text{phys},
\ee
where the first term---given by the radio of the horizon's ``area'' $ 2\pi \ellcc \tl r_+$ and $\ellpl/(2\pi) = 4G$---is the Bekenstein--Hawking entropy of the BTZ balck hole.

\section{Explicit example of regularization and gauge fixing\label{app_gaugefixing}}

In this appendix, we explicitly show how to gauge-fix the Ponzano-Regge model on the twisted cylinder, thereby getting a well-defined finite amplitude. We apply the gauge fixing procedure introduced by Freidel and Louapre in \cite{FreidelLouapre2004b}. It requires specifying a bulk maximal tree $T$ along the edges of the discretization, such that it touches the boundary only once. For each face dual to an edge in this tree, we will remove the ccorresponding $\delta$-distribution. We also need to specify  a maximal tree $T^*$ on the dual cellular decomposition. The group element carried by the edges of this dual tree will be gauge-fixed to the identity $\id$.

\begin{figure}[h!]
	\begin{center}
		\begin{tikzpicture}[scale=3.7]
			\coordinate (a1) at (-0.8,0.4);
			\coordinate (a2) at (-0.2,1);
			\coordinate (a3) at (0.8,0.2);
			\coordinate (o1) at (-0.08,0.6);
			\coordinate (b1) at (-0.8,-0.6);
			\coordinate (b2) at (-0.2,0);
			\coordinate (b3) at (0.8,-0.8);
			\coordinate (o2) at (-0.08,-0.4);
			
			\draw (a1)--(a2)--(a3)--(a1); 
			\draw (o1)--(a1); \draw (o1)--(a2); \draw (o1)--(a3);
			
			\draw[dotted, thick] (b1)--(b2)--(b3); \draw (b3)--(b1);
			\draw[dashed,red,very thick] (o2)--(b1); \draw[dotted,thick] (o2)--(b2); \draw[dotted,thick] (o2)--(b3);
			
			\draw (a1)--(b1); \draw[dotted,thick] (a2)--(b2); \draw (a3)--(b3);
			\draw[dashed,red,very thick] (o1)--(o2);
			
			\draw (a1)--++(0,0.2); \draw (a2)--++(0,0.2); \draw (a3)--++(0,0.2); \draw (o1)--++(0,0.2);
			\draw[dotted,thick] (b1)--++(0,-0.2); \draw[dotted,thick] (b2)--++(0,-0.2); \draw[dotted,thick] (b3)--++(0,-0.2); \draw[dotted,thick] (o2)--++(0,-0.2);
			
		\end{tikzpicture}
	\end{center}
	\caption{Discretization for the case where $N_x=3$, $N_t=2$ and $N_\gamma=0$. The bulk tree $T$ is drawn in red dashes.}
	\label{fig:torus_discretization_3_1}
\end{figure}
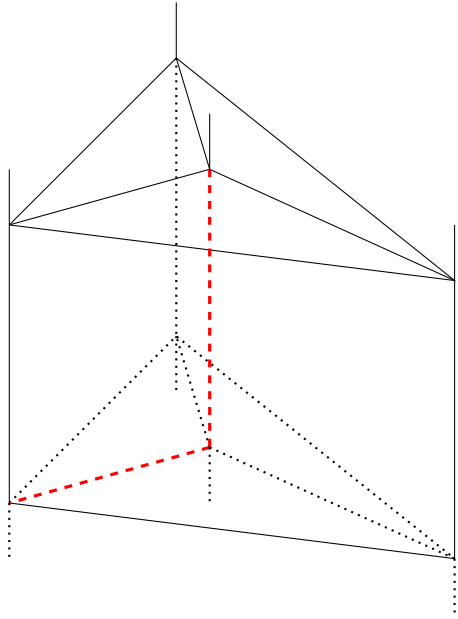

For the sake of simplicity, we look at the case with no twist, $N_\gamma = 0$ and consider a particular discretization of the torus: the case where $N_x = 3$ and $N_t = 2$ as represented on fig.\ref{fig:torus_discretization_3_1}.

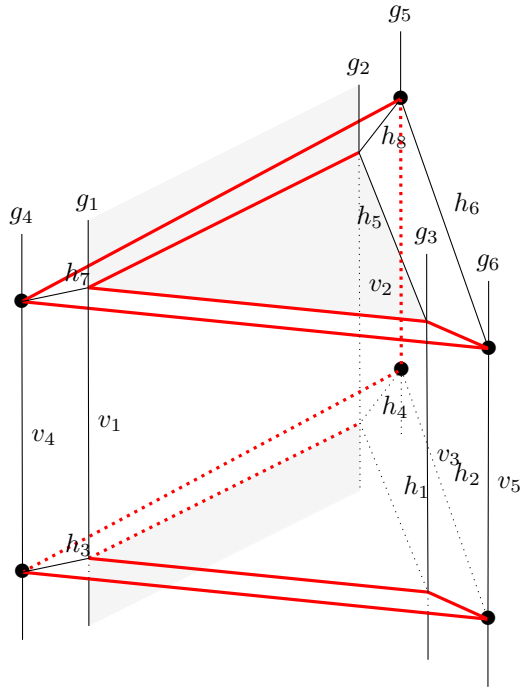
\begin{figure}[h!]
	\begin{center}
		\begin{tikzpicture}[scale=0.45]
			\coordinate (a1) at (7,6); \coordinate (A1) at (5.04,5.59);
			\coordinate (a2) at (15,10); \coordinate (A2) at (16.23,11.58);
			\coordinate (a3) at (17,5); \coordinate (A3) at (18.83,4.2);
			\coordinate (b1) at (7.02,-2); \coordinate (B1) at (5.06,-2.41); 
			\coordinate (b2) at (15.03,2); \coordinate (B2) at (16.25,3.58);
			\coordinate (b3) at (17.04,-3); \coordinate (B3) at (18.81,-3.8);
			
			\draw (A1) node[scale=1.4]{$\bullet$};\draw (A2) node[scale=1.4]{$\bullet$};\draw (A3) node[scale=1.4]{$\bullet$};\draw (B1) node[scale=1.4]{$\bullet$};\draw (B2) node[scale=1.4]{$\bullet$};\draw (B3) node[scale=1.4]{$\bullet$};
			
			\draw[draw opacity=0,fill=gray!8] (a1)--(a2)--(a3)--cycle;
			\draw[draw opacity=0,fill=gray!8] (7,8)--(a1)--(a2)--++(0,2);
			\draw[draw opacity=0,fill=gray!8] (7.02,-4)--(b1)--(b2)--++(0,-2);
			
			\draw[red,very thick] (a1)--(a2); \draw (a2)-- node[midway, above left]{$h_5$} (a3); \draw[red,very thick] (a3)--(a1); \draw[red,very thick] (A1)--(A2); \draw (A2)-- node[midway, above right]{$h_6$}(A3); \draw[red,very thick] (A3)--(A1);
			\draw (A1)--node[midway,above right]{$h_7$}(a1); \draw (A2)--node[pos=0.7, right]{$h_8$}(a2); \draw[red,very thick] (A3)--(a3);
			
			\draw[dotted,red,very thick] (b1)--(b2); \draw[dotted] (b2)-- node[midway, above right]{$h_1$} (b3); \draw[red,very thick] (b3)--(b1); \draw[dotted,red,very thick] (B1)--(B2); \draw[dotted] (B2) -- node[midway, above right]{$h_2$}(B3); \draw[red,very thick] (B3)--(B1); \draw (B1)-- node[midway, above right]{$h_3$} (b1); \draw[red,very thick] (B3)--(b3); \draw[dotted] (B2)-- node[pos = 0.7, right]{$h_4$} (b2);
			
			\draw (a1)--node[midway, right]{$v_1$}(b1); \draw[dotted] (a2)--node[midway, right]{$v_2$}(b2); \draw (a3)--node[midway, right]{$v_3$}(b3); \draw (A1)--node[midway, right]{$v_4$}(B1); \draw (A3)--node[midway, right]{$v_5$}(B3); \draw[dotted,red,very thick] (A2)--(B2);
			
			\draw (a1)--node[pos=1, above]{$g_1$}++(0,2); \draw (a2)--node[pos=1, above]{$g_2$}++(0,2); \draw (a3)--node[pos=1, above]{$g_3$}++(0,2); \draw (A1)--node[pos=1, above]{$g_4$}++(0,2); \draw (A2)--node[pos=1, above]{$g_5$}++(0,2); \draw (A3)--node[pos=1, above]{$g_6$}++(0,2); 
			\draw[dotted] (b1)--++(0,-0.65); \draw (7.02,-2.65)--++(0,-1.35); \draw (B1)--++(0,-2); \draw (B3)--++(0,-2); \draw[dotted] (b3)--++(0,-0.65); \draw (17.04,-3.65)--++(0,-1.35); \draw[dotted] (b2)--++(0,-2); \draw[dotted] (B2)--++(0,-2);

		\end{tikzpicture}
	\end{center}
	\caption{
We draw the dual cellular complex for the considered case with $N_x=3$, $N_t=2$ and $N_\gamma=0$.  The bullets represent  nodes of the boundary spin-network. The maximal tree $T^{*}$ is drawn in red: we integrate over a $\SU(2)$ group element for each edge not in $T^*$. The faces  dual to an edge in the bulk tree $T$ are represented in grey: the $\delta$-distribution corresponding to these faces are removed from the partition function. This gauge-fixing procedure leads to a well-defined finite Ponzano-Regge amplitude for the cylinder.
\label{fig:torus_dual_discretization_3_1}}
\end{figure}

Now, we consider the dual of the discretization, see fig.\ref{fig:torus_dual_discretization_3_1}.  We also draw in red the tree $T^*$.The grey faces correspond to a face dual to an edge in $T$. We recall that the links are oriented. We denote by $h_i$ the group elements associated to the horizontal links, by $v_j$ the group elements for the vertical links in the first time slice, and by $g_k$ the group elements for the vertical links in the second time slice (which loop from top to bottom in fig.\ref{fig:torus_dual_discretization_3_1}).

We start by writing explicitly the partition function with all the $\delta$-distributions: 
\begin{equation}
	\begin{split}
		 \int_{\SU(2)}&dg_{1}dg_{2}dg_{3}dg_{4}dg_{5}dg_{6} dh_1 dh_2 dh_3 dh_4 dh_5 dh_6 dh_7 dh_8 dv_1 dv_2 dv_3 dv_4 dv_5 \\
			& \delta(h_3) \delta(h_1) \delta(h_1 h_4 h_2^{-1}) \delta(h_4 h_3^{-1}) \\
			& \delta(h_3 v_4 h_7^{-1} v_1^{-1}) \delta(v_1 v_2^{-1}) \delta(v_2 h_8) \delta(v_2 h_5^{-1} v_3^{-1} h_{1}) \delta(v_3 v_5^{-1}) \delta(v_1 v_3^{-1}) \\
			& \delta(h_7) \delta(h_7 h_8^{-1}) \delta(h_5 h_8 h_6^{-1}) \\
			& \delta(h_7 g_4 h_3^{-1} g_1^{-1}) \delta(h_8 g_5 h_4^{-1} g_2^{-1}) \delta(h_5 l_2 h_1^{-1}g_{3}^{-1}) \delta(g_6 g_3^{-1}) \delta(g_1 g_3^{-1}) \\
			&\Phi(h_{i'},g_{j'},v_{k'})
	\end{split}
\end{equation}
where the dashed indices $i',j',k'$ in the argument of the boundary wave function indicate group elements associated to the boundary links. 
The delta--functions imply that $h_3 = h_1 = h_7 = h_4 = h_2 = h_8 = v_2=v_1=v_4=v_3=v_5 = h_5=h_6 = \id$ and that all the $g_{j}$ and thus in particular the boundary $g_{j'}$ are equal to each other, thus leaving us in the end with the simple integral:
\begin{equation}
\begin{split}
\int_{\SU(2)} d g\, \Phi(g_{j'}=g,h_{i'}=\id,v_{k'}=\id)  \quad .
\end{split}
\end{equation}

\section{Spin Recoupling and 4-valent Intertwiner Basis}\label{app_spinintertwiner}

The irreducible unitary representations of the $\SU(2)$ Lie group are labeled by half-integers, called spins, $j\in\f12\N$. The corresponding representation space $V^j$ is $(2j+1)$-dimensional and spanned by basis vectors $|j,m\ra$ diagonalising the Casimir operator and the generator $J_{z}$:
\be
\vec{J}^2\,|j,m\ra
=\Big{[}J_{z}^2+\f12\big{(}J_{-}J_{+}+J_{+}J_{-}\big{)}\Big{]}\,|j,m\ra
=j(j+1)\,|j,m\ra
\,,\qquad
J_{z}\,|j,m\ra=m\,|j,m\ra
\,.
\ee
These states can be interpreted geometrically as quantized 3-vectors of length $j$. One can actually define coherent states, \`a la Perelomov, peaked on classical vectors wth minimal spread (e.g. \cite{Livine:2007vk}).

The spin $j$ representation is unitarily equivalent to its conjugate representation and the map is:
\be
\begin{array}{lclcl}
\varsigma: &&V^j& \rightarrow &\overline{V^j} \\
&& |j,m\ra &\mapsto &  D^j(\varsigma)|j,m\ra=(-1)^{j+m}\,|j,-m\ra
\end{array}
\ee
such that $\varsigma^2=(-1)^{2j}\,\id$. This maps applies to the representation of the $\SU(2)$ group elements:
\be
\overline{\la j,n|D^j(g)|j,m\ra}
=
\la j,n|D^j(\varsigma^{-1}) g \varsigma)| j,m\ra\,
\ee
 which is the expression for arbitrary spins of the matrix identity in the fundamental representation, for 2$\times$2 matrices:
\be
\eps^{-1}g\eps=\overline{g}
\,,\qquad
\eps=\varsigma_{(j=\f12)}=\mat{cc}{0& 1 \\ -1 &0}
\,,\qquad
\forall g =\mat{cc}{a & b \\ -\bar{b}&\bar{a}}\,\in\SU(2)
\,.
\ee
This map allows to define the bivalent intertwiners. This is equivalent to identifying the $\SU(2)$-invariant states in tensor products $V^{j_{1}}\otimes V^{j_{2}}$ of two spins. For such a state to exist, the two spins must be equal, $j_{1}=j_{2}=j$:
\be
|\omega_{j}\ra
=
\f1{\sqrt{2j+1}}\sum_{m}\varsigma|j,m\ra \otimes  |j,m\ra
=
\f1{\sqrt{2j+1}}\sum_{m}(-1)^{j-m}\,|j,m\ra \otimes  |j,-m\ra
\,\quad\in V^j\otimes V^j
\,,
\ee
\be
\begin{array}{lclcl}
\iota_{2}: &&  V^j \otimes V^j  & \rightarrow &\C \\
&& |j,m\ra\otimes |j,n\ra &\mapsto &
\la \omega_{j}\,|\,(j,m)(j,n)\ra
\,=\, 
\la j,m|D^j(\varsigma)|j,n\ra
\end{array}
\ee
The action of the $\su(2)$ generators vanishes on those states, $\vec{J}\,|\omega_{j}\ra\,=0$.

Trivalent intertwiners correspond to $\SU(2)$-invariant states in the tensor products of three spins, $V^{j_{1}}\otimes V^{j_{2}}\otimes V^{j_{3}}$. These only exist is the three spins satisfy triangular inequalities, $|j_{2}-j_{3}|\le j_{1}\le j_{2}+j_{3}$ or equivalently any of the other two circular sets of inequalities, and are then unique once the spins are given. The coefficients of those 3-valent intertwiners are the Wigner 3j-symbols, which are defined in terms of the Clebsh-Gordan coefficients or expicitly given as a sum of factorial factors by the Racah formula. These 3-valent intertwiners can be geometrically interpreted as quantized triangles, with the spins giving the three edge lengths.

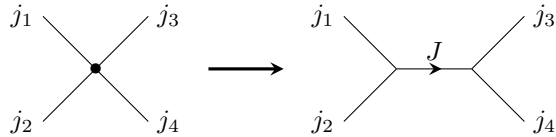
\begin{figure}[h!]
	\begin{center}
		\begin{tikzpicture}[scale=1]
		
\draw (-.7,.7)-- (0,0) node[pos=0,left]{$j_{1}$};
\draw (-.7,-.7)-- (0,0) node[pos=0,left]{$j_{2}$};
\draw (.7,.7)-- (0,0) node[pos=0,right]{$j_{3}$};
\draw (.7,-.7)-- (0,0) node[pos=0,right]{$j_{4}$};
\draw (0,0) node[scale=1]{$\bullet$};

\draw[very thick,->,>=stealth] (1.5,0)--(2.5,0);

\draw (-.7+4,.7)-- (0+4,0) node[pos=0,left]{$j_{1}$};
\draw (-.7+4,-.7)-- (0+4,0) node[pos=0,left]{$j_{2}$};
\draw[decoration={markings,mark=at position 0.6 with {\arrow[scale=1.5,>=stealth]{>}}},postaction={decorate}] (4,0) -- (5,0) node[midway,above]{$J$};
\draw (.7+5,.7)-- (0+5,0) node[pos=0,right]{$j_{3}$};
\draw (.7+5,-.7)-- (0+5,0) node[pos=0,right]{$j_{4}$};
		\end{tikzpicture}
	\end{center}
	\caption{We specify an intertwiner between the four spins $j_{1},..,j_{4}$ by pairing them two by two, say $j_{1}$ with $j_{2}$ and $j_{3}$ with $j_{4}$, and recoupling each pair of spins to an intermediate spin $J$. This defines a basis of 4-valent intertwiner states.}
	\label{4valent}
\end{figure}

Next, 4-valent intertwiners, or equivalently $\SU(2)$-invariant states in the tensor products of four spins, $V^{j_{1}}\otimes V^{j_{2}}\otimes V^{j_{3}}\otimes V^{j_{4}}$, can  be constructed from  3-valent intertwiners. One first chooses a pairing between the spins, say $(12)-(34)$ or we could have chosen $(13)-(24)$ or $(14)-(23)$. Then one recouples the two spins $j_{1}$ and $j_{2}$ to an intermediate spin $J$, also recouples the two other spins $j_{3}$ and $j_{4}$ to the same spin $J$ and finally glues the two 3-valent intertwiners together using the $\varsigma$ map, as illustrated on fig.\ref{4valent}. 
This provides us with the spin basis for 4-valent invariant states:
\be
|\iota^{J}_{(12)(34)}\ra
=
\sum_{m_{i},M} (-1)^{J+m}  |(j_{1}m_{1})(j_{2}m_{2})(j_{3}m_{3})(j_{4}m_{4})\ra 
\la (j_{1}m_{1})(j_{2}m_{2})|J,M \ra
\la (j_{3}m_{3})(j_{4}m_{4})|J,-M \ra
\,,
\ee
where the intermediate spin $J$ ranges from $\max(|j_{1}-j_{2}|,j_{3}-j_{4})$ to $\min(j_{1}+j_{2},j_{3}+j_{4})$, in order to satisfy the triangular inequalities.
One can extend this construction to $n$-valent intertwiners, defining a basis for $\SU(2)$-invariant states living in arbitrary tensor products of irreducible representations. Finally one can define coherent intertwiner states, with the semi-classical interpretation as quantized polygons or polyhedra \cite{Livine:2007vk,Bianchi:2010gc,Freidel:2013fia,Livine:2013tsa}.

\subsection{Intertwiners between spin $\f12$ representations}

Let us write explicitly the basis of 4-valent intertwiners between four fundamental representations, that is between four spins $j_{1}=j_{2}=j_{3}=j_{4}=\f12$. Considering the triangular inequalities,  the Hilbert space $\cH^{(4)}_{\f12}$ of $\SU(2)$-invariant states in the tensor product $\big{(}V^{\f12}\big{)}^{\otimes 4}$ is two-dimensional.

We will use the notations of spin up and spin down for the two states in $V^{\f12}$, 
\be
|\up\ra\,=\,\left|j=\tfrac12,m=\tfrac12\right\ra
\quad\text{and}\quad
|\down\ra\,=\,\left|j=\tfrac12,m=-\tfrac12\right\ra
 \,.\nn
 \ee
First, two spins $\f12$ can recouple to either a spin 0 (scalar) or a spin 1 (vector). The spin 0 state is given by the $\varsigma$ map, while the spin 1 states are given by the corresponding Clebsch-Gordan coefficients:
\be
|0\ra=\f1{\sqrt{2}}\,(|\up\down\ra-|\down\up\ra)
\,,\qquad
|1,+\ra=|\up\up\ra
\,,\quad
|1,0\ra=\f1{\sqrt{2}}\,(|\up\down\ra+|\down\up\ra)
\,,\quad
|1,-\ra=|\down\down\ra
\,.
\ee
Now we choose one pairing, between $(12)-(34)$ or $(13)-(24)$ or $(14)-(23)$, which we can respectively identify as the channels $s$, $t$ or $u$. Choosing one pairing, say starting with  $(12)-(34)$, we define a basis for the 4-valent intertwiners:
\be
|0\ra_s\equiv|0\ra_{(12)-(34)}
\equiv
|0\ra_{12}\otimes |0\ra_{34}
\equiv
\f12\Big{[}
|\up\down\up\down\ra-|\up\down\down\up\ra-|\down\up\up\down\ra+|\down\up\down\up\ra
\Big{]}\,,
\ee
\beq
|1\ra_{(12)-(34)}
&=&
\f1{\sqrt{3}}\Big{[}
|1,+\ra_{12}|1,-\ra_{34}
-|1,0\ra_{12}|1,0\ra_{34}
+|1,-\ra_{12}|1,+\ra_{34}
\Big{]}\\
&=&
\f1{\sqrt{3}}\Big{[}
|\up\up\down\down\ra+|\down\down\up\up\ra
-\f12\big{[}
|\up\down\up\down\ra+|\up\down\down\up\ra+|\down\up\up\down\ra+|\down\up\down\up\ra
\big{]}
\Big{]}
\nn
\eeq
One can similarly define the intertwiner basis corresponding to the other two channels:
\be
|0\ra_t\equiv|0\ra_{(13)-(24)}
=
\f12\Big{[}
|\up\up\down\down\ra-|\up\down\down\up\ra-|\down\up\up\down\ra+|\down\down\up\up\ra
\Big{]}\,,
\ee
\beq
|1\ra_{(13)-(24)}
&=&
\f1{\sqrt{3}}\Big{[}
|1,+\ra_{13}|1,-\ra_{24}
-|1,0\ra_{13}|1,0\ra_{24}
+|1,-\ra_{13}|1,+\ra_{24}
\Big{]}\\
&=&
\f1{\sqrt{3}}\Big{[}
|\up\down\up\down\ra+|\down\up\down\up\ra
-\f12\big{[}
|\up\up\down\down\ra+|\up\down\down\up\ra+|\down\up\up\down\ra+|\down\down\up\up\ra
\big{]}
\Big{]}
\nn
\eeq
\be
|0\ra_u\equiv|0\ra_{(14)-(23)}
=
\f12\Big{[}
|\up\up\down\down\ra-|\up\down\up\down\ra-|\down\up\down\up\ra+|\down\down\up\up\ra
\Big{]}\,,
\ee
\beq
|1\ra_{(14)-(23)}
&=&
\f1{\sqrt{3}}\Big{[}
|1,+\ra_{14}|1,-\ra_{23}
-|1,0\ra_{14}|1,0\ra_{23}
+|1,-\ra_{14}|1,+\ra_{23}
\Big{]}\\
&=&
\f1{\sqrt{3}}\Big{[}
|\up\down\down\up\ra+|\down\up\up\down\ra
-\f12\big{[}
|\up\up\down\down\ra+|\up\down\up\down\ra+|\down\up\down\up\ra+|\down\down\up\up\ra
\big{]}
\Big{]}
\nn
\eeq
From these explicit formulae, one can easily  compute the unitary matrices mapping one channel onto another. This is especially useful when looking at the volume operator \cite{Feller:2015yta}.
In the present work, in order to analyze the twisted torus partition function for spins $\f12$ on the boundary, we are more interested in the decomposition of the identity on the intertwiner space $\cH^{(4)}_{\f12}$ by the $0$-states in the three channels:
\be
\id=\f23
\big{[}
|0\ra_{s}{}_{s}\la 0|
+|0\ra_{t}{}_{t}\la 0|
+|0\ra_{u}{}_{u}\la 0|
\big{]}
\ee
To prove this decomposition of the identity, it is enough to apply it to the orthonormal basis $|0\ra_{s},|1\ra_{s}$. Let us nevertheless not forget that the three states $|0\ra_{s},|0\ra_{t},|0\ra_{u}$ are not independent and form an over-complete basis:
\be
|0\ra_{u}
=
|0\ra_{t}-|0\ra_{s}
\,.
\ee

\subsection{Intertwiners between spin $1$ representations}
\label{app:spinone}

In order to study the twisted torus partition function for boundary spin network states with spins $1$ on all the links, we similarly analyze the Hilbert space $\cH^{(4)}_{1}$ for 4-valent intertwiner between four spin $1$ representations. Since two spins 1 can recouple to either a spin 0 or a spin 1 or a spin 2, this intertwiner space is three-dimensional.

We will use the notation $|+\ra,|0\ra,|-\ra$ for the three basis states  in $V^1$. The recoupling of two spins is given by:
\be
|\varnothing\ra=|J=0\ra=
\f1{\sqrt{3}}\big{(}
|+-\ra-|00\ra+|-+\ra
\big{)}
\ee
\be
|v_{1}\ra=|J=1,M=1\ra
=\f1{\sqrt{2}}(|+0\ra-|0+\ra)
\,,\quad
|v_{0}\ra
=\f1{\sqrt{2}}(|+-\ra-|-+\ra)
\,,\quad
|v_{-1}\ra
=\f1{\sqrt{2}}(|0-\ra-|-0\ra)
\nn
\ee
\beq
&|t_{2}\ra=|J=2,M=2\ra=|++\ra
\,,\quad
|t_{1}\ra=\f1{\sqrt{2}}(|+0\ra+|0+\ra)
\,,\quad
|t_{0}\ra=\f1{\sqrt{6}}(|+-\ra+|-+\ra+2|00\ra)
\,,&
\nn\\
&|t_{-1}\ra=\f1{\sqrt{2}}(|-0\ra+|0-\ra)
\,,\quad
|t_{-2}\ra=|--\ra, \nn&
\eeq
where $v$ stands for vector and $t$ for tensor.
This is the recoupling basis for the 9-dimensional tensor product $V^1\otimes V^1$.

This allows us to define the basis of 4-valent intertwiners. Choosing the spin pairing $(12)-(34)$, we get:
\be
|\iota^0\ra_{(12)(34)}
=
|\varnothing\ra_{12}\otimes|\varnothing\ra_{34}
\,,
\ee

\be
|\iota^1\ra_{(12)(34)}
=
\frac{1}{\sqrt{3}}\Big{[}
|v_{1}\ra_{12}|v_{-1}\ra_{34}
-|v_{0}\ra_{12}|v_{0}\ra_{34}
+|v_{-1}\ra_{12}|v_{1}\ra_{34}
\Big{]}
\,,
\ee
\be
|\iota^2\ra_{(12)(34)}
=
\frac{1}{\sqrt{5}}\Big{[}
|t_{2}\ra_{12}|t_{-2}\ra_{34}
-|t_{1}\ra_{12}|t_{-1}\ra_{34}
+|t_{0}\ra_{12}|t_{0}\ra_{34}
-|t_{-1}\ra_{12}|t_{1}\ra_{34}
+|t_{-2}\ra_{12}|t_{2}\ra_{34}
\Big{]}
\,,
\ee
Using these explicit formulae, one can compute all the relevant scalar products between intertwiner basis states in the three recoupling channels.

What will be interesting for our present work is simply to remark that the three  states $\iota^{s|0},\iota^{t|0},\iota^{u|0}$, with intermediate spin $J=0$ in the three different channels, form a basis of the 4-valent intertwiner space $\cH^{(4)}_{1}$.

\section{Exact computation of the partition function for the 0-spin intertwiner in the s-channel}
\label{app_partitionfunction}

In this appendix, we focus on the exact computation of the Ponzano-Regge partition function given in \eqref{PRformula} for a boundary spin network state on the square lattice, with a homogeneous spin $j$ and the 0-spin intertwiner in the s-channel at every vertex:
\be
\la \text{PR} | \Phi_{j,\iota^{s|0}}  \ra
=
\frac{d_{j}^{N_{t}}}{d_j^{N_t N_x}}
\int_{\SU(2)} \d g \; \chi_{j}(g^{p})^{K}
= 
\frac{1}{d_j^{N_t (N_x-1)}}
\frac{2}{\pi} \int_{0}^{\pi} \d\theta\; \sin^2(\theta) \; \chi_{j}(p \theta)^K
\,,
\nn
\ee
Expanding the character in the $m$-basis, we express this integral over a random walk counting:
\be
\la \text{PR} | \Phi_{j,\iota^{s|0}}  \ra
=
\frac{1}{4\pi d_j^{N_t (N_x-1)}}
\int_{0}^{2\pi} {d}\theta\,
(2-\E^{2\I\theta}-\E^{-2\I\theta})
\sum_{m_{1},..,m_{K}} \E^{2\I\sum_{k=1}^{K}m_{k}p\theta}
\,,
\ee
with the $m$'s running in integer steps from $-j$ to $+j$. We focus on the generic case for $p\ge 3$.  In that case, among the three terms coming from the measure factor $\sin^2\theta$, only the constant term contributes to a non-trivial random walk counting while the other two terms, in $\E^{2\I\theta}$ and $\E^{-2\I\theta}$, give vanishing integrals.
Thus, assuming $p\ge 3$, we see that the dependance on $p$ actually drops out and the renormalized Ponzano-Regge amplitude $d_j^{N_t (N_x-1)}\,\la \text{PR} | \Phi_{j,\iota^{s|0}}  \ra$  is always equal to the following integral:
\begin{equation}
	\cI_{K} = \frac{1}{\pi} \int_{0}^{\pi} \text{d}\theta \chi^{j}\left(\theta\right)^{K} = \frac{1}{\pi} \int_{0}^{\pi} \text{d}\theta \left(\frac{\sin(d_j \theta)}{\sin\theta}\right)^{K}
	\,.
\end{equation}

Using the binomial formula and the following series expansion
\begin{equation}
	\frac{1}{(1-x)^{K}} = \sum_{n=0}^{\infty} \binom{n+K-1}{K-1}x^{n},
\end{equation}
we expand the both sines in the numerator and  in the denominator:
\begin{equation}
	\cI_{K} = \frac{1}{\pi} \sum_{k=0}^{K} \sum_{n=0}^{\infty}
	(-1)^{k}\binom{K}{k} \binom{n+K-1}{K-1}  \int_{0}^{\pi} \d\theta\; \E^{\I\Big[(K-2k)d_j -K + 2n\Big]\theta}.
	\label{eq:I_expression_1}
\end{equation}
where the two binomial coefficients are given by
\begin{equation}
	\begin{split}
			\binom{K}{k} \binom{n+K-1}{K-1} &= \frac{K}{k!(K-k)!}(n+1)(n+2)....(n+K-1) \\
			&= \frac{K}{2^{K-1}k!(K-k)!}(2n+2)(2n+4)....(2n+2K-2).
	\end{split}
	\label{eq:I_binomial_expression}
\end{equation}
The integration over the exponential is straightforward and gives a Kronecker delta:
\begin{equation}
	\int_{0}^{\pi} \d\theta\; \E^{\I \Big[(K-2k)d_j -K + 2n\Big]\theta} = \pi \delta_{0,(K-2k)d_j -K -2n},
\end{equation}
which translates into the constraint for $n$:
\begin{equation}
	2n = (K-2k)d_j -K.
	\label{eq:I_n_constraint}
\end{equation}
Plugging this into  equation \eqref{eq:I_expression_1} leads to an expression of the integral $\cI_{K}$ as a sum: 
\begin{equation}
	\cI_{K}=\frac{K}{2^{K-1}} \sum_{k=0}^{[K/2]} \frac{(-1)^{k}}{k!(K-k)!} \prod_{m=1}^{K-1} \big[(K-2k)d_j + (2m-K)\big].
\end{equation}
It is natural to write this formula as a polynomial in $d_j$ of the form
\begin{equation}
	\cI_{K} = \frac{K}{2^{K-1}} \sum_{m=0}^{K-1} A_{m} d_j^{m},
\end{equation}
where  $A_{m}$ are polynomials in $K$. This coefficients can be computed for $m<K-1$ as:
\begin{equation}
	A_{m}
	= \sum_{k=0}^{K-1} \frac{(-1)^{k}}{k!(K-k)!} (K-2k)^{m} \sum_{n_1<n_2<...<n_{K-1-m}}\prod_{i=1}^{K-1-m}(2n_i-K)
	\,.
\end{equation}
The highest order coefficient $A_{K-1}$ is
\begin{equation}
	\frac{1}{2^{K-1}} A_{K-1} = \frac{1}{2^{K-1}} \sum_{k=0}^{K} (-1)^{k} \frac{(K-2k)^{K-1}}{k! (K-k)!}.
\end{equation}

These are in fact  known numbers, appearing in the Fourier series of powers of the cardinal sine $\frac{1}{n!}\Big(\frac{\sin(\theta)}{\theta}\Big)^{n}$. They can be identified as the value of the integral  $\frac{1}{\pi}\int_{0}^{\infty}\d x\Big(\frac{\sin(x)}{x}\Big)^k$ in \cite{SWANEPOEL2015} or as the coefficients of the Duflo map for $\SU(2)$ in \cite{FreidelMajid2008}. We will refer to them as the Freidel-Majid numbers $C_{n,s}$ after this latter work, defined by:
\begin{equation}
	C_{n,s} = \frac{1}{2^n} \sum_{k=0}^{n} (-1)^{k} \frac{(n-2k)^{s}}{k!(n-k)!}.
\end{equation}

All the coefficients $A_{K-2n}$, for $n\in\N$ vanish and we can focus on the terms  $A_{K-(2n+1)}$. They are expressed in terms of the Freidel-Majid numbers:
\begin{equation}
	\frac{1}{2^{K-1}} A_{K-(2n+1)} = 2C_{K,K-(2n+1)} \times \sum_{n_1<n_2<...<n_{K-1-m}}\prod_{i=1}^{K-1-m}(2n_i-K)
\end{equation}
We do not have an explicit closed formula for the remaining sums. An order by order investigation shows that they are proportional to a Pochhammer coefficient (defined as $(N)_{(n)} = N(N-1)...(N-n+1)$) times a polynomial of degree $(n-1)$, for instance:
\begin{equation}
	\begin{split}
		\frac{1}{2^{K-1}} A_{K-1} &= 2 C_{K,K-1} \,,\\
		\frac{1}{2^{K-1}} A_{K-3} &= -\frac{1}{6} (K)_{(3)} 2 C_{K,K-3} \,,\\
		\frac{1}{2^{K-1}} A_{K-5} &= \frac{1}{360} (K)_{(5)} (5K+2) 2 C_{K,K-5}\,.
	\end{split}
\end{equation}

\section{Computation of the partition function for the spin 1/2 case}
\label{App_onehalf}

\subsection{The no-twist case}
\label{app_onehalf}

Here we compute the PR partition function for a boundary spin network state with homogeneous spin $j=\f12$ and arbitrary intertwiner at the four--valent nodes, in the no-twist case. The background is explained in section \ref{spinhalf}.

Consider  a single time slice, $N_{t}=1$ and no twist, $N_{\gamma}=0$. The trivial intertwiner configuration consists in only s-channel intertwiners. The partition function then sums over all possible u-channel intertwiner insertion along the time slice. For $k$ insertions of u-channel intertwiners, with $k$ between 0 and $N_{x}$, let us write $0\le x_{1}<..<x_{k}\le N_{x}-1$ for the position of those intertwiners. As one can see on fig.\ref{oneslice}, if we ignore the decoupled loops corresponding to the s-channel intertwiners, we have a single loop going through all the u-channel intertwiner insertions, thus leading to a simple expression for the partition function:
\begin{equation}
\la \text{PR} | \Phi_{j = \f12,\iota[\lambda,\rho]}  \ra
=
\sum_{k=0}^{N_x} \binom{N_{x}}{k} \lambda^{N_x-k}\rho^k 
\int_{\SU(2)} \d g\,
\chi_{\f12}(g)^{N_{x}-k}\chi_{\f12}(g^{k})
\,.
\end{equation}
This can be computed exactly by expanding the character as $\chi_{1/2}(\theta)=(\E^{\I\theta}+\E^{-\I\theta})$. We first evaluate the integrals, which are non-vanishing if and only if $N_{x}$ is even. Writing $N_{x}=2M$ with $M\in\N$, we get:
\begin{eqnarray}
&&\int_{\SU(2)} \d g\,
\chi_{\f12}(g)^{N_{x}-k}\chi_{\f12}(g^{k})\\
&&=
\f1{2\pi}\int_{0}^{2\pi}\d\theta\,
\left(1-\f{\E^{2\I\theta}+\E^{-2\I\theta}}2\right)\,
(\E^{\I k\theta}+\E^{-\I k\theta})
\sum_{n=0}^{2M-k}\binom{2M-k}{n}\E^{\I(2M-k-2n)\theta}
\nn\\
&&=
\binom{2M-k}{M}\,
\f{\big{[}2(M+1)-k(k+1)\big{]}}{(M+1)(M-k+1)}
\nn
\end{eqnarray}
Plugging this back into the sum, we get the Ponzano-Regge amplitude:
\begin{eqnarray}
\la \text{PR} | \Phi_{j = \f12,\iota[\lambda,\rho]}  \ra
&=&
\sum_{k=0}^{M+1}
\lambda^{2M-k}\rho^k 
\,
\f{(2M)!}{k!(M+1)!(M-k+1)!}
\,
\bigg{[}2(M+1)-k(k+1)\bigg{]}
\nn\\
&=&
\f2{M+1}\binom{2M}{M}
\lambda^{M-1}(\lambda+\rho)^{M-1}
\bigg{[}\lambda^{2}+\lambda\rho-\f M2\rho^{2}\bigg{]}
\,.
\end{eqnarray}

\subsection{The Haar intertwiner in terms of the symmetric group}
\label{app:permutations}

We are interested in computing the group averaging over $\SU(2)$ of the product of $\SU(2)$ matrix elements in the fundamental spin-$\f12$ representation:
\be
\int_{\SU(2)} \d g\,\prod_{i=1}^{2M}D^{\f12}_{a_{i}b_{i}}(g)\,,
\nn
\ee
which is relevant to the calculation of the Ponzano--Regge partition function for a spin-$\f12$ boundary spin network state as presented in section \ref{spinhalf}.

As it is well-known, we can understand the theory of the representations of $\SU(2)$ and their recoupling (tensor product and decomposition into irreducible representations) in terms of the representations of the symmetric groups using the Young tableaux tools (see e.g. \cite{Livine:2005mw}). Here we are interested in the representation of the symmetric group of $2M$ elements corresponding to the integer partition $2M=M+M$. We consider the Young frame, made of two horizontal rows with $M$ boxes. We consider all possible Young tableaux fitting in that frame, i.e. assignments of integers between 1 and $2M$ to the frame boxes such that numbers are always increasing to the right and downwards, as in the tables below. These Young tableaux are in one-to-one correspondence with basis states of the space of intertwiners between $2M$ spin-$\f12$ representations. The number of  such Young tableaux is the Catalan number $C_{M}=\f1{M+1}\binom{2M}{M}$. We label them $[\nu]$.

\noindent
For $M=1$, we get a single Young tableau corresponding to the unique bivalent intertwiner:
\begin{tabular}{|c|}
\hline
1 \\ 
\hline
2\\
\hline
\end{tabular}

\noindent
For $M=2$, we get two Young tableaux corresponding to the two 4-valent intertwiner states in $\mathrm{Inv}[(V^{\f12})^{4}]$:
\begin{tabular}{|c|c|}
\hline
1 & 2 \\ 
\hline 3 & 4\\
\hline
\end{tabular}
and
\begin{tabular}{|c|c|}
\hline
1 & 3 \\ 
\hline 2 & 4\\
\hline
\end{tabular}

\noindent
For $M=3$, we get five Young tableaux corresponding to the  6-valent intertwiner basis of $\mathrm{Inv}[(V^{\f12})^{6}]$:

\begin{center}
\begin{tabular}{|c|c|c|}
\hline
1 & 2 & 3\\ 
\hline
4 & 5 & 6\\
\hline
\end{tabular}
\hspace*{10mm}
\begin{tabular}{|c|c|c|}
\hline
1 & 3 & 5\\ 
\hline
2 & 4 & 6\\
\hline
\end{tabular}
\hspace*{10mm}
\begin{tabular}{|c|c|c|}
\hline
1 & 2 & 4\\ 
\hline
3 & 5 & 6\\
\hline
\end{tabular}
\hspace*{10mm}
\begin{tabular}{|c|c|c|}
\hline
1 & 3 & 4\\ 
\hline
2 & 5 & 6\\
\hline
\end{tabular}
\hspace*{10mm}
\begin{tabular}{|c|c|c|}
\hline
1 & 2 & 5\\ 
\hline
3 & 4 & 6\\
\hline
\end{tabular}

\end{center}

There is a simple prescription for the matrix elements of the basic $(2M-1)$ permutations $T_{i,i+1}$, which form a basis of the symmetric group. For each Young tableaux, we define the {\it axial distance} $\ell$ between two numbers, here $i$ and $i+1$, as the number of boxes going from one to the other, by counting $+1$ when going right or upwards and counting $-1$ when going left or downwards. Then the matrix elements of  $T_{i,i+1}$ are:
\be
\la [\nu]\,|\,D^{[M,M]}(T_{i,i+1})\,|\,[\nu]\ra =\f1{\ell_{i,i+1}}
\,,\quad
\la [\nu']\,|\,D^{[M,M]}(T_{i,i+1})\,|\,[\nu]\ra =\delta_{ [\nu'], T_{i,i+1}\,[\nu]}\f{\sqrt{\ell^{2}-1}}{\ell}
\,.
\ee
From these, one can compute the matrices representing any permutation $\om$ and their characters $s_{[M,M]}(\om)$ and finally plug them in the formula for the Haar intertwiner:
\be
\int\d g\,\prod_{i=1}^{2M}D^{\f12}_{a_{i}b_{i}}(g)
=
\f{M!}{(2M)!}\sum_{\om\in S_{2M}} s_{[M,M]}[\om] \prod_{i=0}^{2M-1}\delta_{a_{i}b_{\om(i)}}
\,.
\ee
A first remark is that this description of the matrix representation of the permutations works on any partition of $2M$ and its corresponding Young frame. A second remark is that we get a whole lot of interesting sums, using arbitrary representations defined by its corresponding partition $\cP$:
\be
\sum_{\om\in S_{2M}} s_{\cP}[\om] \prod_{i=0}^{2M-1}\delta_{a_{i}b_{\om(i)}}
\,.
\nn
\ee
All of these are $\SU(2)$-covariant and define various $\SU(2)$-invariant projectors. For instance, as explained in \cite{Livine:2005mw}, the integral with a $\SU(2)$-character insertion of spin $J$ corresponds to the partition $2M=(M+J)+(M-J)$:
\be
\int\d g\,\chi_{J}(g)\,\prod_{i=1}^{2M}D^{\f12}_{a_{i}b_{i}}(g)
\propto
\sum_{\om\in S_{2M}} s_{[M+J,M-J]}[\om] \prod_{i=0}^{2M-1}\delta_{a_{i}b_{\om(i)}}
\ee
where the character of the permutations can be computed using the same formulas as above.

We conclude this appendix with explicit formulas for the simplest cases $M=1$ and $M=2$. We compute for two matrix elements, in the $M=1$ case:
\be
\int\d g\,D^{\f12}_{a_{1}b_{1}}(g)\,D^{\f12}_{a_{2}b_{2}}(g)
=
\f12\big{[}
\delta_{a_{1}b_{1}}\delta_{a_{2}b_{2}}-\delta_{a_{1}b_{2}}\delta_{a_{2}b_{1}}
\big{]}
=
\f12\eps_{a_{1}a_{2}}\eps_{b_{1}b_{2}}\,.
\ee
In the $M=2$ case, we represent all the permutations as 2$\times$2 matrices and compute their traces, and we obtain:
\be
\int\d g\,\prod_{i=1}^{4}D^{\f12}_{a_{i}b_{i}}(g)
=
\f1{12}\Big{[}
2\delta^{\id}
+2\big{[}
\delta^{(12)(34)}+\delta^{(13)(24)}+\delta^{(14)(23)}
\big{]}
-\big{[}
\delta^{(123)}+\delta^{(124)}+\delta^{(134)}+\delta^{(234)}
\big{]}
\Big{]}
\,,
\ee
with $\delta^{\id}$ the direct identification of the $a_{i}$ with the $b_{i}$ and the following notations for the double permutations and 3-cycles:
\be
\delta^{(12)(34)}=
\delta_{a_{1}b_{2}}\delta_{a_{2}b_{1}}\delta_{a_{3}b_{4}}\delta_{a_{4}b_{3}}
\,,
\quad
\delta^{(123)}=
\big{[}
\delta_{a_{1}b_{2}}\delta_{a_{2}b_{3}}\delta_{a_{3}b_{1}}
+\delta_{a_{1}b_{3}}\delta_{a_{2}b_{1}}\delta_{a_{3}b_{2}}
\big{]}
\,\delta_{a_{4}b_{4}}
\,.
\nn
\ee
We could also compute  the integral $\int dg\,D^{\f12}(g)^{\otimes 4}$ by decomposing it on the 0-spin intertwiners in the s,t,u channels, obtaining:
\beq
\int\d g\,\prod_{i=1}^{4}D^{\f12}_{a_{i}b_{i}}(g)
&=&
\f1{6}
\big{[}
\eps_{a_{1}a_{2}}\eps_{a_{3}a_{4}}\eps_{b_{1}b_{2}}\eps_{b_{3}b_{4}}+\dots 
\big{]}\\
&=&
\f1{6}\Big{[}
3\delta^{\id}
+\big{[}
\delta^{(12)(34)}+\delta^{(13)(24)}+\delta^{(14)(23)}
\big{]}
-\big{[}\delta^{(12)}+\dots\big{]}
\Big{]}
\nn
\eeq
with terms corresponding to the identification up to a simple permutation, of the type:
$$
\delta^{(12)}=\delta_{a_{1}b_{2}}\delta_{a_{2}b_{1}}\delta_{a_{3}b_{3}}\delta_{a_{4}b_{4}}
\,.
$$
To prove the equality between the two expressions, we can use a very useful identity on bit circuits:
\be
\delta^{(123)}=\delta^{(12)}+\delta^{(13)}+\delta^{(23)}-\delta^{\id}\,.
\ee




\bibliographystyle{bibstyle_aldo}
\bibliography{PonzanoRegge}

\end{document}